\PassOptionsToPackage{unicode}{hyperref}
\PassOptionsToPackage{hyphens}{url}
\documentclass[
  11pt,
]{article}
\usepackage{amsmath,amssymb}
\usepackage{setspace}
\usepackage{iftex}
\ifPDFTeX
  \usepackage[T1]{fontenc}
  \usepackage[utf8]{inputenc}
  \usepackage{textcomp} 
\else 
  \usepackage{unicode-math} 
  \defaultfontfeatures{Scale=MatchLowercase}
  \defaultfontfeatures[\rmfamily]{Ligatures=TeX,Scale=1}
\fi
\usepackage{lmodern}
\ifPDFTeX\else
\fi
\IfFileExists{upquote.sty}{\usepackage{upquote}}{}
\IfFileExists{microtype.sty}{
  \usepackage[]{microtype}
  \UseMicrotypeSet[protrusion]{basicmath} 
}{}
\makeatletter
\@ifundefined{KOMAClassName}{
  \IfFileExists{parskip.sty}{%
    \usepackage{parskip}
  }{
    \setlength{\parindent}{0pt}
    \setlength{\parskip}{6pt plus 2pt minus 1pt}}
}{
  \KOMAoptions{parskip=half}}
\makeatother
\usepackage{xcolor}
\usepackage[margin=1in]{geometry}
\usepackage{longtable,booktabs,array}
\usepackage{calc} 
\usepackage{etoolbox}
\makeatletter
\patchcmd\longtable{\par}{\if@noskipsec\mbox{}\fi\par}{}{}
\makeatother
\IfFileExists{footnotehyper.sty}{\usepackage{footnotehyper}}{\usepackage{footnote}}
\makesavenoteenv{longtable}
\usepackage{graphicx}
\makeatletter
\def\maxwidth{\ifdim\Gin@nat@width>\linewidth\linewidth\else\Gin@nat@width\fi}
\def\maxheight{\ifdim\Gin@nat@height>\textheight\textheight\else\Gin@nat@height\fi}
\makeatother
\setkeys{Gin}{width=\maxwidth,height=\maxheight,keepaspectratio}
\makeatletter
\def\fps@figure{htbp}
\makeatother
\usepackage{amsmath}
\usepackage{booktabs}
\usepackage{enumitem}
\usepackage{siunitx}
\usepackage{threeparttable}
\usepackage{adjustbox}
\usepackage{graphicx}
\DeclareSymbolFont{yhlargesymbols}{OMX}{yhex}{m}{n}
\DeclareMathAccent{\yhwidehat}{\mathord}{yhlargesymbols}{"62}
\usepackage{newunicodechar}
\newunicodechar{−}{\ensuremath{-}}

\usepackage[T1]{fontenc}
\usepackage{lmodern}
\usepackage{xurl}
\PassOptionsToPackage{sort&compress}{natbib}
\makeatletter\AtBeginDocument{\setkeys{Gin}{width=!,height=!,keepaspectratio=false}}\makeatother
\ifLuaTeX
  \usepackage{selnolig}  
\fi
\usepackage[numbers]{natbib}
\IfFileExists{bookmark.sty}{\usepackage{bookmark}}{\usepackage{hyperref}}
\IfFileExists{xurl.sty}{\usepackage{xurl}}{} 
\hypersetup{
  pdftitle={Jeffreys-Type Penalized GEE for Correlated Binary Data with an Odds-Ratio Parameterization},
  hidelinks,
  pdfcreator={LaTeX via pandoc}}

\title{Jeffreys-Type Penalized GEE for Correlated Binary Data with an Odds-Ratio Parameterization}
\author{Anestis Touloumis\\
School of Architecture, Technology and Engineering, University of Brighton, Brighton, UK\\
\texttt{A.Touloumis@brighton.ac.uk} \(\cdot\) ORCID: \href{https://orcid.org/0000-0002-5965-1639}{0000-0002-5965-1639}}
\date{}

\begin{document}
\maketitle
\begin{abstract}
Generalized estimating equations (GEE) are widely used for population-averaged
inference on correlated binary responses, but ordinary GEE can fail under
separation, a situation that is more likely in small-sample, sparse, or
rare-event settings, leading to nonconvergence, infinite or extreme estimates,
and unreliable inference. Existing penalized GEE (PGEE) approaches mitigate
some of these problems but do not generally guarantee finite estimates under
nonindependence working structures and often rely on correlation-coefficient
parameterizations whose admissible range shrinks as fitted probabilities
approach zero or one, forcing the working association toward independence
under separation. We propose a PGEE framework that combines a Jeffreys-prior
penalty with marginalized odds-ratio working parameterizations. The odds-ratio
parameterization avoids this failure, while the penalty, with tunable strength
\(\delta\) and default \(\delta = 1/2\), stabilizes estimation under separation.
Under working independence, PGEE reduces to the Jeffreys-prior penalized
maximum-likelihood estimator, yielding finite estimates for logit, probit,
complementary log-log, and cauchit links. Under nonindependence odds-ratio
structures, where a formal finiteness guarantee is unavailable, PGEE achieves
near-complete empirical convergence even in separated settings. We also
propose one-step and hybrid variants, OPGEE and HPGEE, that reduce
computational cost. Simulations show that all three variants substantially
outperform ordinary GEE under separation while retaining the performance of
ordinary GEE in regular settings. We illustrate the method using a
respiratory-illness trial in which ordinary GEE fails, and provide an
implementation in the \textsf{R} package \texttt{geer}.

\end{abstract}

\setstretch{1.15}
\hypertarget{introduction}{%
\section{Introduction}\label{introduction}}

Repeated-measures and other clustered designs are common in biomedical and epidemiological research, inducing within-subject dependence that violates the independence assumption underlying traditional regression models. When the goal is population-averaged inference, generalized estimating equations (GEE) methodology is a standard tool that combines a marginal mean model with a working association structure indexed by the nuisance parameter vector \(\boldsymbol \alpha\) \citep{Liang1986}. Under standard regularity conditions, and even when the working association is misspecified, GEE yields consistent estimates of the regression parameters \(\boldsymbol \beta\) and of their standard errors, enabling Wald-type tests and confidence intervals for the regression parameters.

For binary responses, GEE can be numerically unreliable in sparse settings and may fail under separation \citep{albert1984, heinze2002}, a condition that becomes more likely when the number of covariates is large relative to the number of observed events, the model includes interaction terms among discrete covariates, or when clusters are small and outcomes are sparse. Separation arises when a covariate or linear combination of covariates perfectly predicts the response (complete separation) or does so for all observations on at least one side of a decision boundary (quasi-complete separation). In such regimes, ordinary GEE may fail to converge or may return extreme estimates and standard errors, yielding unreliable Wald-type inference even when convergence is reported \citep{mondol2019bias, Geroldinger2022, gosho2023}. These difficulties are analogous to those in maximum-likelihood (ML) estimation for binary generalized linear models (GLMs), where separation implies nonexistence of the ML estimator \citep{albert1984}. Jeffreys-type penalization, including Firth's method for logistic regression, is well known to yield finite estimates and improved numerical behavior in these settings \citep{firth1993, heinze2002, Kosmidis2018}.

In marginal models for correlated binary data, the connection between GEE and GLMs is direct under the independence working structure, since the GEE and ML estimators then coincide. Consequently, under complete separation the working-independence GEE estimator does not exist, and under quasi-complete separation it may also fail to exist or produce extreme estimates. This separation-driven numerical failure extends to nonindependence working structures in practice, because standard GEE implementations use GLM-based starting values and iterative updates that fail to converge or produce extreme iterates when fitted probabilities approach 0 or 1.

Recently proposed methods adapt Firth-type adjustments \citep{firth1993} to marginal logistic models, either through adjusted estimating equations \citep{mondol2019bias} or data augmentation \citep{Geroldinger2022}, but these approaches do not, in general, prevent this numerical failure from extending to nonindependence working structures. A systematic comparison of bias-corrected and penalized GEE approaches found that these methods continue to exhibit limitations in sparse or small-sample settings \citep{gosho2023}. Moreover, these proposals are primarily restricted to the logit link and rely on correlation-coefficient working parameterizations for the association structure, which are intrinsically constrained by the marginal probabilities via the Fréchet bounds and may yield inadmissible association estimates \(\widehat{\boldsymbol \alpha}\) \citep{Prentice_Correlated_1988, Chaganty2006}. As fitted probabilities approach 0 or 1 under separation or quasi-complete separation, the feasible range of the working correlation can collapse toward zero, effectively reverting to a working-independence GEE fit precisely when within-subject dependence may be strongest. This collapse is structural, arising directly from the Fréchet bounds, and it is not remedied by penalizing or augmenting the estimating equations.

To address these limitations, we build on \cite{Kosmidis2018} and propose a penalized GEE (PGEE) framework for marginal models with correlated binary responses, where the within-subject association is parameterized through continuity-corrected marginalized odds ratios computed from pooled \(2 \times 2\) tables across subjects \citep{haldane1956}. Unlike the odds-ratio parameterizations of Carey, Zeger and Diggle \citep{CAREY_Modelling_1993} and Lipsitz, Laird and Harrington \citep{LIPSITZ_Generalized_1991}, which estimate association through iterative updates that depend on the fitted marginal probabilities and can therefore fail numerically under separation, the pooled association estimates require no such updates and remain finite under separation and sparsity settings (see Section \ref{sec:or-association}). The pooled odds ratio targets a marginal summary of within-subject association, which is a natural estimand in a population-averaged analysis and consistent with the multinomial approach of Touloumis, Agresti and Kateri \citep{Touloumis2013}, of which the binary setting is a special case.

This paper is complementary to our work on adjusted estimating equations for bias reduction in GEE \citep{Touloumis2026bias}. That framework targets the leading finite-sample bias under standard regularity conditions, assuming that the estimating equations admit a well-behaved solution. Unlike Firth's correction for binary GLMs with the logit link, where the Jeffreys-prior penalty yields a well-characterized first-order bias reduction \citep{firth1993}, an analogous guarantee does not carry over to GEE. In Appendix C.2 of the Supporting Information, we provide a sufficient condition for the Jeffreys-type penalty to act as a first-order bias-reducing adjustment. However, satisfying this condition requires correct specification of the working association structure, which is incompatible with the robustness that motivates GEE and is unlikely to hold in practice for correlated binary data. The objective of PGEE is therefore existence and reliable estimation in the GEE framework under challenging situations such as separation, not bias reduction, and it should be interpreted accordingly rather than as a refinement of the approach of Mondol and Rahman \citep{mondol2019bias}.

We make three contributions. First, we develop a Jeffreys-type PGEE framework with penalty strength \(\delta \in (0,1]\) and a marginalized odds-ratio working association that remains well-defined under separation. Second, we establish finiteness under the independence working structure for common binary links and show that, under standard regularity conditions, PGEE is asymptotically equivalent to ordinary GEE. Although the finiteness result does not extend to PGEE under nonindependence working structures, our simulations suggest near-complete empirical convergence in the separated and sparse regimes that motivated the method. Third, we propose one-step and hybrid approximations and demonstrate their performance through simulations and a respiratory-illness trial. The methods are implemented in the \textsf{R} package \texttt{geer}.

The remainder of the paper is organized as follows. Section \ref{sec:method} presents GEE with marginalized odds-ratio working association structures, defines PGEE using a Jeffreys-type penalty, summarizes existence and finiteness properties, discusses one-step and hybrid approximations, and provides practical guidance and software notes. Section \ref{sec:sim} reports simulation results across separated and regular regimes. Section \ref{sec:application} analyzes a respiratory-illness trial in which ordinary GEE fails. Section \ref{sec:discussion} concludes with implications and future work. Technical derivations, implementation details, and extended numerical results are provided in the Supporting Information.

\hypertarget{sec:method}{%
\section{A Penalized GEE Framework with Odds-Ratio Association}\label{sec:method}}

\hypertarget{sec:gee}{%
\subsection{GEE Framework and Covariance Estimation}\label{sec:gee}}

For subject \(i\) (\(i = 1, \ldots, N\)), let \(Y_{ij} \in \{0, 1\}\) denote the binary response at time \(j\) (\(j = 1, \ldots, n_{i}\)) with success probability \(\pi_{ij} = \Pr \left(Y_{ij} = 1 \mid \mathbf x_{ij}\right)\), where \(\mathbf x_{ij}\) is the associated \(p\)-variate covariate vector. Define the response vector \(\mathbf Y_{i}= \left( Y_{i1}, \ldots, Y_{in_{i}} \right)^{\top}\), the corresponding probability vector \(\boldsymbol \pi_{i}= \left( \pi_{i1}, \ldots, \pi_{in_{i}} \right)^{\top}\), and the \(n_{i} \times p\) covariate matrix \(\mathbf X_{i}= \left( \mathbf{x}_{i1}, \ldots, \mathbf{x}_{in_{i}} \right)^{\top}\). Assume the marginal regression model \(g\left(\pi_{ij}\right) = \eta_{ij} = \mathbf x_{ij}^{\top} \boldsymbol \beta\), where \(g(\cdot)\) is a known link function and \(\boldsymbol \beta\) is the \(p\)-variate regression parameter vector. The GEE estimator \(\widehat{\boldsymbol \beta}_{G}\) solves
\begin{equation}
\mathbf U_{G}= \sum_{i=1}^{N}\mathbf U_{i} = \sum_{i=1}^{N}\mathbf D_{i}^{\top} \mathbf V^{-1}_{i}\mathbf S_{i}= \mathbf{0}_p,
\label{gee_equations}
\end{equation}
where \(\mathbf U_{i} = \mathbf D_{i}^{\top} \mathbf V^{-1}_{i}\mathbf S_{i}\), \(\mathbf S_{i}= \mathbf Y_{i}- \boldsymbol \pi_{i}\), and \(\mathbf D_{i}= \boldsymbol \Delta_{i}\mathbf X_{i}\). Here \(\boldsymbol \Delta_{i}\) is the \(n_{i} \times n_{i}\) diagonal matrix with \(j\)th diagonal element \(\partial \pi_{ij} / \partial \eta_{ij}\). The working covariance matrix \(\mathbf V_i\) is a symmetric \(n_{i} \times n_{i}\) matrix that approximates the true covariance matrix \(\operatorname{cov}\left( \mathbf Y_{i}\right)\). Let \(\pi^{\star}_{ijj^{\prime}} = \Pr(Y_{ij}=Y_{ij^{\prime}}=1 \mid \mathbf x_{ij}, \mathbf{x}_{ij^{\prime}})\) denote the joint success probability under a working assumption depending on \(\boldsymbol \beta\) and the nuisance parameter vector \(\boldsymbol \alpha\). The \((j,j^{\prime})\)th element of \(\mathbf V_i\) is
\begin{equation}
\left[ \mathbf V_i\right]_{jj^{\prime}}=
\begin{cases}
\pi_{ij} \left( 1 - \pi_{ij} \right) & \mbox{if } j = j^{\prime}, \\
\pi^{\star}_{ijj^{\prime}}-\pi_{ij}\pi_{ij^{\prime}} & \mbox{if } j \neq j^{\prime}.
\end{cases}
\label{true_covariance_elements}
\end{equation}
When \(\boldsymbol \alpha\) is unknown, \(\widehat{\boldsymbol \beta}_{G}\) is defined as the solution to \eqref{gee_equations} with \(\boldsymbol \alpha\) replaced by a \(\sqrt{N}\)-consistent estimator \(\widehat{\boldsymbol \alpha}\). The specific form of \(\pi^{\star}_{ijj^{\prime}}\) in terms of \(\boldsymbol \alpha\) under the marginalized odds-ratio parameterization is given in Section \ref{sec:or-association}.

Under standard regularity conditions, \(\widehat{\boldsymbol \beta}_{G}\) is asymptotically \(p\)-variate Normal with mean vector \(\boldsymbol \beta\) and covariance matrix \(\boldsymbol \Sigma_{G} = \boldsymbol \Sigma_{0}^{-1} \boldsymbol \Sigma_{1} \boldsymbol \Sigma_{0}^{-1}\), where \(\boldsymbol \Sigma_{0} = \sum_{i=1}^{N}\mathbf D_{i}^{\top} \mathbf V^{-1}_{i}\mathbf D_{i}\) and \(\boldsymbol \Sigma_{1} = \sum_{i=1}^{N}\mathbf D_{i}^{\top} \mathbf V^{-1}_{i}\operatorname{cov}\left( \mathbf Y_{i}\right)\mathbf V^{-1}_{i}\mathbf D_{i}\). If the association structure is correctly specified, that is \(\mathbf V_i= \operatorname{cov}\left( \mathbf Y_{i}\right)\) for all \(i\), then \(\boldsymbol \Sigma_{G}\) reduces to the model-based (naive) covariance matrix \(\boldsymbol \Sigma_{0}^{-1}\). A consistent estimator of \(\boldsymbol \Sigma_{G}\) is the sandwich (robust) covariance estimator
\begin{equation}
\widehat{\boldsymbol \Sigma}_{R} = \widehat{\boldsymbol \Sigma}_{0}^{-1} \widehat{\boldsymbol \Sigma}_{1} \widehat{\boldsymbol \Sigma}_{0}^{-1},
\label{sandwich}
\end{equation}
where \(\widehat{\boldsymbol \Sigma}_{0}\) and \(\widehat{\boldsymbol \Sigma}_{1}\) are computed by replacing \((\boldsymbol \beta, \boldsymbol \alpha)\) with \((\widehat{\boldsymbol \beta}_{G}, \widehat{\boldsymbol \alpha})\) and replacing \(\operatorname{cov}\left( \mathbf Y_{i}\right)\) with the empirical residual outer product \((\mathbf Y_{i}- \widehat{\boldsymbol \pi}_{i})(\mathbf Y_{i}- \widehat{\boldsymbol \pi}_{i})^{\top}\). In small samples, \(\widehat{\boldsymbol \Sigma}_{R}\) can underestimate the true variability of \(\widehat{\boldsymbol \beta}_{G}\). We therefore use the bias-corrected covariance estimator of Morel, Bokossa and Neerchal \citep{Morel_Small_2003},
\begin{equation}
\widehat{\boldsymbol \Sigma}_{M} = \frac{n^{\star} - 1}{n^{\star} - p}\frac{N}{N - 1} \widehat{\boldsymbol \Sigma}_{R} +
\lambda \xi \widehat{\boldsymbol \Sigma}_{0}^{-1},
\label{morel}
\end{equation}
where \(n^{\star} = \sum_{i=1}^{N}n_i\), \(\lambda = \min \{0.5, p/(N - p)\}\), and \(\xi = \max \left\{1, \operatorname{tr}\left(\widehat{\boldsymbol \Sigma}_{0}^{-1} \widehat{\boldsymbol \Sigma}_{1}\right)/p\right\}\). The additive correction \(\lambda \xi \widehat{\boldsymbol \Sigma}_{0}^{-1}\) ensures that \(\widehat{\boldsymbol \Sigma}_{M}\) remains positive definite whenever \(\widehat{\boldsymbol \Sigma}_{0}^{-1}\) is positive definite, a practically important safeguard in small-sample, separated, and sparse settings where \(\widehat{\boldsymbol \Sigma}_{R}\) may fail to be positive definite. Both \(\widehat{\boldsymbol \Sigma}_{R}\) and \(\widehat{\boldsymbol \Sigma}_{M}\) remain consistent when the working association structure is misspecified, which is why GEE inference is based on them rather than on the model-based estimator \(\widehat{\boldsymbol \Sigma}_{0}^{-1}\). Section \ref{sec:pgee} shows that this robustness carries over to PGEE and Section \ref{sec:sim-variance} compares the three estimators empirically.

\hypertarget{sec:or-association}{%
\subsection{Marginalized Odds-Ratio Parameterization of Working Association}\label{sec:or-association}}

To specify and estimate the working association, we adopt a pairwise odds-ratio parameterization \citep{CAREY_Modelling_1993, LIPSITZ_Generalized_1991} and follow the marginalized contingency-table approach of Touloumis, Agresti and Kateri \citep{Touloumis2013}. Let \(n=\max\{n_1,\ldots,n_N\}\) and \(C=n(n-1)/2\) denote the number of distinct time pairs. For each pair \((j,j^{\prime})\) with \(j<j^{\prime}\), we form a marginalized \(2\times2\) contingency table by pooling the binary responses across all subjects for whom both times \(j\) and \(j^{\prime}\) are observed, ignoring covariates. This pooling combines information across subjects and yields a marginal summary of association, consistent with the moment-based association estimation of Liang and Zeger \citep{Liang1986} and the multinomial approach of Touloumis, Agresti and Kateri \citep{Touloumis2013}.

The working association is characterized by \(\boldsymbol\alpha=\big(\alpha_{12},\alpha_{13},\ldots,\alpha_{1n},\ldots,\alpha_{(n-1)n}\big)^\top\), where \(\alpha_{jj^{\prime}}\) denotes the marginalized odds ratio for time pair \((j,j^{\prime})\). Given \((\boldsymbol \beta,\boldsymbol \alpha)\), the \((j,j^{\prime})\) element of \(\mathbf V_i\) is obtained from \eqref{true_covariance_elements}, where
\[
\pi^\star_{ijj^{\prime}}=
\begin{cases}
\dfrac{f_{ijj^{\prime}}-\sqrt{f_{ijj^{\prime}}^{2}-4\,\alpha_{jj^{\prime}}(\alpha_{jj^{\prime}}-1)\,\pi_{ij}\pi_{ij^{\prime}}}}{2(\alpha_{jj^{\prime}}-1)} & \text{if }\alpha_{jj^{\prime}}\neq 1,\\[8pt]
\pi_{ij}\pi_{ij^{\prime}} & \text{if }\alpha_{jj^{\prime}}=1,
\end{cases}
\]
and \(f_{ijj^{\prime}}=1-(1-\alpha_{jj^{\prime}})(\pi_{ij}+\pi_{ij^{\prime}})\).

We estimate each \(\alpha_{jj^{\prime}}\) by adding \(\zeta>0\) to each cell of the pooled \(2\times2\) table and computing the resulting sample odds ratio. By default, \(\zeta=0.5\) \citep{haldane1956}, which ensures finite estimates in the presence of zero cells. Because the pooled estimates do not require iterative mean-association updates, they remain finite under separation and sparsity.

The most general working structure (unstructured) uses all \(C\) estimated odds ratios, allowing association to vary across time pairs. The independence structure sets \(\boldsymbol \alpha=\mathbf 1_C\), requires no association estimation, and assumes working independence across all time pairs. Intermediate structures estimate fewer than \(C\) odds ratios. For example, an exchangeable odds-ratio structure assumes a common odds ratio \(\alpha\) across all time pairs, which we may estimate by the geometric mean,
\[
\widehat{\alpha}
=\exp\!\left\{\frac{2}{n(n-1)}
\sum_{j=1}^{n-1}\sum_{j^{\prime}=j+1}^{n}\log\!\bigl(\widehat{\alpha}_{jj^{\prime}}\bigr)\right\}.
\]

To choose a working structure, we recommend computing the log odds ratios for all time pairs using the default \(\zeta=0.5\), following the strategy of Touloumis, Agresti and Kateri \citep{Touloumis2013}. If their dispersion across time pairs is modest, an exchangeable odds-ratio structure is a reasonable parsimonious choice. Otherwise, an unstructured odds-ratio structure is preferred when the sample size is moderate to large and the pooled \(2\times2\) tables for the distinct time pairs are not excessively sparse.

\hypertarget{sec:pgee}{%
\subsection{Jeffreys-Type Penalized GEE}\label{sec:pgee}}

To adapt the penalized approach of Kosmidis and Firth \citep{Kosmidis2018} to the GEE framework, we treat equations \eqref{gee_equations} as the analog of likelihood score equations for \(\boldsymbol \beta\) given \(\boldsymbol \alpha\), following Paul and Zhang \citep{Paul_Small_2014}, Geroldinger et al. \citep{Geroldinger2022}, and Mondol and Rahman \citep{mondol2019bias}. For any matrix \(\mathbf{A}\) that depends on a vector \(\mathbf{a}\), write \(\mathcal{D}\mathbf{A}(\mathbf{a})=\partial\,\operatorname{vec}(\mathbf{A})/\partial \mathbf{a}^{\top}\), where \(\operatorname{vec}(\mathbf{A})\) stacks the columns of \(\mathbf{A}\). Then
\[
E \left[ - \mathcal{D}\mathbf U_{G}\left( \boldsymbol \beta\right) \right]
=
\sum_{i=1}^{N}\mathbf D_{i}^{\top} \mathbf V^{-1}_{i}\mathbf D_{i}
= 
\boldsymbol \Sigma_{0}
\]
serves as the GEE analog of the Fisher information. We consider the GEE analog of the power-Jeffreys class
\[
\mathcal{J}_{PG} \propto \left[\operatorname{det} \left( \boldsymbol \Sigma_{0} \right) \right]^{\delta},
\]
where \(\delta\in(0,1]\) controls the penalty strength, with \(\delta=1/2\) corresponding to the GEE analog of the Jeffreys prior. The contribution of the log-penalty to the GEE estimating equations is its gradient with respect to \(\boldsymbol \beta\), with \(\boldsymbol \alpha\) held fixed,
\[
\boldsymbol \Lambda_{PG}
=
\delta \left[ \mathcal{D}\log \left\{ \operatorname{det} \left( \boldsymbol \Sigma_{0} \right) \right\} (\boldsymbol \beta) \right]^{\top}
=
\delta \left[ \mathcal{D}\boldsymbol \Sigma_{0} \left( \boldsymbol \beta\right) \right]^{\top} \operatorname{vec}\left( {\boldsymbol \Sigma_{0}^{-1}} \right),
\]
where \(\mathcal{D}\boldsymbol \Sigma_{0}(\boldsymbol \beta)\) denotes the derivative of \(\boldsymbol \Sigma_{0}\) with respect to \(\boldsymbol \beta\). Equivalently, the \(r\)th element of \(\boldsymbol \Lambda_{PG}\) is \(\delta\operatorname{tr}(\boldsymbol \Sigma_{0}^{-1}\boldsymbol \Sigma_{0r}^{\star})\), where \(\boldsymbol \Sigma_{0r}^{\star}=\partial\boldsymbol \Sigma_{0}/\partial\beta_r\). An explicit expression for \(\mathcal{D}\boldsymbol \Sigma_{0}(\boldsymbol \beta)\) and the derivation of \(\boldsymbol \Lambda_{PG}\) are provided in Appendix B of the Supporting Information. For fixed \(\delta\), the proposed PGEE estimator \(\widehat{\boldsymbol \beta}_{PG}\) solves the adjusted estimating equations
\begin{equation}
\mathbf U_{PG} = \mathbf U_{G}+ \boldsymbol \Lambda_{PG} = \mathbf 0_{p}. 
\label{pgee}
\end{equation}
As in standard GEE methodology, when \(\boldsymbol \alpha\) is unknown, we replace it with a \(\sqrt{N}\)-consistent plug-in estimator, such as the marginalized odds-ratio estimator \(\widehat{\boldsymbol \alpha}\).

\hypertarget{sec:pgee-gee}{%
\subsubsection{Relation to Ordinary GEE}\label{sec:pgee-gee}}

PGEE changes only the estimating function for \(\boldsymbol \beta\): \(\mathbf U_{PG}\) in \eqref{pgee} is the GEE estimating function \(\mathbf U_{G}\) in \eqref{gee_equations} plus the adjustment \(\boldsymbol \Lambda_{PG}\), and \(\mathbf U_{PG}=\mathbf U_{G}\) when \(\delta=0\). The working covariance matrices \(\mathbf V_i\) are computed as in ordinary GEE, and the marginalized odds-ratio estimator \(\widehat{\boldsymbol \alpha}\) of Section \ref{sec:or-association} does not depend on \(\boldsymbol \beta\), so it is identical for GEE and PGEE. Under the independence working structure, \(\mathbf U_{G}\) is the score function of a binomial GLM and \(\boldsymbol \Sigma_{0}\) its Fisher information, so PGEE coincides with the Jeffreys-prior penalized maximum-likelihood estimator of Kosmidis and Firth \citep{Kosmidis2018} (Section \ref{sec:existence}) and, with the logit link and \(\delta=1/2\), with the bias-reduced estimator of Firth \citep{firth1993}.

\hypertarget{sec:pgee-mechanism}{%
\subsubsection{How the Penalty Operates}\label{sec:pgee-mechanism}}

Because \(\boldsymbol \Lambda_{PG}\) is the gradient of \(\delta\log\det(\boldsymbol \Sigma_{0})\) with respect to \(\boldsymbol \beta\), the adjustment pushes the estimates in the direction in which the GEE analog of the Fisher information increases. This is most transparent under working independence, where \(\boldsymbol \Sigma_{0}=\sum_{i=1}^{N}\mathbf X_{i}^{\top}\mathbf W_{i}\mathbf X_{i}\) and \(\mathbf W_{i}\) is the diagonal matrix with \(j\)th diagonal element
\[
w_{ij}=\frac{\left(\partial \pi_{ij}/\partial \eta_{ij}\right)^2}{\pi_{ij}(1-\pi_{ij})},
\]
which equals \(\pi_{ij}(1-\pi_{ij})\) for the logit link. By condition \eqref{condition_existence} in Section \ref{sec:existence}, \(w_{ij}\to 0\) as \(\pi_{ij}\) approaches 0 or 1, so observations whose fitted probabilities approach the boundary contribute little information and \(\det(\boldsymbol \Sigma_{0})\) decreases. The PGEE equations are then the score equations of the penalized log-likelihood \(\ell(\boldsymbol \beta)+\delta\log\det(\boldsymbol \Sigma_{0})\), where \(\ell(\boldsymbol \beta)\) is the binomial log-likelihood; because \(\ell(\boldsymbol \beta)\) is bounded above and the penalty tends to \(-\infty\) as \(\boldsymbol \beta\) diverges \citep{Kosmidis2018}, the penalty acts as a barrier that keeps the estimates finite. Under nonindependence working structures, \(\mathbf U_{G}\) is not in general the gradient of an objective function, so the barrier interpretation is only heuristic, but \(\boldsymbol \Lambda_{PG}\) remains the gradient of the same log-determinant and acts in the same direction. Larger values of \(\delta\) strengthen this effect, moving fitted probabilities further from 0 and 1.

\hypertarget{sec:pgee-asymptotics}{%
\subsubsection{Asymptotic Behavior}\label{sec:pgee-asymptotics}}

The penalty does not vanish as \(N \to \infty\), but it remains bounded, whereas the GEE estimating function grows with \(N\). Both \(\boldsymbol \Sigma_{0}\) and \(\boldsymbol \Sigma_{0r}^{\star}\) in \([\boldsymbol \Lambda_{PG}]_r=\delta\operatorname{tr}(\boldsymbol \Sigma_{0}^{-1}\boldsymbol \Sigma_{0r}^{\star})\) are sums of \(N\) subject-level terms. Under standard regularity conditions, each subject-level term is bounded, so that \(\boldsymbol \Sigma_{0}^{-1}=O(N^{-1})\) and \(\boldsymbol \Sigma_{0r}^{\star}=O(N)\). Hence \(\boldsymbol \Lambda_{PG}=O_p(1)\), and the same argument shows that its derivatives with respect to \(\boldsymbol \beta\) and \(\boldsymbol \alpha\) are \(O_p(1)\) (see Appendix C.3 of the Supporting Information). By contrast, \(\mathbf U_{G}\) is \(O_p(N^{1/2})\) at the true value of \(\boldsymbol \beta\), and its derivative with respect to \(\boldsymbol \beta\) is \(O_p(N)\). The penalty is therefore asymptotically negligible relative to the GEE estimating function, which has three important consequences. First, \(\widehat{\boldsymbol \beta}_{PG}\) and \(\widehat{\boldsymbol \beta}_{G}\) are asymptotically equivalent and share the same asymptotic normal distribution, and the covariance matrices \(\widehat{\boldsymbol \Sigma}_{R}\) and \(\widehat{\boldsymbol \Sigma}_{M}\) in \eqref{sandwich} and \eqref{morel} remain valid after substituting \(\widehat{\boldsymbol \beta}_{PG}\) for \(\widehat{\boldsymbol \beta}_{G}\). Second, whenever both estimators exist, \(\widehat{\boldsymbol \beta}_{PG}-\widehat{\boldsymbol \beta}_{G}=O_p(N^{-1})\) (Corollary 1 in Appendix C.3 of the Supporting Information), which is of smaller order than their \(O_p(N^{-1/2})\) sampling variability, so that the penalty modifies only terms of order \(N^{-1}\), such as the leading bias term. Third, because \(\boldsymbol \Lambda_{PG}\) is linear in \(\delta\), these bounds hold uniformly over \(\delta\in(0,1]\), so that a data-driven choice of \(\delta\) does not affect first-order inference. This parallels the behavior of Jeffreys-type penalties in binomial-response GLMs \citep{Kosmidis2018}. These results describe the behavior of PGEE as \(N\) increases; they do not describe a fixed sample in which separation occurs, where \(\widehat{\boldsymbol \beta}_{G}\) does not exist and the penalty is what provides finite estimates, at least under working independence (Section \ref{sec:existence}).

\hypertarget{sec:pgee-delta}{%
\subsubsection{\texorpdfstring{Choice of \(\delta\)}{Choice of \textbackslash delta}}\label{sec:pgee-delta}}

We recommend \(\delta=1/2\) as the default, with \(\delta=0.1\) as a less strongly penalized alternative, as in penalized likelihood settings with rare or recurrent events \citep{elgmati2015, puhr2017}. The parameter \(\delta\) is the strength of a prior-type penalty rather than a model parameter, and it should not be confused with the degree of separation, which is diagnosed separately, for example with the \textsf{R} package \texttt{detectseparation} \citep{detectseparation}. Its effect on the separated contrasts can be substantial in small samples, yet under separation the data indicate only that these contrasts are large, not how large, so criteria based on fit to the observed data may not discriminate between values of \(\delta\). We therefore recommend checking the default by leave-one-cluster-out cross-validation over a grid of \(\delta\) values and by a sensitivity analysis, as illustrated in Section \ref{sec:application}. Very small values of \(\delta\) should be avoided under separation, because PGEE then approaches the ordinary GEE solution, which does not exist.

\hypertarget{sec:existence}{%
\subsection{Existence and Finiteness Properties}\label{sec:existence}}

Under working independence (\(\boldsymbol \alpha=\mathbf 1_{C}\)), the estimating equations \eqref{gee_equations} reduce to the likelihood score equations for \(\boldsymbol \beta\), so \(\widehat{\boldsymbol \beta}^{I}_{G}\) exists and is finite if and only if the ML estimator \(\widehat{\boldsymbol \beta}_{ML}\) exists and is finite. Under separation, \(\widehat{\boldsymbol \beta}_{ML}\) contains at least one infinite component \citep{albert1984}, and \(\widehat{\boldsymbol \beta}^{I}_{G}\) is likewise infinite. In contrast, under working independence the proposed PGEE equations reduce to the penalized ML score equations of Kosmidis and Firth \citep{Kosmidis2018}. Assume that
\phantomsection\label{sec:existence_ind}
\begin{equation}
\frac{\left(\partial \pi_{ij}/\partial \eta_{ij}\right)^2}{\pi_{ij}(1-\pi_{ij})}
\to 0 \quad \text{as } \eta_{ij}\to \pm\infty, \quad \text{for all } i,j.
\label{condition_existence}
\end{equation}
Condition \eqref{condition_existence} holds for common links including the logit, probit, complementary log-log, and cauchit, and matches the finiteness condition for the Jeffreys-penalized ML estimator in independent binomial-response GLMs \citep{Kosmidis2018}. Hence, provided the design matrix has full column rank and the link function satisfies
\eqref{condition_existence}, \(\widehat{\boldsymbol \beta}^{I}_{PG}\) exists and is finite.

To illustrate, consider the bivariate design of Appendix C.1 of the Supporting Information, with \(\operatorname{logit}(\pi_{ij})=\beta_0+\beta_1 I(j=2)\) for \(i=1,\ldots,N\) and \(j=1,2\), and let \(s_j=\sum_{i=1}^{N}Y_{ij}\) denote the number of successes at occasion \(j\). Because all subjects share the same design, ordinary GEE gives \(\widehat{\pi}_{ij}=s_j/N\) under any working structure, so that \(\widehat{\boldsymbol \beta}_{G}\) is infinite whenever the time-by-response \(2\times2\) table has a zero cell. Under working independence, PGEE instead gives
\[
\widehat{\pi}_{ij}=\frac{s_j+\delta}{N+2\delta},\qquad j=1,2;
\]
that is, it adds \(\delta\) to each cell of the table, so that its estimates are finite for every configuration. With \(\delta=1/2\) this is the continuity correction of Haldane \citep{haldane1956}. Appendix C.4 of the Supporting Information derives this closed form and reports tables with no zero cell, one zero cell (quasi-complete separation), two zero cells (complete separation), and no successes: ordinary GEE fails whenever a cell is empty, whereas PGEE remains finite under the independence and exchangeable working structures and the logit and probit links.

\phantomsection\label{sec:existence_nonind}
A general finiteness guarantee is not available when \(\boldsymbol \alpha\neq \mathbf{1}_{C}\), because \(\pi^{\star}_{ijj'}\) depends nonlinearly on \(\boldsymbol \beta\) and extreme marginalized odds-ratio estimates can affect the conditioning of \(\boldsymbol \Sigma_{0}\). Establishing formal conditions would require stronger assumptions on the dependence structure than GEE methodology usually imposes. We therefore treat finiteness under nonindependence working structures as an open problem. Empirically, PGEE achieved near-complete convergence across the nonindependence structures considered in our simulations, with the few failures linked to very sparse pooled \(2\times2\) tables. Practical remedies are discussed in Section \ref{sec:workflow}.

\hypertarget{sec:cc}{%
\subsection{Failure of Correlation-Coefficient Parameterizations under Separation}\label{sec:cc}}

Existing penalized GEE approaches, including those of Mondol and Rahman \citep{mondol2019bias} and Geroldinger et al. \citep{Geroldinger2022}, rely on correlation-coefficient working structures and estimate the association parameters using the method of moments of Liang and Zeger \citep{Liang1986}.

Under separation, fitted probabilities are driven toward their boundary values. For binary responses, this makes the Pearson residuals shrink toward zero, because observations with \(Y_{ij}=1\) have \(\widehat{\pi}_{ij}\to 1\), whereas observations with \(Y_{ij}=0\) have \(\widehat{\pi}_{ij}\to 0\). Consequently, moment-based estimates of the working correlation can be driven close to zero. In Appendix C.1 of the Supporting Information, we provide a simple example with bivariate correlated responses and an exchangeable working correlation matrix showing that the common correlation parameter \(\rho\) converges to \(0\) regardless of the true within-subject dependence. Appendix C.4 shows the same behavior across other \(2\times2\) configurations: the estimated correlation is close to zero whenever the table has a zero cell, and with an estimated dispersion parameter the fit fails outright, whereas the odds-ratio PGEE remains finite in every configuration. The fitted working structure becomes effectively indistinguishable from independence precisely in the regimes where modeling within-subject dependence remains important, undermining the efficiency gains that motivate the use of nonindependence working structures. When implementations estimate a free dispersion parameter \(\phi\) rather than fixing it at its canonical value of one for binary responses, \(\phi\) can also be driven toward zero, inducing ill-conditioning of \(\mathbf V^{-1}_{i}\) and further numerical instability.

In contrast, odds-ratio parameterizations estimated from marginalized \(2\times 2\) tables yield finite association estimates under the continuity correction of Haldane \citep{haldane1956} with \(\zeta = 0.5\). The pooled odds ratio does not depend on the fitted marginal probabilities and is therefore not subject to this correlation-parameter collapse. Beyond collapse to independence, correlation-coefficient estimates can also violate the Fréchet bounds implied by the fitted marginal probabilities, rendering the working covariance matrix inadmissible for binary responses.

\hypertarget{one-step-approximations}{%
\subsection{One-step Approximations}\label{one-step-approximations}}

Having established existence and finiteness of \(\widehat{\boldsymbol \beta}^{I}_{PG}\) under condition \eqref{condition_existence}, we propose two computationally efficient alternatives to the fully iterated PGEE estimator. The one-step PGEE (OPGEE) estimator \(\widehat{\boldsymbol \beta}_{OPG}\) performs a single Fisher scoring iteration of the penalized estimating equations
\eqref{pgee} from \(\widehat{\boldsymbol \beta}^{I}_{PG}\),
\begin{equation}
\widehat{\boldsymbol \beta}_{OPG}= \widehat{\boldsymbol \beta}^{I}_{PG}- \boldsymbol \Sigma_{0}^{-1}\!\left(\widehat{\boldsymbol \beta}^{I}_{PG}; \widehat{\boldsymbol \alpha}\right)\,
\mathbf U_{PG}\!\left(\widehat{\boldsymbol \beta}^{I}_{PG}; \widehat{\boldsymbol \alpha}\right),
\label{opgee}
\end{equation}
where the update is well defined when \(\boldsymbol \Sigma_{0}(\widehat{\boldsymbol \beta}^{I}_{PG};\widehat{\boldsymbol \alpha})\) is nonsingular. Repeated application of \eqref{opgee} yields the fully iterated PGEE estimator upon convergence. The hybrid one-step PGEE (HPGEE) estimator \(\widehat{\boldsymbol \beta}_{HPG}\) instead performs a single Fisher scoring step of the ordinary GEE equations
\eqref{gee_equations} from \(\widehat{\boldsymbol \beta}^{I}_{PG}\),
\begin{equation}
\widehat{\boldsymbol \beta}_{HPG}= \widehat{\boldsymbol \beta}^{I}_{PG}- \boldsymbol \Sigma_{0}^{-1}\!\left(\widehat{\boldsymbol \beta}^{I}_{PG}; \widehat{\boldsymbol \alpha}\right)\,
\mathbf U_{G}\!\left(\widehat{\boldsymbol \beta}^{I}_{PG}; \widehat{\boldsymbol \alpha}\right),
\label{hybrid_gee}
\end{equation}
so that iterating \eqref{hybrid_gee} recovers the ordinary GEE solution. Because \(\widehat{\boldsymbol \beta}^{I}_{PG}\) is finite even under separation (Section \ref{sec:existence_ind}), both one-step estimators may exist in settings where the traditional GEE starting value \(\widehat{\boldsymbol \beta}^{I}_{G}\) fails. By arguments analogous to those of Lipsitz et al. \citep{Lipsitz1994} for the one-step GEE estimator,
both are asymptotically equivalent to their fully iterated counterparts. We recommend \(\widehat{\boldsymbol \beta}_{OPG}\) when fully iterated PGEE fitting is slow or fails to converge, and \(\widehat{\boldsymbol \beta}_{HPG}\) when a reliable one-step approximation to ordinary GEE is desired.

\hypertarget{practical-workflow-and-implementation}{%
\subsection{Practical Workflow and Implementation}\label{practical-workflow-and-implementation}}

\phantomsection\label{sec:workflow}

When ordinary GEE fails to converge or yields extreme estimates, we recommend the following workflow. First, fit PGEE under the independence working structure with the default \(\delta=1/2\). Next, estimate continuity-corrected marginalized odds ratios from pooled \(2\times2\) tables and select a working odds-ratio structure as described in Section \ref{sec:or-association}. Finally, refit PGEE under the selected structure and use \(\widehat{\boldsymbol \Sigma}_{M}\) for inference.

If fully iterated PGEE fitting encounters convergence difficulties, compute \(\widehat{\boldsymbol \beta}_{OPG}\), \(\widehat{\boldsymbol \beta}_{HPG}\), or both. Conveniently, \(\widehat{\boldsymbol \beta}_{HPG}\) can be obtained using standard GEE software. One first computes \(\widehat{\boldsymbol \beta}^{I}_{PG}\) via the penalized likelihood approach of Kosmidis and Firth \citep{Kosmidis2018}, then performs a single GEE Fisher scoring step from \(\widehat{\boldsymbol \beta}^{I}_{PG}\) under the desired working structure. If one-step approximations themselves fail to converge because of extreme estimated odds ratios, one may revert to a simpler working structure (exchangeable or, if needed, independence), bound the odds-ratio estimates to a conservative interval such as \([0.01, 100]\), or increase the continuity constant \(\zeta\). Finally, check the choice of \(\delta\) by leave-one-cluster-out cross-validation over a grid of values in \((0,1]\), scoring the responses of each held-out cluster by the log score, and report a brief sensitivity analysis across \(\delta\) and a small set of plausible working association structures. If the cross-validation curve is flat, as in Section \ref{sec:application}, we recommend retaining \(\delta=1/2\). Section \ref{sec:application} illustrates these steps in the respiratory-illness trial.

The \textsf{R} package \texttt{geer}, available from Comprehensive R Archive Network (CRAN) at \url{https://CRAN.R-project.org/package=geer}, implements ordinary GEE and the proposed PGEE, OPGEE, and HPGEE estimators via the function \texttt{geewa\_binary()}, which supports odds-ratio working parameterizations and allows the penalty strength \(\delta\) and continuity-correction constant \(\zeta\) to be set through
\texttt{geer\_control()}.

\hypertarget{sec:sim}{%
\section{Simulation Study}\label{sec:sim}}

We evaluate the proposed procedures in two regimes: complete and quasi-complete separation, where ordinary GEE can fail or produce numerically unstable solutions, and regular settings, where ordinary GEE is expected to perform well. In the separation regimes, the relevant criteria are the empirical power for the separation-inducing covariate and the empirical rejection rate for the irrelevant covariate. In regular settings, the goal is to verify that the proposed estimators perform comparably to ordinary GEE.

For each regression parameter \(\beta\), we compare the ordinary GEE estimator \(\widehat{\boldsymbol \beta}_{G}\), the robust bias-reduced estimator \(\widehat{\boldsymbol \beta}_{RBR}\) \citep{Touloumis2026bias}, and the proposed penalized estimators \(\widehat{\boldsymbol \beta}_{PG}\), \(\widehat{\boldsymbol \beta}_{OPG}\), and \(\widehat{\boldsymbol \beta}_{HPG}\). All methods are implemented under marginalized odds-ratio working association structures: independence, exchangeable, and unstructured. Penalization strength is set to \(\delta\in\{0.1,0.5\}\). The continuity correction constant for the marginalized odds-ratio estimator is set to its default value \(\zeta = 0.5\) throughout. Existing penalized GEE approaches, including those of Mondol and Rahman \citep{mondol2019bias} and Geroldinger et al. \citep{Geroldinger2022}, are restricted to the logit link and cannot be applied to the probit model used here. Moreover, as established in Section \ref{sec:cc}, correlation-coefficient parameterizations degenerate under separation, reducing those methods to working independence and precluding a meaningful comparison in the separation regimes. In the regular regime, ordinary GEE serves as the natural reference. None of the working structures coincides with the data-generating mechanism: in the quasi-complete separation and regular regimes, the responses are generated by a latent-threshold model whose latent correlation matrix is neither diagonal nor exchangeable, and the true pairwise odds ratios vary with the covariates, whereas the marginalized odds ratios are pooled over subjects. The simulations therefore evaluate inference under a misspecified working association.

We consider \(N \in \{30, 50, 100, 500\}\) subjects with cluster size \(J=4\). For each configuration, we generate \(B=10,000\) datasets. We record convergence proportion (CP), Monte Carlo bias, empirical standard error (ESE), the ratio of the mean bias-corrected estimated standard error to the empirical standard error (RE), the empirical coverage of nominal 95\% Wald intervals based on the bias-corrected covariance matrix (EC), and mean squared error (MSE). Values of RE close to 1 indicate good agreement between the empirical and bias-corrected estimated standard errors; values below 1 indicate that the bias-corrected standard errors underestimate the empirical variability. In the regular regime, we additionally record the prediction mean squared error (PMSE), the mean over replications of the average squared difference between the fitted and true marginal probabilities over the observed subject-occasions \citep{Geroldinger2022}, and the simulated relative efficiency (SRE) relative to \(\widehat{\boldsymbol \beta}^{I}_{G}\), where values greater than 1 indicate higher efficiency than \(\widehat{\boldsymbol \beta}^{I}_{G}\). For every method, we treat nonconvergence within 500 iterations, extreme regression parameter estimates (absolute value \(>100\)), and failures of the covariance computation (for example, nonsymmetric or nonpositive-definite estimates) as failures when computing CP, and all other summaries are computed over the replications in which the fit did not fail. The threshold of 100 is large relative to the true parameter values in all scenarios considered, so results are unlikely to be sensitive to the precise choice of cutoff.

\hypertarget{separation-regimes}{%
\subsection{Separation Regimes}\label{separation-regimes}}

For the separation regimes, we used two covariate configurations:

\begin{itemize}[leftmargin=*]
\item \textbf{Scenario A (time-stationary $x_1$):} $x_{1ij}=x_{1i}\overset{iid}{\sim}\text{Categorical}(1/3,1/3,1/3)$ and $x_{2ij}=x_{2i}\overset{iid}{\sim} \mathcal{U}(0,1)$.
\item \textbf{Scenario B (time-varying $x_1$):} $x_{1ij}\overset{iid}{\sim}\text{Categorical}(1/3,1/3,1/3)$ and $x_{2ij}=x_{2i} \overset{iid}{\sim} \mathcal{U}(0,1)$.
\end{itemize}

In Scenario A, \(x_{1i}\) is drawn once per subject and held constant across occasions, whereas in Scenario B, \(x_{1ij}\) is drawn independently at each occasion. Under complete separation, Scenario A therefore produces identical responses within each subject, which maximizes the sparsity of the pooled \(2\times2\) tables used to estimate the marginalized odds ratios and makes it the more demanding configuration; Scenario B permits heterogeneous within-subject responses.

\hypertarget{complete-separation}{%
\subsubsection{Complete Separation}\label{complete-separation}}

To induce complete separation, we generate covariates under Scenario A or B and set
\(Y_{ij}=I(x_{1ij}\ge 2)\), so that the responses are a deterministic function of \(x_{1ij}\). For each dataset, we fit the probit marginal mean model
\begin{equation}
\Phi^{-1}(\pi_{ij})=\beta_0+\beta_1x_{1ij}+\beta_2x_{2ij},
\label{sim_model}
\end{equation}
where \(\Phi(\cdot)\) denotes the standard normal cumulative distribution function.

This is the most demanding regime, targeting estimator existence, numerical reliability, and discrimination between the separating (\(x_1\)) and irrelevant covariates (\(x_2\)). Table \ref{tab:sim_separation} (panel A) reports the convergence proportions. Ordinary GEE failed in every replication of every configuration. RBR failed in every replication for \(N\in\{100,500\}\). For \(N\in\{30,50\}\) it satisfied the convergence criterion in up to 99.41\% of replications under Scenario A and up to 66.17\% under Scenario B, but every such solution was numerically degenerate: the largest bias-corrected standard error was at least \(2.97\times10^{6}\) and \(|\widehat\beta_1|\) ranged from 16.1 to 53.9, reflecting fitted probabilities that are numerically 0 or 1. The results of ordinary GEE and RBR are therefore not summarized further. In contrast, PGEE, OPGEE, and HPGEE converged in all replications under all working association structures and yielded well-defined covariance estimates throughout.

Because the responses are a deterministic function of \(x_{1ij}\), simulation summaries such as bias, ESE, MSE, and PMSE are not informative in this setting. We therefore focus on whether inference correctly reflects the data-generating rule. Tests of \(H_0:\beta_1=0\) reject in \(100\%\) of replications for all proposed penalized estimators across all sample sizes, working structures, and penalty values. For \(H_0:\beta_2=0\), the null hypothesis is never rejected under Scenario B, whereas under Scenario A rejection rates remain negligible, with a maximum of \(0.65\%\) across all configurations. The penalized procedures therefore recognize the separating covariate without spuriously identifying the irrelevant one.

\hypertarget{quasi-complete-separation}{%
\subsubsection{Quasi-complete Separation}\label{quasi-complete-separation}}

To generate correlated responses under a quasi-complete separation scheme, we first generate binary outcomes from model \eqref{sim_model} with \((\beta_0,\beta_1,\beta_2)^\top=(-0.5,1,-3)^\top\) via the latent-threshold approach implemented in \texttt{SimCorMultRes} \citep{Touloumis2016a}, using the latent correlation
matrix
\[
\begin{pmatrix}
1   & 0.5 & 0.8 & 0.2\\
0.5 & 1   & 0.5 & 0.8\\
0.8 & 0.5 & 1   & 0.5\\
0.2 & 0.8 & 0.5 & 1
\end{pmatrix}.
\]
This matrix reflects a dependence structure with moderate-to-strong short-range and weaker longer-range correlations. After generating responses and covariates, we enforce quasi-complete separation by setting \(Y_{ij}=0\) whenever \(x_{1ij}=1\) and \(Y_{ij}=1\) whenever \(x_{1ij}=3\), retaining the simulated value only when \(x_{1ij}=2\). As in the complete separation regime, ordinary GEE failed in every replication, and RBR failed in between 73.01\% and 100\% of replications, depending on the scenario, sample size, and working structure (Table \ref{tab:sim_separation}, panel A). Their results are therefore not summarized further.

All proposed estimators achieve near-complete empirical convergence across both scenarios, working structures, and penalty values, with a minimum CP of 99.87\% (Table \ref{tab:sim_separation}, panel A). Empirical rejection rates for \(H_0:\beta_1=0\) are at or near \(100\%\) across all configurations (minimum empirical rejection rate \(99.74\%\)). Thus, all proposed penalized estimators correctly identify the separation-inducing covariate.

Table \ref{tab:sim_separation} (panel B) summarizes the results for \(\beta_2\) under the unstructured working structure at \(N\in\{30,100\}\); full results for all working structures and sample sizes are in Tables S1--S4 of the Supporting Information. Across both scenarios, performance improves systematically with sample size, with absolute bias, ESE, and MSE declining, RE moving toward 1, and EC approaching the nominal level. The stronger penalty \(\delta=0.5\) is most beneficial at the smallest sample sizes, with differences between penalty values becoming negligible by \(N=500\). PGEE and OPGEE behave almost identically throughout, while HPGEE tends to yield slightly higher EC at the cost of generally larger variability and MSE; for example, under Scenario A with \(N=30\), \(\delta=0.5\), and the unstructured working structure, EC is 80.96\% for PGEE and 85.63\% for HPGEE, with MSE of 1.6724 and 9.7373, respectively (Table \ref{tab:sim_separation}, panel B; Table S1). RE values below 1 at small sample sizes (between 0.31 and 0.77 at \(N=30\); Tables S1 and S3) indicate that the bias-corrected standard errors underestimate the empirical variability, producing intervals that are too narrow. The undercoverage at small sample sizes therefore reflects both these underestimated standard errors and finite-sample bias in the point estimates induced by the enforcement procedure, which shifts the intervals away from the true parameter value.

Scenario A is considerably more demanding than Scenario B at small sample sizes: its more homogeneous within-subject responses inflate variability and MSE and lead to more pronounced undercoverage, particularly under working independence with the weaker penalty, where MSE is substantially larger than under the exchangeable and unstructured working structures. Under Scenario B, bias and variability are smaller, EC is closer to nominal even at small sample sizes, and the choice of working structure matters less. Under Scenario A, all three working structures under-cover at \(N=30\) and \(N=50\) (Table S1); at \(N=30\), the unstructured structure under-covers slightly more than independence with \(\delta=0.5\) (EC of 80.96\% versus 82.51\% for PGEE), whereas the ordering is reversed with \(\delta=0.1\) (86.04\% versus 84.46\%).

\hypertarget{regular-regime}{%
\subsection{Regular Regime}\label{regular-regime}}

To assess finite-sample performance in a regular setting where separation does not occur, we generated marginal success probabilities from model \eqref{sim_model} with \((\beta_{0},\beta_{1},\beta_{2})=(0,0.5,0.5)\). Covariates were generated as \(x_{1ij}=x_{1i}\overset{iid}{\sim}\mathcal N(0,0.5^{2})\) and \(\mathbf x_{2i}\sim\mathcal N(\mathbf 0_{4},\Sigma_{x})\), where \(\Sigma_{x}\) has a compound-symmetry structure with variance \(0.5^{2}\) and correlation \(0.8\). The covariates were generated once for each \(N\) and held fixed across replications, so that the results are conditional on the covariate design. Conditional on the covariates, correlated binary responses were generated via \texttt{SimCorMultRes} \citep{Touloumis2016a} using the latent correlation matrix
\[
\mathbf R=
\begin{pmatrix}
1 & 0.85 & 0.50 & 0.15\\
0.85 & 1 & 0.85 & 0.50\\
0.50 & 0.85 & 1 & 0.85\\
0.15 & 0.50 & 0.85 & 1
\end{pmatrix},
\]
reflecting a dependence structure with strong short-range and weaker long-range correlations. We considered a complete-data scenario and a monotone MCAR \citep{rubin1976inference} missing-data scenario, in which, at each of occasions 2, 3, and 4, a subject still under observation dropped out with probability 0.2, independently of the responses and covariates, so that 51.2\% of subjects were observed at all four occasions. Table \ref{tab:sim_regular} summarizes the results under the unstructured working structure with complete data at \(N\in\{30,100\}\); full results are in Tables S5--S8 (complete data) and Tables S9--S12 (MCAR dropout) of the Supporting Information.

Under complete data, all methods converged in essentially all replications (minimum CP 99.99\%; Tables S5--S8). ESE and PMSE decreased with \(N\) for every method and working structure, and absolute bias generally decreased with \(N\) for both parameters. PGEE with \(\delta=0.5\) reduced the absolute bias of ordinary GEE in every configuration except one: the time-varying covariate \(x_2\) under the unstructured working structure, where RBR, PGEE, and OPGEE all had larger absolute bias than ordinary GEE at every sample size and HPGEE had the smallest (Table \ref{tab:sim_regular}). Elsewhere, RBR usually gave the smallest absolute bias (in 16 of the 24 combinations of parameter, working structure, and sample size; Tables S5--S8), and HPGEE tracked ordinary GEE closely. Efficiency gains relative to the independence working structure were pronounced under the unstructured working structure for both covariates, where the largest SRE values were observed, up to 2.21 for \(\beta_2\) at \(N=30\) (Table \ref{tab:sim_regular}). Empirical coverage was near nominal under the independence and exchangeable working structures, with mild undercoverage under the unstructured working structure for \(x_2\) at small \(N\) (EC between 92.98\% and 93.20\% at \(N=30\); Table \ref{tab:sim_regular}). Under monotone MCAR dropout, ESE and PMSE increased for every method, working structure, and sample size, and absolute bias increased in most configurations (Tables S9--S12), but the qualitative ordering of bias reduction and the coverage patterns were preserved, with EC between 92.20\% and 95.62\%. Efficiency gains under the exchangeable and unstructured working structures remained pronounced. Overall, in regular settings the proposed penalized estimators retained the numerical reliability, empirical coverage, and predictive performance of ordinary GEE, with modest small-sample bias reductions that diminished as \(N\) increased.

\begin{table}[!htbp]
\caption{Simulation results under separation, based on 10,000 replications per configuration. Panel A: convergence proportion (CP, \%) of each method, given as the range over the independence, exchangeable, and unstructured working structures for ordinary GEE and RBR, and as the minimum over methods, working structures, and $\delta\in\{0.1,0.5\}$ for the penalized estimators. Under complete separation with $N\in\{30,50\}$, every RBR fit counted as converged was numerically degenerate (largest bias-corrected standard error at least $2.97\times 10^{6}$; Section \ref{sec:sim}). Panel B: bias, ratio of the mean bias-corrected standard error to the empirical standard error (RE), and empirical coverage (EC, \%) of nominal 95\% Wald intervals for $\beta_2=-3$ under quasi-complete separation, unstructured working structure. Full results are in Tables S1--S4 of the Supporting Information.}
\centering
\adjustbox{max width=\textwidth}{%
\begin{tabular}{lllrrrr}
\multicolumn{7}{l}{\textit{Panel A: convergence proportion (\%)}}\\
\toprule
Regime & Scenario & Method & $N=30$ & $N=50$ & $N=100$ & $N=500$\\
\midrule
Complete & A & GEE & 0.00 & 0.00 & 0.00 & 0.00 \\
 &  & RBR & 99.36--99.41 & 76.53--76.55 & 0.00 & 0.00 \\
 &  & PGEE, OPGEE, HPGEE & 100.00 & 100.00 & 100.00 & 100.00 \\
Complete & B & GEE & 0.00 & 0.00 & 0.00 & 0.00 \\
 &  & RBR & 66.14--66.17 & 1.05--1.09 & 0.00 & 0.00 \\
 &  & PGEE, OPGEE, HPGEE & 100.00 & 100.00 & 100.00 & 100.00 \\
\addlinespace[2pt]
Quasi-complete & A & GEE & 0.00 & 0.00 & 0.00 & 0.00 \\
 &  & RBR & 25.55--26.99 & 10.53--12.78 & 0.28--1.03 & 0.02--0.44 \\
 &  & PGEE, OPGEE, HPGEE & 99.87 & 99.98 & 99.98 & 100.00 \\
Quasi-complete & B & GEE & 0.00 & 0.00 & 0.00 & 0.00 \\
 &  & RBR & 10.48--12.61 & 0.12--0.20 & 0.00 & 0.00 \\
 &  & PGEE, OPGEE, HPGEE & 100.00 & 100.00 & 100.00 & 100.00 \\
\bottomrule
\end{tabular}}

\vspace{2ex}
\adjustbox{max width=\textwidth}{%
\begin{tabular}{lllrrrrrr}
\multicolumn{9}{l}{\textit{Panel B: quasi-complete separation, $\beta_2=-3$, unstructured working structure}}\\
\toprule
 & & & \multicolumn{3}{c}{Scenario A} & \multicolumn{3}{c}{Scenario B}\\
\cmidrule(lr){4-6}\cmidrule(lr){7-9}
$N$ & Method & $\delta$ & Bias & RE & EC & Bias & RE & EC\\
\midrule
30 & PGEE & 0.5 & 0.5273 & 0.77 & 80.96 & $-$0.0812 & 0.74 & 90.08 \\
 & OPGEE & 0.5 & 0.4800 & 0.71 & 81.39 & $-$0.0845 & 0.73 & 90.18 \\
 & HPGEE & 0.5 & $-$0.6218 & 0.45 & 85.63 & $-$0.3553 & 0.71 & 92.00 \\
 & PGEE & 0.1 & $-$0.6340 & 0.38 & 86.04 & $-$0.3895 & 0.60 & 92.40 \\
 & OPGEE & 0.1 & $-$0.5747 & 0.40 & 85.97 & $-$0.3888 & 0.61 & 92.42 \\
 & HPGEE & 0.1 & $-$1.0669 & 0.33 & 87.93 & $-$0.4644 & 0.57 & 93.20 \\
\addlinespace[2pt]
100 & PGEE & 0.5 & 0.1228 & 0.88 & 90.53 & $-$0.0227 & 0.94 & 93.96 \\
 & OPGEE & 0.5 & 0.1250 & 0.88 & 90.66 & $-$0.0225 & 0.94 & 93.97 \\
 & HPGEE & 0.5 & $-$0.0829 & 0.89 & 93.09 & $-$0.0847 & 0.94 & 93.95 \\
 & PGEE & 0.1 & $-$0.0578 & 0.90 & 93.19 & $-$0.0771 & 0.94 & 94.00 \\
 & OPGEE & 0.1 & $-$0.0452 & 0.90 & 93.15 & $-$0.0769 & 0.94 & 94.01 \\
 & HPGEE & 0.1 & $-$0.0886 & 0.90 & 93.40 & $-$0.0896 & 0.94 & 94.09 \\
\bottomrule
\end{tabular}}
\label{tab:sim_separation}
\end{table}

\begin{table}[!htbp]
\caption{Simulation results in the regular regime with complete data, unstructured working structure, based on 10,000 replications per configuration: bias ($\times 100$), empirical coverage (EC, \%) of nominal 95\% Wald intervals, and simulated relative efficiency (SRE) relative to ordinary GEE under working independence, for $\beta_1=0.5$ (cluster-level covariate $x_1$) and $\beta_2=0.5$ (time-varying covariate $x_2$), and prediction mean squared error (PMSE). Full results, including the MCAR scenario, are in Tables S5--S12 of the Supporting Information.}
\centering
\adjustbox{max width=\textwidth}{%
\begin{tabular}{lllrrrrrrr}
\toprule
 & & & \multicolumn{3}{c}{$\beta_1$} & \multicolumn{3}{c}{$\beta_2$} & \\
\cmidrule(lr){4-6}\cmidrule(lr){7-9}
$N$ & Method & $\delta$ & Bias & EC & SRE & Bias & EC & SRE & PMSE\\
\midrule
30 & GEE &  & 2.0357 & 94.82 & 1.13 & $-$2.1734 & 92.98 & 2.05 & 0.01040 \\
 & RBR &  & $-$0.3608 & 95.42 & 1.24 & $-$4.6161 & 93.09 & 2.21 & 0.00980 \\
 & PGEE & 0.5 & $-$0.1808 & 95.51 & 1.24 & $-$4.4345 & 93.11 & 2.15 & 0.00985 \\
 & OPGEE & 0.5 & 0.1730 & 95.44 & 1.23 & $-$2.4021 & 93.20 & 2.10 & 0.00991 \\
 & HPGEE & 0.5 & 2.3815 & 94.79 & 1.12 & $-$0.2847 & 93.08 & 1.99 & 0.01046 \\
 & PGEE & 0.1 & 1.5835 & 95.01 & 1.15 & $-$2.6278 & 93.00 & 2.07 & 0.01029 \\
\addlinespace[2pt]
100 & GEE &  & 0.6661 & 94.38 & 1.10 & $-$0.7629 & 94.08 & 1.58 & 0.00318 \\
 & RBR &  & $-$0.1819 & 94.66 & 1.14 & $-$1.5977 & 94.29 & 1.62 & 0.00311 \\
 & PGEE & 0.5 & 0.0556 & 94.65 & 1.13 & $-$1.3564 & 94.12 & 1.60 & 0.00313 \\
 & OPGEE & 0.5 & 0.0839 & 94.66 & 1.13 & $-$0.9515 & 94.14 & 1.60 & 0.00313 \\
 & HPGEE & 0.5 & 0.6958 & 94.40 & 1.10 & $-$0.3578 & 94.16 & 1.58 & 0.00318 \\
 & PGEE & 0.1 & 0.5430 & 94.46 & 1.11 & $-$0.8823 & 94.12 & 1.58 & 0.00317 \\
\bottomrule
\end{tabular}}
\label{tab:sim_regular}
\end{table}

\hypertarget{sec:sim-variance}{%
\subsection{Variance Estimation under Misspecified Working Association}\label{sec:sim-variance}}

For every fit in Sections 3.1.2 and 3.2, we also computed standard errors from the model-based estimator \(\widehat{\boldsymbol \Sigma}_{0}^{-1}\) and the sandwich estimator \(\widehat{\boldsymbol \Sigma}_{R}\), in addition to the bias-corrected estimator \(\widehat{\boldsymbol \Sigma}_{M}\). Table S17 of the Supporting Information reports, for \(N\in\{30,100,500\}\), the empirical coverage of nominal 95\% Wald intervals and the ratio of the mean estimated to the empirical standard error for PGEE with \(\delta=0.5\) and, in the regular regime, ordinary GEE.

First, model-based standard errors were unreliable under the misspecified working structures and did not improve as \(N\) increased: in the regular regime under working independence, model-based intervals covered between 79.68\% and 84.87\% of the time for both ordinary GEE and PGEE, and under quasi-complete separation in Scenario A they over-covered under the exchangeable and unstructured working structures (96.20\% to 98.74\%) and under-covered under working independence (83.39\% to 85.23\%). Second, the sandwich estimator \(\widehat{\boldsymbol \Sigma}_{R}\) under-covered at small sample sizes, most severely under quasi-complete separation (between 71.17\% and 78.37\% for PGEE in Scenario A at \(N=30\)). Third, the bias-corrected estimator \(\widehat{\boldsymbol \Sigma}_{M}\) gave coverage at least as high as \(\widehat{\boldsymbol \Sigma}_{R}\) in every method and configuration, and improved the coverage of PGEE at \(N=30\) by between 1.69 and 10.54 percentage points. In the regular regime, it gave coverage between 92.98\% and 95.51\% for both ordinary GEE and PGEE. Under quasi-complete separation, it still under-covered at \(N=30\) (between 80.96\% and 90.88\%), because its standard errors remained too small (Section \ref{sec:sim}), and its coverage approached the nominal level as \(N\) increased (between 93.99\% and 94.64\% at \(N=500\)).

The results for OPGEE closely matched those for PGEE, the results for HPGEE followed the same patterns, and the same conclusions held under MCAR dropout. These results support basing inference on \(\widehat{\boldsymbol \Sigma}_{M}\), as done throughout the paper.

\hypertarget{sec:application}{%
\section{Application to a Respiratory Illness Trial}\label{sec:application}}

Stokes, Davis and Koch \citep{stokes2012} presented data from a randomized clinical trial evaluating the efficacy of an active treatment relative to placebo in reducing the risk of respiratory illness. We restrict attention to the 55 patients enrolled at Clinic 2, which exhibits complete or quasi-complete separation under several model specifications, providing a practical setting in which ordinary GEE is known to fail. Respiratory status was recorded at baseline and at four scheduled follow-up visits (\(j=1,2,3,4\)). The binary response was coded \(1\) for good respiratory status and \(0\) for poor respiratory status.

The mean model for the probability of good respiratory status of patient \(i\) at visit \(j\), denoted \(\pi_{ij}\), is specified via a probit link, consistent with the simulation study of Section \ref{sec:sim},
\begin{equation}
\Phi^{-1} \left( \pi_{ij} \right) = \beta_0 + \beta_1 \, trt_i + \beta_2 \, g_i + \beta_3 \, j + \beta_4 \, a_i + \beta_5 \, b_i + \beta_6 \left(trt_i \times g_i\right) + \beta_7 \left(j \times a_i\right).
\label{respiratory_model}
\end{equation}
The same model was selected and substantive conclusions were unchanged under a logit link, as reported in Appendix E.2 of the Supporting Information. Covariates included treatment assignment (\(trt_i\), coded 1 for active treatment and 0 for placebo), gender (\(g_i\), coded 1 for male and 0 for female), baseline age (\(a_i\)), and baseline respiratory status (\(b_i\)).

For model selection, we fitted a candidate mean model containing all main effects and pairwise interactions among treatment, gender, visit, baseline age, and baseline status, using ordinary GEE under the independence working structure. This candidate model failed to converge within \(500\) iterations and produced extremely large coefficient magnitudes, of order \(10^{12}\) in absolute value. A separation check using the \textsf{R} package \texttt{detectseparation} \citep{detectseparation} confirmed complete separation, implying that no finite GEE estimates exist. The separation arises because all four female patients assigned to active treatment maintained good respiratory status at all four follow-up visits, so that none of their \(16\) observations recorded poor status. Hence ordinary GEE is not a viable option for this dataset, and we proceed with the proposed PGEE methodology.

We performed model selection using PGEE under the independence working structure with \(\delta=0.5\), starting from the same candidate model. The penalized estimating equations converged after one iteration. Backward elimination at the \(10\%\) significance level, using Wald tests with bias-corrected standard errors, yielded model \eqref{respiratory_model}. Repeating the procedure with \(\delta=0.1\) resulted in the same selected model. Moreover, replacing the independence working structure by the exchangeable or unstructured working structures produced the same selected model for both \(\delta\) values. Consistent with Section \ref{sec:cc}, PGEE with correlation-coefficient working parameterizations produced inadmissible association estimates under every nonindependence working structure considered (Appendix E.3 of the Supporting Information), which further supports the odds-ratio parameterization under separation.

\begin{table}[!htbp]
\caption{PGEE estimates for the respiratory illness trial (Clinic 2) with $\delta=0.5$ under odds-ratio parameterizations.}
\centering
\adjustbox{max width=\textwidth}{%
\begin{tabular}{lrrrrrrrrrr}
  \toprule
  & \multicolumn{3}{c}{Independence}
  & \multicolumn{3}{c}{Exchangeable}
  & \multicolumn{3}{c}{Unstructured}\\
  \cmidrule(lr){2-4}\cmidrule(lr){5-7}\cmidrule(lr){8-10}
  Parameter
  & Estimate & SE & $p$
  & Estimate & SE & $p$
  & Estimate & SE & $p$\\
  \midrule
  $\beta_0$ & $-$1.1848 & 0.7797 & 0.1286
            & $-$1.3384 & 0.7559 & 0.0766
            & $-$1.2292 & 0.7164 & 0.0862 \\
  $\beta_1$ &    2.3727 & 0.5196 & $<$0.0001
            &    2.3341 & 0.5096 & $<$0.0001
            &    2.3229 & 0.5129 & $<$0.0001 \\
  $\beta_2$ &    0.2927 & 0.4345 & 0.5005
            &    0.2795 & 0.4458 & 0.5307
            &    0.3130 & 0.4383 & 0.4751 \\
  $\beta_3$ &    0.3542 & 0.2204 & 0.1080
            &    0.3994 & 0.2044 & 0.0507
            &    0.3509 & 0.1957 & 0.0730 \\
  $\beta_4$ &    0.0203 & 0.0162 & 0.2103
            &    0.0250 & 0.0155 & 0.1082
            &    0.0195 & 0.0146 & 0.1803 \\
  $\beta_5$ &    0.6650 & 0.3175 & 0.0362
            &    0.6604 & 0.3194 & 0.0387
            &    0.6976 & 0.3136 & 0.0261 \\
  $\beta_6$ & $-$1.7077 & 0.6293 & 0.0067
            & $-$1.6485 & 0.6228 & 0.0081
            & $-$1.6341 & 0.6221 & 0.0086 \\
  $\beta_7$ & $-$0.0106 & 0.0053 & 0.0449
            & $-$0.0119 & 0.0049 & 0.0145
            & $-$0.0100 & 0.0046 & 0.0295 \\
  \bottomrule
\end{tabular}}
\label{results_dataset}
\end{table}

Ordinary GEE also failed to converge for model \eqref{respiratory_model} under every working structure, whereas PGEE converged under the independence, exchangeable, and unstructured odds-ratio working structures for both \(\delta\in\{0.1,0.5\}\). Table \ref{results_dataset} reports PGEE estimates for \(\delta=0.5\). Results for \(\delta=0.1\) and for the one-step OPGEE and HPGEE approximations are reported in Appendix E.1 of the Supporting Information.

Estimates of \(\boldsymbol \beta\) were similar across the three working structures and both \(\delta\) values. Standard errors were in some instances smaller for \(\delta=0.5\) than for \(\delta=0.1\), consistent with patterns observed in Section \ref{sec:sim}. The estimated marginalized odds ratios ranged from \(3.56\) to \(19.04\), indicating substantial within-subject association. This variability helps explain the slightly smaller standard errors under the unstructured working structure.

To assess the choice of \(\delta\) in a data-driven way, we performed leave-one-cluster-out cross-validation for model \eqref{respiratory_model}. For each of 13 values of \(\delta\) between \(0.01\) and \(1\) and each working structure, PGEE was fitted to the data of 54 patients, and the four responses of the held-out patient were scored by the log score \(-\sum_{j=1}^{4}\{y_{ij}\log\widehat{\pi}_{ij}+(1-y_{ij})\log(1-\widehat{\pi}_{ij})\}\). The mean log score per patient was almost constant in \(\delta\), varying by at most \(0.010\) under each working structure. The Brier score, repeated stratified five-fold cross-validation, and QIC \citep{Pan2001} were similarly uninformative (Appendix E.4 of the Supporting Information). This flatness is a consequence of separation. Held-out responses of the four female patients on active treatment, which were all good, were predicted better under weaker penalization, whereas those of the remaining 51 patients were predicted better under stronger penalization, and the two contributions nearly canceled. Because the data do not discriminate between values of \(\delta\) in \((0,1]\), we retained the default \(\delta=1/2\).

The substantive conclusions were stable across odds-ratio working association structures. In the selected illustrative model, the strongest signals were baseline respiratory status, the treatment-by-gender interaction, and the visit-by-age interaction. Good baseline respiratory status was associated with an increased probability of good respiratory follow-up status (\(\widehat{\beta}_5>0\)). The active treatment was beneficial for both females and males, though the effect was attenuated in males due to the negative treatment-by-gender interaction (\(\widehat{\beta}_6<0\)). The estimated age effect at visit \(j\) on the probit scale is \(\widehat{\beta}_4+j\widehat{\beta}_7\), which is negative at later visits (\(j=3,4\)), suggesting that older patients are less likely to have good respiratory status as the trial progresses.

\hypertarget{sec:discussion}{%
\section{Discussion and Future Work}\label{sec:discussion}}

The difficulties that separation poses for ordinary GEE in practice have been documented across a range of biomedical and epidemiological settings, and existing penalized approaches address some but not all of these difficulties. The proposed PGEE framework advances this literature in two respects: it combines a Jeffreys-type penalty with a marginalized odds-ratio parameterization that remains well-defined under separation, and, unlike existing penalized GEE approaches, it is not restricted to the logit link but accommodates the probit, complementary log-log, and cauchit links. Unlike correlation-coefficient parameterizations, which are subject to collapse under separation or extreme sparsity (Section \ref{sec:cc}), marginalized odds ratios estimated from pooled \(2 \times 2\) tables do not suffer from this collapse, allowing PGEE to incorporate nonindependence structures with potential efficiency gains relative to working independence. Sections \ref{sec:sim} and \ref{sec:application} provide empirical evidence of this in regimes where ordinary GEE fails. Because the penalty contribution remains bounded while the GEE estimating function grows at rate \(N^{1/2}\), PGEE shares the same asymptotic distribution as ordinary GEE, and standard sandwich-based inference applies without modification (Section \ref{sec:pgee} and Appendix C.3 of the Supporting Information). The function \texttt{geewa\_binary()} of the \textsf{R} package \texttt{geer} implements the proposed estimators.

Across the separation scenarios, PGEE, OPGEE, and HPGEE achieved near-complete empirical convergence whereas ordinary GEE failed in every replication and RBR \citep{Touloumis2026bias} either failed or returned numerically degenerate solutions (Table \ref{tab:sim_separation}), consistent with the finiteness guarantee under the independence working structure. Under complete separation, the separation-inducing covariate was identified in all replications, while the irrelevant covariate was rejected in at most 0.65\% of replications, well below the nominal level. Under quasi-complete separation, PGEE and OPGEE generally exhibited smaller absolute bias than HPGEE, and point estimation improved systematically with sample size. In regular regimes and under monotone MCAR dropout, all penalized estimators closely matched ordinary GEE in convergence, coverage, and prediction. Across all regimes, the bias-corrected sandwich estimator \(\widehat{\boldsymbol \Sigma}_{M}\) gave coverage at least as high as the sandwich estimator \(\widehat{\boldsymbol \Sigma}_{R}\), although it still under-covered at \(N=30\) under quasi-complete separation, whereas model-based standard errors were unreliable under the misspecified working associations (Section \ref{sec:sim-variance}). The recommended workflow, described in Section \ref{sec:method} and illustrated in Section \ref{sec:application}, employs \(\widehat{\boldsymbol \Sigma}_{M}\) for inference with PGEE.

Four directions stand out for future development. First, the most immediate theoretical gap is establishing formal finiteness conditions for PGEE under nonindependence odds-ratio working parameterizations, addressing a gap noted more broadly by Geroldinger et al. \citep{Geroldinger2022}. Progress would likely require uniform control of \(\boldsymbol \Sigma_{0}\) as \(\widehat{\boldsymbol \alpha}\) becomes extreme, which may be achievable by imposing bounded odds-ratio estimates, as suggested in the practical workflow of Section \ref{sec:workflow}. Second, the respiratory trial shows that, under separation, criteria based on fit to the observed data, whether cross-validated predictive scores or QIC, carry little information about the penalty strength \(\delta\). A simulation study of cross-validated \(\delta\) across separated and regular regimes, or choices of \(\delta\) informed by external information on plausible effect sizes, would clarify when data-driven selection of \(\delta\) is informative. Third, extending penalized estimating-equation ideas to multinomial or ordinal correlated outcomes is a natural further step. In multinomial or ordinal settings, a Jeffreys-type penalty may not guarantee finite regression-parameter estimates, and alternative penalties may need to be developed. Fourth, the \texttt{geer} package also provides \texttt{geewa()}, which implements a Jeffreys-prior penalty for correlated continuous and count outcomes using a correlation-coefficient parameterization. Establishing the finiteness and asymptotic behavior of these estimators would extend the theoretical basis of penalized GEE beyond binary responses.

\hypertarget{supporting-information}{%
\section{Supporting Information}\label{supporting-information}}

Appendices A--F, Tables S1--S18, and Figure S1 are given in the Supporting Information, which follows the references in this preprint.

\hypertarget{acknowledgements}{%
\section*{Acknowledgements}\label{acknowledgements}}
\addcontentsline{toc}{section}{Acknowledgements}

The author thanks Ioannis Kosmidis for helpful discussions, and the Handling Editor and an anonymous reviewer for comments that improved the paper. For the purpose of open access, the author has applied a Creative Commons Attribution (CC-BY) license to any Author Accepted Manuscript version arising from this submission.

\hypertarget{funding}{%
\section*{Funding}\label{funding}}
\addcontentsline{toc}{section}{Funding}

This research received no specific grant from any funding agency in the public, commercial, or not-for-profit sectors.

\hypertarget{data-availability-statement}{%
\section*{Data Availability Statement}\label{data-availability-statement}}
\addcontentsline{toc}{section}{Data Availability Statement}

The respiratory illness trial data analyzed in Section \ref{sec:application} were published by Stokes, Davis and Koch \citep{stokes2012} and are distributed as the dataset \texttt{respiratory} in the \textsf{R} package \texttt{geer}. The package is available from CRAN at \url{https://CRAN.R-project.org/package=geer}, and version 0.1.1, which was used for all results, is available from GitHub at \url{https://github.com/AnestisTouloumis/geer}. The \textsf{R} scripts that reproduce all results in the paper and the Supporting Information are provided as ancillary files with this preprint.

  \bibliography{bibliography.bib}

\end{document}


\maketitle

\hypertarget{sec:guide}{%
\section{Overview and Notation}\label{sec:guide}}

This Supporting Information includes technical derivations for the Jeffreys-type penalty, additional theoretical remarks, extended simulation results, and supplementary analyses for the respiratory illness trial. \ref{sec:deriv} derives the Jeffreys-type adjustment \(\Lambda_{PG}\) used in the main paper. \ref{sec:additional} provides additional theoretical results, illustrating why correlation-coefficient parameterizations can behave poorly under separation, recording a sufficient condition related to first-order bias reduction, establishing the asymptotic equivalence of \(\widehat{\boldsymbol \beta}_{PG}\) and \(\widehat{\boldsymbol \beta}_{G}\), and illustrating, across \(2\times2\) configurations, the finiteness of PGEE and the behavior of correlation-coefficient parameterizations. \ref{sec:simtables} reports extended simulation results. \ref{sec:example} reports additional numerical results for the respiratory illness trial, including sensitivity to the penalty parameter and link function and a cross-validation assessment of the penalty strength. \ref{sec:rcode} provides \textsf{R} code to reproduce the analysis and documents the computing environment.

The following notation is used throughout and is consistent with Section 2 of the main paper. For matrix derivatives, we use the narrow definition of \citet{Magnus2010}. If \(\mathbf{a}=(a_1,\ldots,a_c)^{\top}\) is a vector function of \(\mathbf{b}=(b_1,\ldots,b_d)^{\top}\), then \(\partial \mathbf{a}/\partial \mathbf{b}^{\top}\) is the \(c\times d\) matrix with \((i,j)\)th element \(\partial a_i/\partial b_j\). If \(\mathbf{A}\) is a matrix function of \(\mathbf{a}\), then \(\mathcal{D}\mathbf{A}(\mathbf{a})=\partial\,\operatorname{vec}(\mathbf{A})/\partial \mathbf{a}^{\top}\), where \(\operatorname{vec}(\mathbf{A})\) stacks the columns of \(\mathbf{A}\).

\hypertarget{sec:deriv}{%
\section{Derivation of the Jeffreys-Type Penalty Adjustment}\label{sec:deriv}}

We derive the Jeffreys-type adjustment \(\boldsymbol \Lambda_{PG}\) under an odds-ratio parameterization of \(\boldsymbol \alpha\). \ref{sec:fisherinfo} establishes the GEE Fisher-information analog \(\boldsymbol \Sigma_{0}\), \ref{sec:dsigma0} derives \(\mathcal{D}\boldsymbol \Sigma_{0}(\boldsymbol \beta)\), and \ref{sec:lambdapg} combines these results to obtain \(\boldsymbol \Lambda_{PG}\) in closed form.

\hypertarget{sec:fisherinfo}{%
\subsection{\texorpdfstring{GEE Fisher-information Analog \(\boldsymbol \Sigma_{0}\)}{GEE Fisher-information Analog \textbackslash boldsymbol \textbackslash Sigma\_\{0\}}}\label{sec:fisherinfo}}

The derivative of \(\mathbf U_{i}\) with respect to \(\boldsymbol \beta\) is given by
\[
\begin{aligned}
\mathcal{D}\mathbf U_{i} \!\left( \boldsymbol \beta\right)
& = 
  \left[\left( \mathbf S_{i}^{\top} \mathbf V^{-1}_{i}\right) \otimes \mathbf I_{p}\right] 
  \mathcal{D}\mathbf D_{i}^{\top} \!\left( \boldsymbol \beta\right)
  +
  \left( \mathbf S_{i}^{\top} \otimes \mathbf D_{i}^{\top} \right)
  \mathcal{D}\mathbf V^{-1}_{i}\!\left( \boldsymbol \beta\right)
  +
  \mathbf D_{i}^{\top} \mathbf V^{-1}_{i}\mathcal{D}\mathbf S_{i}\!\left( \boldsymbol \beta\right) \notag\\[4pt]
& = 
 \left[\left( \mathbf S_{i}^{\top} \mathbf V^{-1}_{i}\right) \otimes \mathbf I_{p}\right]
 \mathcal{D}\mathbf D_{i}^{\top} \!\left( \boldsymbol \beta\right)
+ 
 \left( \mathbf S_{i}^{\top} \otimes \mathbf D_{i}^{\top} \right) 
 \mathcal{D}\mathbf V^{-1}_{i}\!\left( \boldsymbol \beta\right) 
- 
 \mathbf D_{i}^{\top} \mathbf V^{-1}_{i}\mathbf D_{i}.
\end{aligned}
\]
Since \(E(\mathbf S_{i}) = \mathbf{0}_{n_i}\), the first two terms vanish in expectation. Therefore,
\[
E \!\left[ -\,\mathcal{D}\mathbf U_{G} \!\left( \boldsymbol \beta\right) \right]
=
E \!\left[
  -\,\sum_{i=1}^{N}
  \mathcal{D}\mathbf U_{i} \!\left( \boldsymbol \beta\right)
\right]
=
\sum_{i=1}^{N}\mathbf D_{i}^{\top} \mathbf V^{-1}_{i}\mathbf D_{i}
= 
\boldsymbol \Sigma_{0}.
\]
This identifies \(\boldsymbol \Sigma_{0}\) as the GEE Fisher-information analog entering the Jeffreys-type penalty.

\hypertarget{sec:dsigma0}{%
\subsection{\texorpdfstring{Derivative of \(\boldsymbol \Sigma_{0}\) with Respect to \(\boldsymbol \beta\)}{Derivative of \textbackslash boldsymbol \textbackslash Sigma\_\{0\} with Respect to \textbackslash boldsymbol \textbackslash beta}}\label{sec:dsigma0}}

Note that
\begin{align*}
\mathcal{D}\!\left( \mathbf D_{i}^{\top} \mathbf V^{-1}_{i}\mathbf D_{i}\right) \!\left( \boldsymbol \pi_{i}\right)
= {} & 
    \left[ \mathbf I_{p}\otimes \left( \mathbf D_{i}^{\top} \mathbf V^{-1}_{i}\right) \right]
    \mathcal{D}\mathbf D_{i}\!\left( \boldsymbol \pi_{i}\right)
    +
    \left[ \left( \mathbf D_{i}^{\top} \mathbf V^{-1}_{i}\right) \otimes \mathbf I_{p}\right]
    \mathcal{D}\mathbf D_{i}^{\top} \!\left( \boldsymbol \pi_{i}\right) \notag\\
    & \quad +
    \left( \mathbf D_{i}\otimes \mathbf D_{i}\right)^{\top}
    \mathcal{D}\mathbf V^{-1}_{i}\!\left( \boldsymbol \pi_{i}\right) \notag\\[4pt]
= {} &
    \left[ \mathbf I_{p}\otimes \left( \mathbf D_{i}^{\top} \mathbf V^{-1}_{i}\right) \right]
    \!\left( \mathbf D_{i}^{\top} \otimes \mathbf I_{n_{i}}\right)
    \mathbf K_{n_{i}}\boldsymbol \Delta_{i}^{\star}
    +
    \left[
        \left( \mathbf D_{i}^{\top} \mathbf V^{-1}_{i}\right) \otimes \mathbf I_{p}
    \right]
    \!\left( \mathbf I_{n_{i}}\otimes \mathbf D_{i}^{\top} \right)
    \mathbf K_{n_{i}}\boldsymbol \Delta_{i}^{\star} \notag\\
    & \quad +
    \left( \mathbf D_{i}\otimes \mathbf D_{i}\right)^{\top}
    \left( \mathbf V^{-1}_{i}\otimes \mathbf V^{-1}_{i}\right)
    \mathcal{D}\mathbf V_i\!\left( \boldsymbol \pi_{i}\right) \notag\\[4pt]
= {} &
    \left( \mathbf D_{i}\otimes \mathbf D_{i}\right)^{\top}
    \!\left[
        \left( \mathbf V^{-1}_{i}\oplus \mathbf V^{-1}_{i}\right)
        \mathbf K_{n_{i}}\boldsymbol \Delta_{i}^{\star}
        -
        \left( \mathbf V^{-1}_{i}\otimes \mathbf V^{-1}_{i}\right)
        \mathcal{D}\mathbf V_i\!\left( \boldsymbol \pi_{i}\right)
    \right].
\end{align*}
Applying the chain rule \(\mathcal{D}(\cdot)(\boldsymbol \beta) = \mathcal{D}(\cdot)(\boldsymbol \pi_{i})\,\mathbf D_{i}\) and summing over subjects gives
\[
\mathcal{D}\boldsymbol \Sigma_{0} \!\left( \boldsymbol \beta\right)
= 
\sum_{i=1}^{N}
\left( \mathbf D_{i}\otimes \mathbf D_{i}\right)^{\top}
    \!\left[
        \left( \mathbf V^{-1}_{i}\oplus \mathbf V^{-1}_{i}\right)
        \mathbf K_{n_{i}}\boldsymbol \Delta_{i}^{\star}
        -
        \left( \mathbf V^{-1}_{i}\otimes \mathbf V^{-1}_{i}\right)
        \mathcal{D}\mathbf V_i\!\left( \boldsymbol \pi_{i}\right)
    \right]
    \mathbf D_{i}.
\]
Here \(\mathbf K_{n_{i}}= \sum_{j=1}^{n_i} \left[ \left( \mathbf e_{ij} \mathbf e_{ij}^{T} \right) \otimes \mathbf e^{T}_{ij}\right]\), \(\mathbf{e}_{ij}\) is an \(n_i\)-variate vector whose \(j\)th element is equal to one and all others are equal to zero. Note that \(\oplus\) denotes the Kronecker sum, so that \(\mathbf V^{-1}_{i}\oplus \mathbf V^{-1}_{i}= \mathbf I_{n_{i}}\otimes \mathbf V^{-1}_{i}+ \mathbf V^{-1}_{i}\otimes \mathbf I_{n_{i}}\), and \(\boldsymbol \Delta_{i}^{\star}\) is an \(n_i \times n_i\) diagonal matrix with entries
\[
\left[\boldsymbol \Delta_{i}^{\star}\right]_{jj}=\frac{\partial \left[\boldsymbol \Delta_{i}
\right]_{jj}}{\partial \pi_{ij}}=\frac{\partial^2 \pi_{ij}\bigr/\partial 
\eta_{ij}^2}{\partial \pi_{ij}\bigr/\partial \eta_{ij}}.
\]
Here \(\boldsymbol \Delta_{i}\) is the \(n_i \times n_i\) diagonal matrix with \(j\)th diagonal entry \(\partial \pi_{ij} / \partial \eta_{ij}\), where \(\eta_{ij}\) denotes the linear predictor for subject \(i\) at time \(j\), so that \(\pi_{ij} = g^{-1}(\eta_{ij})\) for link function \(g\).

To specify \(\mathcal{D}\mathbf V_i(\boldsymbol \pi_{i})\) under marginalized odds-ratio working association, we define the block matrices \(\mathbf g^{(1)}_{i}\) and \(\mathbf g^{(2)}_{i}\) below, which collect derivatives of the working joint probabilities with respect to the marginal means.

The derivative of \(\mathbf V_i\) with respect to \(\boldsymbol \pi_{i}\) under an odds-ratio parameterization is
\[
\mathcal{D}\mathbf V_i\!\left( \boldsymbol \pi_{i}\right) =
  \mathbf K_{n_{i}}
    +
    \mathbf g^{(1)}_{i} 
    +
    \mathbf g^{(2)}_{i}
    -
    \boldsymbol \pi_{i}\otimes \mathbf I_{n_{i}}
    -
    \mathbf I_{n_{i}}\otimes \boldsymbol \pi_{i}.
\]
Here,
\[
\mathbf g^{(1)}_{i} =  
\begin{bmatrix}
        \mathbf g^{(1)}_{i1} \\
        \mathbf g^{(1)}_{i2}\\
        \vdots\\
        \mathbf g^{(1)}_{i{n_i}}
    \end{bmatrix},
\qquad
\mathbf g^{(2)}_{i} =  
\begin{bmatrix}
        \mathbf g^{(2)}_{i1} \\
        \mathbf g^{(2)}_{i2}\\
        \vdots\\
        \mathbf g^{(2)}_{i{n_i}}
    \end{bmatrix},
\]
are block matrices built from \(n_i\) submatrices \(\mathbf g^{(1)}_{ir}\), \(\mathbf g^{(2)}_{ir}\) each of dimension \(n_i \times n_i\) with elements
\[
\bigl[ \mathbf g^{(1)}_{ir} \bigr]_{jk}
=
    \begin{cases}
        \dfrac{\partial \pi_{ijr}}{\partial \pi_{ij}}, & \text{if } j = k \neq r,
        \\[6pt] 
        0, &  \text{otherwise},
    \end{cases}
\qquad
\bigl[ \mathbf g^{(2)}_{ir} \bigr]_{jk}
=
    \begin{cases}
        \dfrac{\partial \pi_{ijr}}{\partial \pi_{ir}}, & \text{if } k = r \neq j,
        \\[6pt] 
        0, &  \text{otherwise}.
    \end{cases}
\]
where \(\pi_{ijj^{\prime}}\) is the working assumption about the joint probability \(\Pr\left(Y_{ij}=1 , Y_{ij^{\prime}}=1 \mid \mathbf x_{ij}, \mathbf x_{ij^{\prime}}\right)\). Moreover,
\[
\frac{{\partial \pi}_{ijj^{\prime}}}{\partial \pi_{ij^{\prime}}}
=
    \frac{1}{2}
   +
    \frac{1}{2}
    \frac{
        \pi_{ij}
        +
        \pi_{ij^{\prime}}
        -
        \alpha_{ijj^{\prime}}
        \left(
            \pi_{ij}
            -
            \pi_{ij^{\prime}}
        \right)
        -
        1
    }{
        f_{ijj^{\prime}} 
        - 
        2 \!\left(
            \alpha_{ijj^{\prime}} - 1 
        \right)
        \pi_{ijj^{\prime}}
    },
\]
where \(\alpha_{ijj^{\prime}}\) denotes the odds ratio for the time pair \((j,j^{\prime})\), written here with subject index \(i\) for generality. Under the marginalized contingency-table estimator used in the main paper, \(\alpha_{ijj^{\prime}}\) reduces to the pooled \(\alpha_{jj^{\prime}}\), which is common across subjects.

\hypertarget{sec:lambdapg}{%
\subsection{\texorpdfstring{Closed-Form Expression for \(\boldsymbol \Lambda_{PG}\)}{Closed-Form Expression for \textbackslash boldsymbol \textbackslash Lambda\_\{PG\}}}\label{sec:lambdapg}}

Since the Jeffreys-prior analog is proportional to \(\{\det(\boldsymbol \Sigma_{0})\}^{\delta}\), the derivative of its log-penalty with respect to \(\boldsymbol \beta\) is obtained by differentiating \(\delta\log\det(\boldsymbol \Sigma_{0})\) with respect to \(\boldsymbol \beta\). For \(r=1,\ldots,p\), let \(\boldsymbol \Sigma_{0r}^{\star}\) denote the symmetric \(p\times p\) matrix with \((k,k^{\prime})\)th element
\[
\bigl[\boldsymbol \Sigma_{0r}^{\star}\bigr]_{kk^{\prime}}=\frac{\partial
\bigl[\boldsymbol \Sigma_{0}\bigr]_{kk^{\prime}}}{\partial \beta_r},
\qquad k,k^{\prime}=1,\ldots,p.
\]
Symmetry of \(\boldsymbol \Sigma_{0r}^{\star}\) follows from the symmetry of \(\boldsymbol \Sigma_{0}\), since \(\partial[\boldsymbol \Sigma_{0}]_{kk^{\prime}}/\partial\beta_r =\partial[\boldsymbol \Sigma_{0}]_{k^{\prime}k}/\partial\beta_r\) for all \(k, k^{\prime}\). Then the \(r\)th element of \(\boldsymbol \Lambda_{PG}\) is
\[
\bigl[\boldsymbol \Lambda_{PG}\bigr]_r
=\delta\,\operatorname{tr}\!\bigl(\boldsymbol \Sigma_{0}^{-1}\boldsymbol \Sigma_{0r}^{\star}\bigr)
=\delta\,\bigl[\operatorname{vec}\left( {\boldsymbol \Sigma_{0r}^{\star}} \right)\bigr]^{\top}\operatorname{vec}\left( {\boldsymbol \Sigma_{0}^{-1}} \right),
\]
which leads to
\[
\boldsymbol \Lambda_{PG}
=\delta\,\bigl[\mathcal{D}\boldsymbol \Sigma_{0}(\boldsymbol \beta)\bigr]^{\top}\operatorname{vec}\left( {\boldsymbol \Sigma_{0}^{-1}} \right),
\]
as stated in the main text.

\hypertarget{sec:additional}{%
\section{Additional Theoretical Remarks}\label{sec:additional}}

We provide four supplementary remarks below.

\hypertarget{sec:ccfails}{%
\subsection{Failure of Correlation-Coefficient Parameterizations under Separation}\label{sec:ccfails}}

Under a correlation‑coefficient parameterization, the working covariance matrix
\(\mathbf V_i\) is
\[
\mathbf V_i=\phi\,\mathbf A_{i}^{1/2}\,\mathbf R(\boldsymbol \rho)\,\mathbf A_{i}^{1/2},
\]
where \(\phi\) is the dispersion parameter, \(\mathbf A_{i}\) is an \(n_i\times n_i\) diagonal matrix with \(j\)th diagonal element \(\pi_{ij}\bigl(1-\pi_{ij}\bigr)\), and \(\mathbf R(\boldsymbol \rho)\) is the common correlation matrix across subjects, fully specified by the parameter vector \(\boldsymbol \rho\) (association parameterized via correlation coefficients rather than odds ratios). Under this parameterization, the PGEE estimator solves the PGEE equations, with \(\mathcal{D}\mathbf V_i(\boldsymbol \pi_{i})\) entering \(\mathcal{D}\boldsymbol \Sigma_{0}(\boldsymbol \beta)\) via \ref{sec:dsigma0}. In particular,
\[
\mathcal{D}\mathbf V_i(\boldsymbol \pi_{i})=
-
\bigl(\mathbf V_i\oplus \mathbf V_i\bigr)\,\mathbf K_{n_{i}}\,\mathbf A_{i}^{\star},
\]
and \(\mathbf A_{i}^{\star}\) is an \(n_i\times n_i\) diagonal matrix whose \(j\)th diagonal entry
is
\[
\bigl[\mathbf A_{i}^{\star}\bigr]_{jj}=\frac{\pi_{ij}-0.5}{\pi_{ij}\bigl(1-\pi_{ij}\bigr)}.
\]
We estimate \(\phi\) and \(\boldsymbol \rho\) by the method‑of‑moments \citep{Liang1986}.

To illustrate the failure of correlation-coefficient parameterizations, consider the marginal model
\[
\log\!\left(\frac{\pi_{ij}}{1-\pi_{ij}}\right)=\beta_0+\beta_1\,I(j=2),
\]
for all \(i=1,\ldots,N\) and \(j=1,2\), where \(I(A)\) is the indicator of event \(A\). To construct separable data, let \(\mathbf Y_i=(0,1)^{\top}\) for all \(i\), so that the time indicator \(I(j=2)\) perfectly predicts the response and induces complete separation.

First, consider PGEE under working independence with \(\delta=1/2\) and dispersion parameter \(\phi>0\) to be estimated. The parameters \(\pi_{1}=\pi_{i1}=1-\pi_{i2}\) and \(\phi\) are obtained by iterating
\[
\phi^{(m)}=\frac{N}{N-1}\,\frac{\pi^{(m-1)}_{1}}{1-\pi^{(m-1)}_{1}}
\qquad\text{and}\qquad
\pi^{(m)}_{1}=\frac{\phi^{(m)}/2}{N+\phi^{(m)}},
\]
with \(\phi^{(0)}=1\) and \(\pi^{(0)}_{1}=\frac{0.5}{N+1}\), where \(\phi^{(m)}\) and \(\pi_1^{(m)}\) denote the \(m\)th iterates. One can show that \((\pi^{(m)}_1,\phi^{(m)})\to(0,0)\) as \(m\to\infty\), implying that the estimated \(\mathbf V_i\) becomes nearly singular (ill‑conditioned), while \(\beta_0=\log\!\bigl(\pi_1/(1-\pi_1)\bigr)\to -\infty\) and \(\beta_1=-2\,\beta_0\to\infty\).

Since for binary responses \(\phi\) is sometimes fixed at \(1\) rather than estimated, we consider this case separately. Then \(\pi_{1}\) and the correlation parameter \(\rho\) are obtained by iterating
\[
\rho^{(m)}=-\frac{N}{N-2}\,\frac{\pi^{(m-1)}_{1}}{1-\pi^{(m-1)}_{1}}
\qquad\text{and}\qquad
\pi^{(m)}_{1}=\frac{\bigl(1-\rho^{(m)}\bigr)/2}{N+1-\rho^{(m)}}.
\]
For \(N\ge 3\), one can show \(-(N-2)^{-1}\le \rho^{(m)}<0\) for \(m\ge 1\). Therefore, for \(N\ge 10\), \(-0.125\le \widehat{\rho}<0\). Although the data imply perfect negative association in the sense that \(\mathbf Y_i=(0,1)^{\top}\) for all \(i\), the method‑of‑moments estimate of the exchangeable correlation remains close to zero for moderate \(N\), illustrating the tendency of correlation-coefficient parameterizations to collapse toward working independence under separation. While the PGEE estimator under this parameterization converges in this particular example, the resulting estimates are nearly indistinguishable from those under working independence, even for small \(N\).

More generally, under separation, moment-based correlation-parameterized working structures tend to collapse toward the working independence model, and estimating the dispersion \(\phi\) can render \(\mathbf V_i\) nearly singular. These considerations motivate the use of marginalized odds-ratio parameterizations in the separated and sparse regimes that are the focus of the main paper.

\hypertarget{sec:bias}{%
\subsection{A Sufficient Condition for First-Order Bias Reduction}\label{sec:bias}}

We record a sufficient condition under which the Jeffreys‑type adjustment \(\boldsymbol \Lambda_{PG}\) coincides, to first order, with a bias‑reducing adjustment for the ordinary GEE estimator. If
\[
\boldsymbol \Lambda_{G}-\boldsymbol \Lambda_{PG}=O\!\left(N^{-1/2}\right),
\]
then \(\widehat{\boldsymbol \beta}_{PG}\) has reduced bias relative to \(\widehat{\boldsymbol \beta}_{G}\) \citep[see][]{Touloumis2026bias}. This condition is sufficient but not necessary. It is generally not expected to hold for correlated binary data because \(\boldsymbol \Lambda_{PG}\) depends only on \(\boldsymbol \Sigma_{0}\) and its derivative, whereas \(\boldsymbol \Lambda_{G}\), the ideal bias-reducing adjustment for ordinary GEE, involves expectations of the first and second derivatives of the GEE equations with respect to \(\boldsymbol \beta\).

\hypertarget{sec:asymp}{%
\subsection{\texorpdfstring{Asymptotic Equivalence of \(\widehat{\boldsymbol \beta}_{PG}\) and \(\widehat{\boldsymbol \beta}_{G}\)}{Asymptotic Equivalence of \textbackslash widehat\{\textbackslash boldsymbol \textbackslash beta\}\_\{PG\} and \textbackslash widehat\{\textbackslash boldsymbol \textbackslash beta\}\_\{G\}}}\label{sec:asymp}}

We adopt the regularity conditions of Theorem 2 of \citet{Liang1986}. Let \(\mathcal{B}\) be a compact set with the true value \(\boldsymbol \beta_0\) in its interior, and let \(\mathcal{A}\subset(0,\infty)^{C}\) be a compact set with \(\boldsymbol \alpha^{\ast}\) in its interior, where \(\boldsymbol \alpha^{\ast}\) is the probability limit of \(\widehat{\boldsymbol \alpha}\). We further assume the following conditions.

(C1) The cluster sizes satisfy \(n_i\le n<\infty\) for all \(i\), and the covariate vectors \(\mathbf x_{ij}\) lie in a compact set.

(C2) The inverse link \(h=g^{-1}\) is three times continuously differentiable with \(h^{\prime}(\eta)>0\). Together with (C1), this implies that for \(\boldsymbol \beta\in\mathcal{B}\) the linear predictors are bounded, so that \(\epsilon\le\pi_{ij}\le 1-\epsilon\) for some \(\epsilon>0\).

(C3) There exist constants \(0<\lambda_{\min}\le\lambda_{\max}<\infty\) such that the eigenvalues of \(\mathbf V_i(\boldsymbol \beta,\boldsymbol \alpha)\) lie in \([\lambda_{\min},\lambda_{\max}]\) for all \(i\) and all \((\boldsymbol \beta,\boldsymbol \alpha)\in\mathcal{B}\times\mathcal{A}\).

(C4) There exists \(c>0\) such that the smallest eigenvalue of \(N^{-1}\boldsymbol \Sigma_{0}(\boldsymbol \beta,\boldsymbol \alpha)\) is at least \(c\) for all \((\boldsymbol \beta,\boldsymbol \alpha)\in\mathcal{B}\times\mathcal{A}\) and all sufficiently large \(N\).

Conditions (C1), (C2) and (C4) are standard in GEE asymptotics. Condition (C3) is stated explicitly because, for \(n_i\ge 3\), a pairwise odds-ratio specification does not by itself guarantee that the resulting \(\mathbf V_i\) is positive definite.

\textbf{Lemma 1.} Under (C1)--(C4), there exists a constant \(K<\infty\), not depending on \(N\) or on \(\delta\in(0,1]\), such that for all \((\boldsymbol \beta,\boldsymbol \alpha)\in\mathcal{B}\times\mathcal{A}\), \(r,s=1,\ldots,p\), and \(k=1,\ldots,C\),
\[
\bigl|[\boldsymbol \Lambda_{PG}]_r\bigr|\le K,\qquad
\left|\frac{\partial[\boldsymbol \Lambda_{PG}]_r}{\partial\beta_s}\right|\le K,\qquad
\left|\frac{\partial[\boldsymbol \Lambda_{PG}]_r}{\partial\alpha_k}\right|\le K .
\]

\textit{Proof.} Write \(\mathbf G_{i}=\mathbf D_{i}^{\top}\mathbf V^{-1}_{i}\mathbf D_{i}\), so that \(\boldsymbol \Sigma_{0}=\sum_{i=1}^{N}\mathbf G_{i}\), and let \(\|\cdot\|\) denote the spectral norm. By (C1) and (C2), the elements of \(\mathbf D_{i}=\boldsymbol \Delta_{i}\mathbf X_{i}\) and their first and second derivatives with respect to \(\boldsymbol \beta\) are bounded uniformly in \(i\) and over \(\mathcal{B}\), because they involve \(h^{\prime}\), \(h^{\prime\prime}\) and \(h^{\prime\prime\prime}\) evaluated on a bounded set and multiplied by bounded covariates. The diagonal elements of \(\mathbf V_i\) are \(\pi_{ij}(1-\pi_{ij})\), and the off-diagonal elements are \(\pi^{\star}_{ijj^{\prime}}-\pi_{ij}\pi_{ij^{\prime}}\), where \(\pi^{\star}_{ijj^{\prime}}\) is the admissible root of the quadratic equation \((\alpha_{jj^{\prime}}-1)x^{2}-f_{ijj^{\prime}}x+\alpha_{jj^{\prime}}\pi_{ij}\pi_{ij^{\prime}}=0\) (Section 2.2 of the main paper). Its discriminant \(f_{ijj^{\prime}}^{2}-4\alpha_{jj^{\prime}}(\alpha_{jj^{\prime}}-1)\pi_{ij}\pi_{ij^{\prime}}\) is strictly positive for \(\pi_{ij},\pi_{ij^{\prime}}\in(0,1)\) and \(\alpha_{jj^{\prime}}>0\), so the root is simple. By the implicit function theorem, \(\pi^{\star}_{ijj^{\prime}}\) is therefore an infinitely differentiable function of \((\pi_{ij},\pi_{ij^{\prime}},\alpha_{jj^{\prime}})\), including at \(\alpha_{jj^{\prime}}=1\). On the compact set \([\epsilon,1-\epsilon]^{2}\times\mathcal{A}\), it and its derivatives are bounded. Hence the elements of \(\mathbf V_i\), and their derivatives up to second order in \(\boldsymbol \beta\) and first order in \(\boldsymbol \alpha\), are bounded uniformly in \(i\). By (C3), \(\|\mathbf V^{-1}_{i}\|\le\lambda_{\min}^{-1}\), and derivatives of \(\mathbf V^{-1}_{i}\) are bounded via \(\partial\mathbf V^{-1}_{i}=-\mathbf V^{-1}_{i}(\partial\mathbf V_i)\mathbf V^{-1}_{i}\). It follows that there is a constant \(K_1\) such that \(\mathbf G_{i}\), \(\partial\mathbf G_{i}/\partial\beta_r\), \(\partial^{2}\mathbf G_{i}/\partial\beta_r\partial\beta_s\), \(\partial\mathbf G_{i}/\partial\alpha_k\) and \(\partial^{2}\mathbf G_{i}/\partial\beta_r\partial\alpha_k\) all have spectral norm at most \(K_1\), uniformly in \(i\) and over \(\mathcal{B}\times\mathcal{A}\). Summing over subjects, \(\|\boldsymbol \Sigma_{0r}^{\star}\|\le NK_1\), \(\|\partial\boldsymbol \Sigma_{0r}^{\star}/\partial\beta_s\|\le NK_1\), \(\|\partial\boldsymbol \Sigma_{0}/\partial\alpha_k\|\le NK_1\) and \(\|\partial\boldsymbol \Sigma_{0r}^{\star}/\partial\alpha_k\|\le NK_1\), while (C4) gives \(\|\boldsymbol \Sigma_{0}^{-1}\|\le (cN)^{-1}\). Hence
\[
\bigl|[\boldsymbol \Lambda_{PG}]_r\bigr|=\delta\bigl|\operatorname{tr}(\boldsymbol \Sigma_{0}^{-1}\boldsymbol \Sigma_{0r}^{\star})\bigr| \le\delta p\,\frac{K_1}{c}.
\]
Since \(\partial\boldsymbol \Sigma_{0}^{-1}/\partial\beta_s=-\boldsymbol \Sigma_{0}^{-1}\boldsymbol \Sigma_{0s}^{\star}\boldsymbol \Sigma_{0}^{-1}\),
\[
\frac{\partial[\boldsymbol \Lambda_{PG}]_r}{\partial\beta_s}
=\delta\left\{\operatorname{tr}\!\left(\boldsymbol \Sigma_{0}^{-1}\frac{\partial\boldsymbol \Sigma_{0r}^{\star}}{\partial\beta_s}\right)
-\operatorname{tr}\!\left(\boldsymbol \Sigma_{0}^{-1}\boldsymbol \Sigma_{0s}^{\star}\boldsymbol \Sigma_{0}^{-1}\boldsymbol \Sigma_{0r}^{\star}\right)\right\},
\]
which is bounded in absolute value by \(\delta p\,(K_1/c+K_1^{2}/c^{2})\). The derivative with respect to \(\alpha_k\) has the same form, with \(\partial\boldsymbol \Sigma_{0}/\partial\alpha_k\) in place of \(\boldsymbol \Sigma_{0s}^{\star}\), and obeys the same bound. The result follows with \(K=p\,(K_1/c+K_1^{2}/c^{2})\), since \(\delta\le 1\). \(\square\)

Because \(\widehat{\boldsymbol \alpha}\overset{p}{\to}\boldsymbol \alpha^{\ast}\), the event \(\widehat{\boldsymbol \alpha}\in\mathcal{A}\) has probability tending to one, and Lemma 1 therefore implies that \(\boldsymbol \Lambda_{PG}(\boldsymbol \beta,\widehat{\boldsymbol \alpha})\) and its first derivatives are \(O_p(1)\) uniformly over \(\boldsymbol \beta\in\mathcal{B}\). The PGEE estimator \(\widehat{\boldsymbol \beta}_{PG}\) solves
\[
\mathbf{U}_{PG}(\boldsymbol \beta, \widehat{\boldsymbol{\alpha}}) 
= 
\mathbf{U}_G(\boldsymbol \beta, \widehat{\boldsymbol \alpha}) + \boldsymbol \Lambda_{PG}(\boldsymbol \beta, \widehat{\boldsymbol \alpha}) 
= \mathbf{0}_p.
\]
Let \(\boldsymbol \beta_0\) denote the true value of \(\boldsymbol \beta\). Consistency of \(\widehat{\boldsymbol \beta}_{PG}\) follows from the \(O_p(1)\) condition on \(\boldsymbol \Lambda_{PG}\) together with the identifiability conditions of Theorem 2 of \citet{Liang1986}, which ensure that \(\widehat{\boldsymbol \beta}_{PG}\overset{p}{\to} \boldsymbol \beta_0\). A Taylor expansion of \(\mathbf{U}_{PG}(\widehat{\boldsymbol \beta}_{PG}, \widehat{\boldsymbol \alpha})\) around \(\boldsymbol \beta_0\) gives
\[
\sqrt{N}(\widehat{\boldsymbol \beta}_{PG}- \boldsymbol \beta_0)
=
-\left[
\frac{1}{N}
\frac{\partial \mathbf{U}_{PG}(\boldsymbol \beta^*, \widehat{\boldsymbol \alpha})}{\partial \boldsymbol \beta^{\top}}
\right]^{-1}
\frac{1}{\sqrt{N}}
\mathbf{U}_{PG}(\boldsymbol \beta_0, \widehat{\boldsymbol \alpha}),
\]
where \(\boldsymbol \beta^*\) lies between \(\widehat{\boldsymbol \beta}_{PG}\) and \(\boldsymbol \beta_0\). By Lemma 1, \(\boldsymbol \Lambda_{PG}\) and its first derivative with respect to \(\boldsymbol \beta\) are \(O_p(1)\) uniformly over \(\mathcal{B}\), and in particular at \(\boldsymbol \beta_0\) and at \(\boldsymbol \beta^*\). Hence
\[
\frac{1}{\sqrt{N}}\boldsymbol \Lambda_{PG}(\boldsymbol \beta_0, \widehat{\boldsymbol \alpha}) = o_p(1)
\]
and
\[
\frac{1}{N}
\frac{\partial \boldsymbol \Lambda_{PG}(\boldsymbol \beta^*, \widehat{\boldsymbol \alpha})}{\partial \boldsymbol \beta^{\top}}
= o_p(1).
\]
Therefore,
\[
\frac{1}{\sqrt{N}}\mathbf{U}_{PG}(\boldsymbol \beta_0, \widehat{\boldsymbol \alpha})
=
\frac{1}{\sqrt{N}}\mathbf{U}_G(\boldsymbol \beta_0, \widehat{\boldsymbol \alpha}) + o_p(1)
\]
and
\[
\frac{1}{N}
\frac{\partial \mathbf{U}_{PG}(\boldsymbol \beta^*, \widehat{\boldsymbol \alpha})}{\partial \boldsymbol \beta^{\top}}
=
\frac{1}{N}
\frac{\partial \mathbf{U}_{G}(\boldsymbol \beta^*, \widehat{\boldsymbol \alpha})}{\partial \boldsymbol \beta^{\top}}
+ o_p(1).
\]
The conclusion then follows from Theorem 2 of \citet{Liang1986}, establishing that \(\widehat{\boldsymbol \beta}_{PG}\) and \(\widehat{\boldsymbol \beta}_{G}\) are asymptotically equivalent and that the sandwich covariance estimators \(\widehat{\boldsymbol \Sigma}_{R}\) and \(\widehat{\boldsymbol \Sigma}_{M}\) remain valid for inference after substituting \(\widehat{\boldsymbol \beta}_{PG}\) for \(\widehat{\boldsymbol \beta}_{G}\).

\textbf{Corollary 1.} Suppose, in addition, that \(\widehat{\boldsymbol \beta}_{G}\) exists with probability tending to one. Then \(\widehat{\boldsymbol \beta}_{PG}-\widehat{\boldsymbol \beta}_{G}=O_p(N^{-1})\), uniformly in \(\delta\in(0,1]\).

\textit{Proof.} Subtracting \(\mathbf{U}_G(\widehat{\boldsymbol \beta}_{G},\widehat{\boldsymbol \alpha})=\mathbf{0}_p\) from \(\mathbf{U}_G(\widehat{\boldsymbol \beta}_{PG},\widehat{\boldsymbol \alpha})+\boldsymbol \Lambda_{PG}(\widehat{\boldsymbol \beta}_{PG},\widehat{\boldsymbol \alpha})=\mathbf{0}_p\), and applying the mean value theorem to each component of \(\mathbf{U}_G\), gives
\[
\mathbf{0}_p=\mathbf{H}_N\,(\widehat{\boldsymbol \beta}_{PG}-\widehat{\boldsymbol \beta}_{G})+\boldsymbol \Lambda_{PG}(\widehat{\boldsymbol \beta}_{PG},\widehat{\boldsymbol \alpha}),
\]
where the \(j\)th row of \(\mathbf{H}_N\) is the \(j\)th row of \(\partial\mathbf{U}_G/\partial\boldsymbol \beta^{\top}\), evaluated at a point \(\boldsymbol \beta^{\dagger}_j\) between \(\widehat{\boldsymbol \beta}_{PG}\) and \(\widehat{\boldsymbol \beta}_{G}\). Both estimators are consistent for \(\boldsymbol \beta_0\), so under the conditions of Theorem 2 of \citet{Liang1986}, \(N^{-1}\mathbf{H}_N\) converges in probability to a nonsingular matrix, the negative of the limit of \(N^{-1}\boldsymbol \Sigma_{0}(\boldsymbol \beta_0,\boldsymbol \alpha^{\ast})\). Therefore
\[
\widehat{\boldsymbol \beta}_{PG}-\widehat{\boldsymbol \beta}_{G}=-\bigl(N^{-1}\mathbf{H}_N\bigr)^{-1}N^{-1}\boldsymbol \Lambda_{PG}(\widehat{\boldsymbol \beta}_{PG},\widehat{\boldsymbol \alpha})=O_p(N^{-1})
\]
by Lemma 1. The bound is uniform in \(\delta\in(0,1]\) because \(K\) does not depend on \(\delta\). \(\square\)

Two remarks follow. First, because the bounds in Lemma 1 and Corollary 1 do not depend on \(\delta\), replacing a fixed \(\delta\) by a data-dependent value in \((0,1]\), such as one chosen by cross-validation (\ref{sec:example-cv}), leaves the first-order asymptotic distribution of \(\widehat{\boldsymbol \beta}_{PG}\) unchanged. Second, conditions (C2) and (C4) describe the behavior of PGEE as \(N\) increases. They do not describe a fixed sample in which separation occurs. In this case, fitted probabilities approach 0 or 1, \(\widehat{\boldsymbol \beta}_{G}\) does not exist, and the role of \(\boldsymbol \Lambda_{PG}\) is to provide finite estimates at least under working independence (Section 2.4 of the main paper).

\hypertarget{sec:twobytwo}{%
\subsection{\texorpdfstring{Finiteness across \(2\times2\) Configurations}{Finiteness across 2\textbackslash times2 Configurations}}\label{sec:twobytwo}}

Consider again the design of \ref{sec:ccfails}, with \(N\) subjects, two occasions, and the marginal model \(g(\pi_{ij})=\beta_0+\beta_1 I(j=2)\), and let \(s_j=\sum_{i=1}^{N}Y_{ij}\) for \(j=1,2\). The marginal probabilities \(\boldsymbol \pi_{i}=(\pi_1,\pi_2)^{\top}\) and the design matrix \(\mathbf X_{i}\) are common to all subjects. For the odds-ratio working structures of Section 2.2 of the main paper, so is \(\mathbf V_i\), because the marginalized odds ratios are pooled over subjects. Writing \(\mathbf D\) and \(\mathbf V\) for the common values of \(\mathbf D_{i}\) and \(\mathbf V_i\), the GEE estimating equations reduce to
\[
\mathbf U_{G}=\mathbf D^{\top}\mathbf V^{-1}\sum_{i=1}^{N}\bigl(\mathbf Y_{i}-\boldsymbol \pi_{i}\bigr)=\mathbf D^{\top}\mathbf V^{-1}\bigl\{(s_1,s_2)^{\top}-N(\pi_1,\pi_2)^{\top}\bigr\}=\mathbf 0_2 .
\]
Because \(\mathbf D\) is a nonsingular \(2\times2\) matrix whenever \(0<\pi_1,\pi_2<1\), the unique solution is \(\widehat{\pi}_j=s_j/N\), for any working structure. Hence \(\widehat{\boldsymbol \beta}_{G}\) is finite if and only if \(0<s_1<N\) and \(0<s_2<N\), that is, if and only if the time-by-response \(2\times2\) table has no zero cell.

Under working independence, the PGEE equations are the score equations of the penalized log-likelihood \(\ell(\boldsymbol \beta)+\delta\log\det(\boldsymbol \Sigma_{0})\) (Section 2.3 of the main paper). Here \(\boldsymbol \Sigma_{0}=N\mathbf X_{i}^{\top}\mathbf W_{i}\mathbf X_{i}\) with \(\mathbf W_{i}=\operatorname{diag}(w_1,w_2)\) and \(w_j=(\partial\pi_j/\partial\eta_j)^2/\{\pi_j(1-\pi_j)\}\). Since the design matrix, with rows \((1,0)\) and \((1,1)\), has unit determinant, \(\det(\boldsymbol \Sigma_{0})=N^{2}w_1w_2\). For the logit link, \(w_j=\pi_j(1-\pi_j)\). Because the map from \(\boldsymbol \beta\) to \((\pi_1,\pi_2)\) is one-to-one, the penalized log-likelihood can then be written as
\[
\sum_{j=1}^{2}\bigl\{(s_j+\delta)\log\pi_j+(N-s_j+\delta)\log(1-\pi_j)\bigr\}+2\delta\log N,
\]
which is maximized at \(\widehat{\pi}_j=(s_j+\delta)/(N+2\delta)\). Thus PGEE adds \(\delta\) to each cell of the table and is finite for every configuration, with
\[
\widehat{\beta}_0=\log\frac{s_1+\delta}{N-s_1+\delta},\qquad
\widehat{\beta}_1=\log\frac{(s_2+\delta)(N-s_1+\delta)}{(N-s_2+\delta)(s_1+\delta)}.
\]
For other links and for nonindependence working structures, \(\det(\boldsymbol \Sigma_{0})\) depends on the link derivatives and on \(\mathbf V\), and no closed form is available.

\ref{supp:twobytwo} reports \(\widehat{\beta}_1\) for \(N=20\) subjects in four configurations. Each configuration is defined by the numbers of subjects with response patterns \((Y_{i1},Y_{i2})=(0,0)\), \((0,1)\), \((1,0)\), and \((1,1)\). These counts are the cells of the \(2\times2\) table of \(Y_{i1}\) against \(Y_{i2}\), from which the marginalized odds ratio is estimated (Section 2.2 of the main paper). Whether \(\widehat{\boldsymbol \beta}_{G}\) is finite, however, depends only on the time-by-response \(2\times2\) table, which records the numbers of failures and successes at each occasion, because \(\widehat{\pi}_j=s_j/N\). In terms of that table:

\begin{itemize}
\item A, $(8,4,2,6)$: no zero cell;
\item B, $(14,6,0,0)$: one zero cell, because there are no successes at the first occasion (quasi-complete separation);
\item C, $(0,20,0,0)$: two zero cells, because there are no successes at the first occasion and no failures at the second (complete separation, as in \ref{sec:ccfails});
\item D, $(20,0,0,0)$: two zero cells, because there are no successes at either occasion.
\end{itemize}

Ordinary GEE and PGEE with \(\delta\in\{0.1,0.5,1\}\) were fitted with \texttt{geewa\_binary()} under the independence and exchangeable odds-ratio structures and the logit and probit links. With two occasions, the exchangeable and unstructured structures coincide. As in Section 3 of the main paper, a fit was classed as failed if it did not converge or if any estimate exceeded 100 in absolute value.

Under independence with the logit link, the PGEE estimates agree with the closed form above to the reported precision. Ordinary GEE failed in every configuration with a zero cell, under both links and both structures, whereas all 48 PGEE fits converged with finite estimates and standard errors. In configuration D, \(\widehat{\beta}_1=0\), and the PGEE estimates of \(\beta_0\) ranged from \(-5.303\) (logit, independence, \(\delta=0.1\)) to \(-1.616\) (probit, exchangeable, \(\delta=1\)). In the separated configurations, larger values of \(\delta\) moved the estimates further from the boundary. Relative to independence, the exchangeable structure gave similar estimates in B and estimates of smaller magnitude in C. The estimated marginalized odds ratios were 2.23 in B and 0.02 in C. The same configurations show how the correlation-coefficient parameterization of \ref{sec:ccfails} behaves beyond the single configuration considered there. \ref{supp:twobytwo_cc} reports PGEE with \(\delta=0.5\) under an exchangeable correlation-coefficient working structure, fitted with \texttt{geewa()} and the moment estimators of \citet{Liang1986}, with the dispersion parameter \(\phi\) either estimated or fixed at 1, together with the range of correlations compatible with the fitted marginal probabilities \citep{Prentice_Correlated_1988}. In configuration A, the estimated correlation (0.431 with \(\phi\) estimated and 0.453 with \(\phi=1\)) was close to the sample correlation of the two responses (0.408). In the configurations with a zero cell, the behavior was different. With \(\phi\) estimated, the fit failed in C and D under both links and for every \(\delta\in\{0.1,0.5,1\}\), because the moment estimate of the correlation fell outside \((-1,1)\), and in B the estimated correlation was at most 0.008. With \(\phi=1\), the fits converged, but the estimated correlation was at most 0.056 in absolute value in B, C, and D for every \(\delta\), as in the analytic example of \ref{sec:ccfails}. This collapse is not imposed by the admissible range alone: in C, for example, correlations down to \(-1\) are compatible with the fitted marginal probabilities, yet the estimate was \(-0.028\) under the logit link. It arises because the Pearson residuals on which the moment estimator is based shrink toward zero as the fitted probabilities approach 0 or 1. The fitted probabilities also narrow the admissible range itself, for example to correlations of at most 0.025 in C and at least \(-0.025\) in D under the logit link. By contrast, any odds ratio in \((0,\infty)\) is compatible with any pair of marginal probabilities, the marginalized odds ratios do not depend on the fitted probabilities, and the odds-ratio PGEE fits of \ref{supp:twobytwo} were finite in all four configurations.

The results in \ref{supp:twobytwo} and \ref{supp:twobytwo_cc} are reproduced by script \texttt{04\_twobytwo\_configurations.R} in the accompanying zip folder.

\hypertarget{sec:simtables}{%
\section{Extended Simulation Results}\label{sec:simtables}}

This Appendix reports full simulation results corresponding to the designs described in the main paper. The main manuscript highlights the key comparisons and conclusions. The tables here provide the full results across scenarios and working structures.

\hypertarget{quasi-complete-separation}{%
\subsection{Quasi-complete Separation}\label{quasi-complete-separation}}

\ref{tab:simqs_beta2m3_N30N50} -- \ref{tab:simqs2_N100N500} report full simulation summaries for the quasi-complete separation regime described in the main paper. \ref{tab:simqs_beta2m3_N30N50} and \ref{tab:simqs_beta2m3_N100N500} correspond to Scenario A (time-stationary \(x_1\)) and \ref{tab:simqs2_N30N50} and \ref{tab:simqs2_N100N500} to Scenario B (time-varying \(x_1\)), for \(N\in\{30,50\}\) and \(N\in\{100,500\}\), respectively. Columns report Bias (natural scale), empirical standard error (ESE), the ratio of the mean bias-corrected standard error to the ESE (RE), mean squared error (MSE), empirical coverage of nominal 95\% Wald intervals (EC, \%), and convergence proportion (CP, \%), for \(\delta\in\{0.5,0.1\}\) and working odds-ratio structures. All summaries are computed under the failure rule of Section 3 of the main paper, applied to every method, from 10,000 replications per configuration.

\begin{table}[!htbp]
\centering
\caption{Simulation results for $\beta_2=-3$ in $\Pr(Y_{ij}=1\mid \mathbf x_i)=\Phi(\beta_0+\beta_1 x_{1ij}+\beta_2 x_{2i})$ with time-stationary $x_1$, for sample sizes $N\in\{30,50\}$.
Columns report Bias (natural scale), empirical standard error (ESE), ratio of the mean bias-corrected standard error to the ESE (RE), mean squared error (MSE), empirical coverage of nominal 95\% Wald intervals (EC, \%), and convergence proportion (CP, \%), for $\delta\in\{0.5,0.1\}$ and working odds-ratio structures.}
\label{tab:simqs_beta2m3_N30N50}
\small
\setlength{\tabcolsep}{4pt}
\renewcommand{\arraystretch}{1.15}
\begin{tabular*}{\textwidth}{@{\extracolsep{\fill}} c l c rrrrrr}
\toprule
$N$ & Method & $\delta$ & Bias & ESE & RE & MSE & EC & CP \\
\midrule
\multicolumn{9}{c}{\textit{Independence}}\\
\addlinespace[1pt]
30 & PGEE & 0.5 & -0.4374 & 2.4944 & 0.47 & 6.4136 & 82.51 & 99.99 \\
 & HPGEE & 0.5 & -0.8346 & 3.1673 & 0.43 & 10.7283 & 84.17 & 99.99 \\
 & PGEE & 0.1 & -1.1414 & 4.7621 & 0.31 & 23.9799 & 84.46 & 99.90 \\
 & HPGEE & 0.1 & -1.3093 & 5.2451 & 0.31 & 29.2248 & 85.35 & 99.87 \\
\addlinespace[2pt]
50 & PGEE & 0.5 & -0.2163 & 1.3286 & 0.70 & 1.8120 & 88.99 & 100.00 \\
 & HPGEE & 0.5 & -0.3946 & 1.5190 & 0.67 & 2.4631 & 90.35 & 100.00 \\
 & PGEE & 0.1 & -0.4207 & 2.0682 & 0.50 & 4.4543 & 90.62 & 99.99 \\
 & HPGEE & 0.1 & -0.4596 & 2.0343 & 0.52 & 4.3496 & 90.93 & 99.98 \\
\addlinespace[2pt]

\multicolumn{9}{c}{\textit{Exchangeable}}\\
\addlinespace[1pt]
30 & PGEE & 0.5 & 0.4170 & 1.2088 & 0.76 & 1.6351 & 83.39 & 100.00 \\
 & HPGEE & 0.5 & -0.8072 & 3.1350 & 0.45 & 10.4796 & 85.44 & 99.99 \\
 & OPGEE & 0.5 & 0.3146 & 1.3616 & 0.69 & 1.9531 & 83.89 & 100.00 \\
 & PGEE & 0.1 & -0.8142 & 3.7465 & 0.38 & 14.6993 & 85.61 & 99.91 \\
 & HPGEE & 0.1 & -1.3195 & 5.2371 & 0.34 & 29.1684 & 87.87 & 99.87 \\
 & OPGEE & 0.1 & -0.8196 & 3.5661 & 0.41 & 13.3888 & 85.81 & 99.91 \\
\addlinespace[2pt]
50 & PGEE & 0.5 & 0.1873 & 0.9789 & 0.79 & 0.9933 & 86.31 & 100.00 \\
 & HPGEE & 0.5 & -0.3842 & 1.5112 & 0.68 & 2.4314 & 90.05 & 100.00 \\
 & OPGEE & 0.5 & 0.1431 & 1.0037 & 0.78 & 1.0279 & 86.81 & 100.00 \\
 & PGEE & 0.1 & -0.3373 & 1.9862 & 0.51 & 4.0587 & 90.10 & 100.00 \\
 & HPGEE & 0.1 & -0.4633 & 2.0372 & 0.53 & 4.3647 & 90.78 & 99.98 \\
 & OPGEE & 0.1 & -0.3362 & 1.9657 & 0.51 & 3.9771 & 90.08 & 100.00 \\
\addlinespace[2pt]

\multicolumn{9}{c}{\textit{Unstructured}}\\
\addlinespace[1pt]
30 & PGEE & 0.5 & 0.5273 & 1.1808 & 0.77 & 1.6724 & 80.96 & 100.00 \\
 & HPGEE & 0.5 & -0.6218 & 3.0579 & 0.45 & 9.7373 & 85.63 & 99.99 \\
 & OPGEE & 0.5 & 0.4800 & 1.2946 & 0.71 & 1.9062 & 81.39 & 100.00 \\
 & PGEE & 0.1 & -0.6340 & 3.7137 & 0.38 & 14.1933 & 86.04 & 99.92 \\
 & HPGEE & 0.1 & -1.0669 & 5.1002 & 0.33 & 27.1505 & 87.93 & 99.88 \\
 & OPGEE & 0.1 & -0.5747 & 3.4690 & 0.40 & 12.3642 & 85.97 & 99.92 \\
\addlinespace[2pt]
50 & PGEE & 0.5 & 0.2868 & 0.9564 & 0.80 & 0.9969 & 85.34 & 100.00 \\
 & HPGEE & 0.5 & -0.2347 & 1.4417 & 0.69 & 2.1334 & 90.29 & 100.00 \\
 & OPGEE & 0.5 & 0.2794 & 0.9559 & 0.80 & 0.9917 & 85.61 & 100.00 \\
 & PGEE & 0.1 & -0.1991 & 1.7896 & 0.55 & 3.2425 & 90.32 & 100.00 \\
 & HPGEE & 0.1 & -0.3051 & 2.2567 & 0.46 & 5.1860 & 90.96 & 100.00 \\
 & OPGEE & 0.1 & -0.1612 & 1.7113 & 0.56 & 2.9546 & 90.04 & 100.00 \\
\bottomrule
\end{tabular*}
\end{table}

\begin{table}[!htbp]
\centering
\caption{Simulation results for $\beta_2=-3$ in $\Pr(Y_{ij}=1\mid \mathbf x_i)=\Phi(\beta_0+\beta_1 x_{1ij}+\beta_2 x_{2i})$ with time-stationary $x_1$, for sample sizes $N\in\{100,500\}$.
Columns report Bias (natural scale), empirical standard error (ESE), ratio of the mean bias-corrected standard error to the ESE (RE), mean squared error (MSE), empirical coverage of nominal 95\% Wald intervals (EC, \%), and convergence proportion (CP, \%), for $\delta\in\{0.5,0.1\}$ and working odds-ratio structures.}
\label{tab:simqs_beta2m3_N100N500}
\small
\setlength{\tabcolsep}{4pt}
\renewcommand{\arraystretch}{1.15}
\begin{tabular*}{\textwidth}{@{\extracolsep{\fill}} c l c rrrrrr}
\toprule
$N$ & Method & $\delta$ & Bias & ESE & RE & MSE & EC & CP \\
\midrule
\multicolumn{9}{c}{\textit{Independence}}\\
\addlinespace[1pt]
100 & PGEE & 0.5 & -0.1010 & 0.7585 & 0.88 & 0.5856 & 92.53 & 100.00 \\
 & HPGEE & 0.5 & -0.1705 & 0.7898 & 0.88 & 0.6529 & 92.85 & 100.00 \\
 & PGEE & 0.1 & -0.1621 & 0.7858 & 0.89 & 0.6438 & 92.90 & 100.00 \\
 & HPGEE & 0.1 & -0.1764 & 0.7928 & 0.88 & 0.6596 & 92.95 & 100.00 \\
\addlinespace[2pt]
500 & PGEE & 0.5 & -0.0133 & 0.3090 & 0.98 & 0.0957 & 94.33 & 100.00 \\
 & HPGEE & 0.5 & -0.0251 & 0.3110 & 0.98 & 0.0974 & 94.36 & 100.00 \\
 & PGEE & 0.1 & -0.0232 & 0.3106 & 0.98 & 0.0970 & 94.35 & 100.00 \\
 & HPGEE & 0.1 & -0.0255 & 0.3110 & 0.98 & 0.0974 & 94.37 & 100.00 \\
\addlinespace[2pt]

\multicolumn{9}{c}{\textit{Exchangeable}}\\
\addlinespace[1pt]
100 & PGEE & 0.5 & 0.0555 & 0.7040 & 0.87 & 0.4987 & 90.81 & 100.00 \\
 & HPGEE & 0.5 & -0.1668 & 0.7911 & 0.88 & 0.6537 & 92.50 & 100.00 \\
 & OPGEE & 0.5 & 0.0467 & 0.7041 & 0.87 & 0.4980 & 91.03 & 100.00 \\
 & PGEE & 0.1 & -0.1342 & 0.7764 & 0.89 & 0.6208 & 92.62 & 100.00 \\
 & HPGEE & 0.1 & -0.1770 & 0.7949 & 0.89 & 0.6632 & 92.82 & 100.00 \\
 & OPGEE & 0.1 & -0.1322 & 0.7747 & 0.89 & 0.6176 & 92.63 & 100.00 \\
\addlinespace[2pt]
500 & PGEE & 0.5 & 0.0134 & 0.3050 & 0.97 & 0.0932 & 94.16 & 100.00 \\
 & HPGEE & 0.5 & -0.0243 & 0.3108 & 0.98 & 0.0972 & 94.40 & 100.00 \\
 & OPGEE & 0.5 & 0.0133 & 0.3049 & 0.97 & 0.0931 & 94.18 & 100.00 \\
 & PGEE & 0.1 & -0.0184 & 0.3097 & 0.98 & 0.0962 & 94.44 & 100.00 \\
 & HPGEE & 0.1 & -0.0254 & 0.3108 & 0.98 & 0.0972 & 94.51 & 100.00 \\
 & OPGEE & 0.1 & -0.0181 & 0.3095 & 0.98 & 0.0961 & 94.44 & 100.00 \\
\addlinespace[2pt]

\multicolumn{9}{c}{\textit{Unstructured}}\\
\addlinespace[1pt]
100 & PGEE & 0.5 & 0.1228 & 0.6907 & 0.88 & 0.4922 & 90.53 & 99.98 \\
 & HPGEE & 0.5 & -0.0829 & 0.7686 & 0.89 & 0.5976 & 93.09 & 99.99 \\
 & OPGEE & 0.5 & 0.1250 & 0.6859 & 0.88 & 0.4860 & 90.66 & 99.99 \\
 & PGEE & 0.1 & -0.0578 & 0.7576 & 0.90 & 0.5773 & 93.19 & 99.99 \\
 & HPGEE & 0.1 & -0.0886 & 0.7708 & 0.90 & 0.6020 & 93.40 & 99.99 \\
 & OPGEE & 0.1 & -0.0452 & 0.7517 & 0.90 & 0.5671 & 93.15 & 99.99 \\
\addlinespace[2pt]
500 & PGEE & 0.5 & 0.0288 & 0.3022 & 0.97 & 0.0922 & 93.99 & 100.00 \\
 & HPGEE & 0.5 & -0.0076 & 0.3077 & 0.98 & 0.0948 & 94.42 & 100.00 \\
 & OPGEE & 0.5 & 0.0293 & 0.3020 & 0.98 & 0.0920 & 93.98 & 100.00 \\
 & PGEE & 0.1 & -0.0023 & 0.3067 & 0.98 & 0.0941 & 94.46 & 100.00 \\
 & HPGEE & 0.1 & -0.0086 & 0.3077 & 0.98 & 0.0948 & 94.46 & 100.00 \\
 & OPGEE & 0.1 & -0.0014 & 0.3065 & 0.98 & 0.0939 & 94.48 & 100.00 \\
\bottomrule
\end{tabular*}
\end{table}

\begin{table}[!htbp]
\centering
\caption{Simulation results for $\beta_2=-3$ in $\Pr(Y_{ij}=1\mid \mathbf x_i)=\Phi(\beta_0+\beta_1 x_{1ij}+\beta_2 x_{2i})$ with time-varying $x_1$, for $N\in\{30,50\}$.
Columns report Bias (natural scale), empirical standard error (ESE), ratio of the mean bias-corrected standard error to the ESE (RE), mean squared error (MSE), empirical coverage of nominal 95\% Wald confidence intervals (EC, \%), and convergence proportion (CP, \%), for $\delta\in\{0.5,0.1\}$ and working odds-ratio structures.}
\label{tab:simqs2_N30N50}
\small
\setlength{\tabcolsep}{4pt}
\renewcommand{\arraystretch}{1.15}
\begin{tabular*}{\textwidth}{@{\extracolsep{\fill}} c l c rrrrrr}
\toprule
$N$ & Method & $\delta$ & Bias & ESE & RE & MSE & EC & CP \\
\midrule

\multicolumn{9}{c}{\textit{Independence}}\\
\addlinespace[1pt]
30 & PGEE & 0.5 & -0.1161 & 1.3046 & 0.74 & 1.7154 & 90.88 & 100.00 \\
 & HPGEE & 0.5 & -0.3799 & 1.5482 & 0.71 & 2.5411 & 92.39 & 100.00 \\
 & PGEE & 0.1 & -0.4157 & 1.8619 & 0.61 & 3.6395 & 92.91 & 100.00 \\
 & HPGEE & 0.1 & -0.4895 & 2.0779 & 0.57 & 4.5572 & 93.48 & 100.00 \\
\addlinespace[2pt]
50 & PGEE & 0.5 & -0.0730 & 0.8717 & 0.88 & 0.7652 & 92.49 & 100.00 \\
 & HPGEE & 0.5 & -0.2113 & 0.9422 & 0.88 & 0.9323 & 93.47 & 100.00 \\
 & PGEE & 0.1 & -0.2006 & 0.9400 & 0.88 & 0.9238 & 93.60 & 100.00 \\
 & HPGEE & 0.1 & -0.2307 & 0.9576 & 0.88 & 0.9702 & 93.80 & 100.00 \\
\addlinespace[2pt]

\multicolumn{9}{c}{\textit{Exchangeable}}\\
\addlinespace[1pt]
30 & PGEE & 0.5 & -0.1113 & 1.2922 & 0.74 & 1.6822 & 90.55 & 100.00 \\
 & HPGEE & 0.5 & -0.3743 & 1.5433 & 0.71 & 2.5217 & 92.29 & 100.00 \\
 & OPGEE & 0.5 & -0.1112 & 1.2952 & 0.74 & 1.6899 & 90.56 & 100.00 \\
 & PGEE & 0.1 & -0.4124 & 1.8581 & 0.61 & 3.6225 & 92.70 & 100.00 \\
 & HPGEE & 0.1 & -0.4847 & 2.0749 & 0.57 & 4.5401 & 93.30 & 100.00 \\
 & OPGEE & 0.1 & -0.4107 & 1.8565 & 0.61 & 3.6153 & 92.69 & 100.00 \\
\addlinespace[2pt]
50 & PGEE & 0.5 & -0.0719 & 0.8671 & 0.88 & 0.7571 & 92.44 & 100.00 \\
 & HPGEE & 0.5 & -0.2081 & 0.9388 & 0.87 & 0.9247 & 93.33 & 100.00 \\
 & OPGEE & 0.5 & -0.0713 & 0.8670 & 0.88 & 0.7567 & 92.44 & 100.00 \\
 & PGEE & 0.1 & -0.1989 & 0.9370 & 0.87 & 0.9176 & 93.47 & 100.00 \\
 & HPGEE & 0.1 & -0.2279 & 0.9545 & 0.87 & 0.9629 & 93.69 & 100.00 \\
 & OPGEE & 0.1 & -0.1981 & 0.9364 & 0.88 & 0.9161 & 93.46 & 100.00 \\
\addlinespace[2pt]

\multicolumn{9}{c}{\textit{Unstructured}}\\
\addlinespace[1pt]
30 & PGEE & 0.5 & -0.0812 & 1.2935 & 0.74 & 1.6796 & 90.08 & 100.00 \\
 & HPGEE & 0.5 & -0.3553 & 1.5462 & 0.71 & 2.5170 & 92.00 & 100.00 \\
 & OPGEE & 0.5 & -0.0845 & 1.2969 & 0.73 & 1.6892 & 90.18 & 100.00 \\
 & PGEE & 0.1 & -0.3895 & 1.8572 & 0.60 & 3.6008 & 92.40 & 100.00 \\
 & HPGEE & 0.1 & -0.4644 & 2.0727 & 0.57 & 4.5116 & 93.20 & 100.00 \\
 & OPGEE & 0.1 & -0.3888 & 1.8524 & 0.61 & 3.5826 & 92.42 & 100.00 \\
\addlinespace[2pt]
50 & PGEE & 0.5 & -0.0576 & 0.8672 & 0.87 & 0.7553 & 92.14 & 100.00 \\
 & HPGEE & 0.5 & -0.1975 & 0.9385 & 0.87 & 0.9198 & 93.22 & 100.00 \\
 & OPGEE & 0.5 & -0.0579 & 0.8667 & 0.87 & 0.7546 & 92.18 & 100.00 \\
 & PGEE & 0.1 & -0.1871 & 0.9366 & 0.87 & 0.9123 & 93.36 & 100.00 \\
 & HPGEE & 0.1 & -0.2171 & 0.9538 & 0.87 & 0.9568 & 93.64 & 100.00 \\
 & OPGEE & 0.1 & -0.1867 & 0.9358 & 0.87 & 0.9106 & 93.39 & 100.00 \\
\bottomrule
\end{tabular*}
\end{table}

\begin{table}[!htbp]
\centering
\caption{Simulation results for $\beta_2=-3$ in $\Pr(Y_{ij}=1\mid \mathbf x_i)=\Phi(\beta_0+\beta_1 x_{1ij}+\beta_2 x_{2i})$ with time-varying $x_1$, for $N\in\{100,500\}$.
Columns report Bias (natural scale), empirical standard error (ESE), ratio of the mean bias-corrected standard error to the ESE (RE), mean squared error (MSE), empirical coverage of nominal 95\% Wald confidence intervals (EC, \%), and convergence proportion (CP, \%), for $\delta\in\{0.5,0.1\}$ and working odds-ratio structures.}
\label{tab:simqs2_N100N500}
\small
\setlength{\tabcolsep}{4pt}
\renewcommand{\arraystretch}{1.15}
\begin{tabular*}{\textwidth}{@{\extracolsep{\fill}} c l c rrrrrr}
\toprule
$N$ & Method & $\delta$ & Bias & ESE & RE & MSE & EC & CP \\
\midrule

\multicolumn{9}{c}{\textit{Independence}}\\
\addlinespace[1pt]
100 & PGEE & 0.5 & -0.0288 & 0.5800 & 0.94 & 0.3372 & 93.93 & 100.00 \\
 & HPGEE & 0.5 & -0.0914 & 0.6001 & 0.94 & 0.3685 & 93.93 & 100.00 \\
 & PGEE & 0.1 & -0.0834 & 0.5970 & 0.95 & 0.3634 & 94.03 & 100.00 \\
 & HPGEE & 0.1 & -0.0962 & 0.6014 & 0.95 & 0.3709 & 94.06 & 100.00 \\
\addlinespace[2pt]
500 & PGEE & 0.5 & -0.0057 & 0.2490 & 0.99 & 0.0620 & 94.63 & 100.00 \\
 & HPGEE & 0.5 & -0.0173 & 0.2506 & 0.99 & 0.0631 & 94.64 & 100.00 \\
 & PGEE & 0.1 & -0.0154 & 0.2503 & 0.99 & 0.0629 & 94.66 & 100.00 \\
 & HPGEE & 0.1 & -0.0177 & 0.2506 & 0.99 & 0.0631 & 94.64 & 100.00 \\
\addlinespace[2pt]

\multicolumn{9}{c}{\textit{Exchangeable}}\\
\addlinespace[1pt]
100 & PGEE & 0.5 & -0.0290 & 0.5771 & 0.94 & 0.3338 & 93.95 & 100.00 \\
 & HPGEE & 0.5 & -0.0901 & 0.5975 & 0.94 & 0.3651 & 93.97 & 100.00 \\
 & OPGEE & 0.5 & -0.0286 & 0.5769 & 0.94 & 0.3337 & 93.95 & 100.00 \\
 & PGEE & 0.1 & -0.0828 & 0.5945 & 0.95 & 0.3602 & 94.08 & 100.00 \\
 & HPGEE & 0.1 & -0.0950 & 0.5987 & 0.95 & 0.3675 & 94.10 & 100.00 \\
 & OPGEE & 0.1 & -0.0824 & 0.5943 & 0.95 & 0.3599 & 94.08 & 100.00 \\
\addlinespace[2pt]
500 & PGEE & 0.5 & -0.0059 & 0.2478 & 0.99 & 0.0614 & 94.64 & 100.00 \\
 & HPGEE & 0.5 & -0.0171 & 0.2494 & 0.99 & 0.0625 & 94.61 & 100.00 \\
 & OPGEE & 0.5 & -0.0058 & 0.2478 & 0.99 & 0.0614 & 94.63 & 100.00 \\
 & PGEE & 0.1 & -0.0154 & 0.2491 & 0.99 & 0.0623 & 94.67 & 100.00 \\
 & HPGEE & 0.1 & -0.0175 & 0.2494 & 0.99 & 0.0625 & 94.61 & 100.00 \\
 & OPGEE & 0.1 & -0.0153 & 0.2491 & 0.99 & 0.0623 & 94.68 & 100.00 \\
\addlinespace[2pt]

\multicolumn{9}{c}{\textit{Unstructured}}\\
\addlinespace[1pt]
100 & PGEE & 0.5 & -0.0227 & 0.5770 & 0.94 & 0.3334 & 93.96 & 100.00 \\
 & HPGEE & 0.5 & -0.0847 & 0.5973 & 0.94 & 0.3639 & 93.95 & 100.00 \\
 & OPGEE & 0.5 & -0.0225 & 0.5768 & 0.94 & 0.3332 & 93.97 & 100.00 \\
 & PGEE & 0.1 & -0.0771 & 0.5943 & 0.94 & 0.3591 & 94.00 & 100.00 \\
 & HPGEE & 0.1 & -0.0896 & 0.5985 & 0.94 & 0.3662 & 94.09 & 100.00 \\
 & OPGEE & 0.1 & -0.0769 & 0.5941 & 0.94 & 0.3588 & 94.01 & 100.00 \\
\addlinespace[2pt]
500 & PGEE & 0.5 & -0.0048 & 0.2477 & 0.99 & 0.0614 & 94.64 & 100.00 \\
 & HPGEE & 0.5 & -0.0160 & 0.2493 & 0.99 & 0.0624 & 94.60 & 100.00 \\
 & OPGEE & 0.5 & -0.0047 & 0.2477 & 0.99 & 0.0614 & 94.62 & 100.00 \\
 & PGEE & 0.1 & -0.0143 & 0.2490 & 0.99 & 0.0622 & 94.58 & 100.00 \\
 & HPGEE & 0.1 & -0.0165 & 0.2493 & 0.99 & 0.0624 & 94.64 & 100.00 \\
 & OPGEE & 0.1 & -0.0143 & 0.2490 & 0.99 & 0.0622 & 94.58 & 100.00 \\
\bottomrule
\end{tabular*}
\end{table}

\hypertarget{regular-regime}{%
\subsection{Regular Regime}\label{regular-regime}}

\ref{tab:app_regular_N30N50}--\ref{tab:app_regular_beta2_N100N500_missing} report the full simulation summaries for the regular regime (no separation) described in the main paper. We report results under two data conditions. First, \ref{tab:app_regular_N30N50}--\ref{tab:app_regular_beta2_N100N500} correspond to complete data. Second, \ref{tab:app_regular_beta1_N30N50_missing}-\ref{tab:app_regular_beta2_N100N500_missing} correspond to the monotone MCAR dropout scenario considered in the main text. Columns report Bias (scaled by \(\times 100\)), empirical standard error (ESE), the ratio of the mean bias-corrected standard error to the ESE (RE), simulated relative efficiency (SRE), empirical coverage (EC) of nominal \(95\%\) Wald confidence intervals based on the bias-corrected covariance matrix, convergence proportion (CP), and prediction mean squared error (PMSE), computed over the observed subject-occasions. The covariates are generated once for each \(N\) and held fixed across the 10,000 replications. Unlike the quasi-complete separation tables above, bias in the following tables is scaled by \(\times 100\) throughout to reflect the smaller magnitude of estimation error in the regular regime.

\begin{table}[p]
\centering
\caption{Simulation results for estimation of $\beta_1=0.5$ in
$\Pr(Y_{ij}=1 \mid \mathbf{x}_{ij})=\Phi(\beta_{0}+\beta_{1}x_{1i}+\beta_{2}x_{2ij})$.
Columns report Bias ($\times 100$), empirical standard error (ESE), ratio of the mean bias-corrected standard error to the ESE (RE),
simulated relative efficiency (SRE), empirical coverage of the 95\% Wald confidence intervals based on the bias-corrected covariance matrix (EC),
convergence proportion (CP), and prediction mean squared error (PMSE), across sample sizes $N \in \{30, 50\}$ and working odds-ratio structures.}
\label{tab:app_regular_N30N50}
\footnotesize
\setlength{\tabcolsep}{2.8pt}
\renewcommand{\arraystretch}{1.0}
\begin{tabular*}{\textwidth}{@{\extracolsep{\fill}} c l c r r r r r r r @{}}
\toprule
\multicolumn{1}{c}{$N$} &
\multicolumn{1}{c}{Method} &
\multicolumn{1}{c}{$\delta$} &
\multicolumn{1}{c}{Bias ($\times 100$)} &
\multicolumn{1}{c}{ESE} &
\multicolumn{1}{c}{RE} &
\multicolumn{1}{c}{SRE} &
\multicolumn{1}{c}{EC} &
\multicolumn{1}{c}{CP} &
\multicolumn{1}{c}{PMSE} \\
\midrule
\multicolumn{10}{c}{\textit{Independence}}\\
\addlinespace[1pt]
30 & GEE &  & 3.7474 & 0.4166 & 0.9923 & 1.00 & 94.69 & 100.00 & 0.01265 \\
 & RBR &  & 0.8693 & 0.3956 & 1.0311 & 1.12 & 95.29 & 100.00 & 0.01173 \\
 & PGEE & 0.5 & 2.5345 & 0.4071 & 1.0099 & 1.05 & 94.94 & 100.00 & 0.01223 \\
 & HPGEE & 0.5 & 3.7345 & 0.4165 & 0.9927 & 1.00 & 94.70 & 100.00 & 0.01265 \\
 & PGEE & 0.1 & 3.5004 & 0.4147 & 0.9958 & 1.01 & 94.72 & 100.00 & 0.01257 \\
 & HPGEE & 0.1 & 3.7462 & 0.4166 & 0.9924 & 1.00 & 94.69 & 100.00 & 0.01265 \\
\addlinespace[2pt]
50 & GEE &  & 2.6144 & 0.3116 & 0.9993 & 1.00 & 94.98 & 100.00 & 0.00744 \\
 & RBR &  & 0.8230 & 0.3010 & 1.0264 & 1.08 & 95.47 & 100.00 & 0.00707 \\
 & PGEE & 0.5 & 1.9334 & 0.3072 & 1.0105 & 1.03 & 95.24 & 100.00 & 0.00729 \\
 & HPGEE & 0.5 & 2.6108 & 0.3116 & 0.9995 & 1.00 & 94.99 & 100.00 & 0.00744 \\
 & PGEE & 0.1 & 2.4767 & 0.3107 & 1.0016 & 1.01 & 95.03 & 100.00 & 0.00741 \\
 & HPGEE & 0.1 & 2.6140 & 0.3116 & 0.9994 & 1.00 & 94.98 & 100.00 & 0.00744 \\
\addlinespace[2pt]
\multicolumn{10}{c}{\textit{Exchangeable}}\\
\addlinespace[1pt]
30 & GEE &  & 3.3711 & 0.4142 & 1.0042 & 1.01 & 94.80 & 100.00 & 0.01151 \\
 & RBR &  & 0.7960 & 0.3955 & 1.0401 & 1.12 & 95.19 & 100.00 & 0.01076 \\
 & PGEE & 0.5 & 1.3052 & 0.3976 & 1.0382 & 1.11 & 95.22 & 100.00 & 0.01091 \\
 & HPGEE & 0.5 & 3.5664 & 0.4145 & 1.0044 & 1.01 & 94.82 & 100.00 & 0.01155 \\
 & OPGEE & 0.5 & 1.5129 & 0.3981 & 1.0378 & 1.10 & 95.19 & 100.00 & 0.01096 \\
 & PGEE & 0.1 & 2.9471 & 0.4108 & 1.0110 & 1.03 & 94.90 & 100.00 & 0.01139 \\
 & HPGEE & 0.1 & 3.5767 & 0.4147 & 1.0041 & 1.01 & 94.80 & 100.00 & 0.01156 \\
 & OPGEE & 0.1 & 3.1586 & 0.4113 & 1.0107 & 1.03 & 94.90 & 100.00 & 0.01143 \\
\addlinespace[2pt]
50 & GEE &  & 2.4717 & 0.3107 & 1.0065 & 1.01 & 95.01 & 100.00 & 0.00679 \\
 & RBR &  & 0.8401 & 0.3012 & 1.0312 & 1.08 & 95.35 & 100.00 & 0.00649 \\
 & PGEE & 0.5 & 1.2633 & 0.3029 & 1.0279 & 1.06 & 95.38 & 100.00 & 0.00657 \\
 & HPGEE & 0.5 & 2.5205 & 0.3109 & 1.0064 & 1.01 & 95.00 & 100.00 & 0.00681 \\
 & OPGEE & 0.5 & 1.3142 & 0.3031 & 1.0275 & 1.06 & 95.34 & 100.00 & 0.00659 \\
 & PGEE & 0.1 & 2.2261 & 0.3091 & 1.0108 & 1.02 & 95.06 & 100.00 & 0.00675 \\
 & HPGEE & 0.1 & 2.5220 & 0.3109 & 1.0063 & 1.00 & 95.01 & 100.00 & 0.00681 \\
 & OPGEE & 0.1 & 2.2782 & 0.3093 & 1.0104 & 1.02 & 95.06 & 100.00 & 0.00677 \\
\addlinespace[2pt]
\multicolumn{10}{c}{\textit{Unstructured}}\\
\addlinespace[1pt]
30 & GEE &  & 2.0357 & 0.3928 & 1.0027 & 1.13 & 94.82 & 100.00 & 0.01040 \\
 & RBR &  & -0.3608 & 0.3756 & 1.0385 & 1.24 & 95.42 & 99.99 & 0.00980 \\
 & PGEE & 0.5 & -0.1808 & 0.3761 & 1.0389 & 1.24 & 95.51 & 99.99 & 0.00985 \\
 & HPGEE & 0.5 & 2.3815 & 0.3941 & 1.0007 & 1.12 & 94.79 & 100.00 & 0.01046 \\
 & OPGEE & 0.5 & 0.1730 & 0.3776 & 1.0362 & 1.23 & 95.44 & 100.00 & 0.00991 \\
 & PGEE & 0.1 & 1.5835 & 0.3894 & 1.0098 & 1.15 & 95.01 & 99.99 & 0.01029 \\
 & HPGEE & 0.1 & 2.3860 & 0.3942 & 1.0005 & 1.12 & 94.77 & 100.00 & 0.01046 \\
 & OPGEE & 0.1 & 1.9366 & 0.3908 & 1.0074 & 1.14 & 94.87 & 100.00 & 0.01035 \\
\addlinespace[2pt]
50 & GEE &  & 1.7795 & 0.2966 & 1.0029 & 1.11 & 94.90 & 100.00 & 0.00611 \\
 & RBR &  & 0.2876 & 0.2880 & 1.0267 & 1.18 & 95.26 & 100.00 & 0.00587 \\
 & PGEE & 0.5 & 0.5016 & 0.2888 & 1.0251 & 1.17 & 95.28 & 100.00 & 0.00591 \\
 & HPGEE & 0.5 & 1.8316 & 0.2970 & 1.0023 & 1.10 & 94.88 & 100.00 & 0.00612 \\
 & OPGEE & 0.5 & 0.5493 & 0.2893 & 1.0243 & 1.17 & 95.21 & 100.00 & 0.00592 \\
 & PGEE & 0.1 & 1.5198 & 0.2950 & 1.0073 & 1.12 & 94.94 & 100.00 & 0.00607 \\
 & HPGEE & 0.1 & 1.8297 & 0.2970 & 1.0023 & 1.10 & 94.88 & 100.00 & 0.00612 \\
 & OPGEE & 0.1 & 1.5705 & 0.2954 & 1.0066 & 1.12 & 94.96 & 100.00 & 0.00608 \\
\bottomrule
\end{tabular*}
\end{table}

\begin{table}[p]
\centering
\caption{Simulation results for estimation of $\beta_1=0.5$ in
$\Pr(Y_{ij}=1 \mid \mathbf{x}_{ij})=\Phi(\beta_{0}+\beta_{1}x_{1i}+\beta_{2}x_{2ij})$.
Columns report Bias ($\times 100$), empirical standard error (ESE), ratio of the mean bias-corrected standard error to the ESE (RE),
simulated relative efficiency (SRE), empirical coverage of the 95\% Wald confidence intervals based on the bias-corrected covariance matrix (EC),
convergence proportion (CP), and prediction mean squared error (PMSE), across sample sizes $N \in \{100, 500\}$ and working odds-ratio structures.}
\label{tab:app_regular_N100N500}
\footnotesize
\setlength{\tabcolsep}{2.8pt}
\renewcommand{\arraystretch}{1.0}
\begin{tabular*}{\textwidth}{@{\extracolsep{\fill}} c l c r r r r r r r @{}}
\toprule
\multicolumn{1}{c}{$N$} &
\multicolumn{1}{c}{Method} &
\multicolumn{1}{c}{$\delta$} &
\multicolumn{1}{c}{Bias ($\times 100$)} &
\multicolumn{1}{c}{ESE} &
\multicolumn{1}{c}{RE} &
\multicolumn{1}{c}{SRE} &
\multicolumn{1}{c}{EC} &
\multicolumn{1}{c}{CP} &
\multicolumn{1}{c}{PMSE} \\
\midrule
\multicolumn{10}{c}{\textit{Independence}}\\
\addlinespace[1pt]
100 & GEE &  & 1.1683 & 0.2083 & 0.9950 & 1.00 & 94.77 & 100.00 & 0.00379 \\
 & RBR &  & 0.1944 & 0.2043 & 1.0100 & 1.04 & 95.10 & 100.00 & 0.00369 \\
 & PGEE & 0.5 & 0.8259 & 0.2068 & 1.0008 & 1.02 & 94.93 & 100.00 & 0.00375 \\
 & HPGEE & 0.5 & 1.1676 & 0.2083 & 0.9950 & 1.00 & 94.77 & 100.00 & 0.00379 \\
 & PGEE & 0.1 & 1.0995 & 0.2080 & 0.9961 & 1.00 & 94.80 & 100.00 & 0.00378 \\
 & HPGEE & 0.1 & 1.1683 & 0.2083 & 0.9950 & 1.00 & 94.77 & 100.00 & 0.00379 \\
\addlinespace[2pt]
500 & GEE &  & 0.3303 & 0.0938 & 0.9941 & 1.00 & 94.60 & 100.00 & 0.00076 \\
 & RBR &  & 0.1212 & 0.0933 & 0.9976 & 1.01 & 94.75 & 100.00 & 0.00076 \\
 & PGEE & 0.5 & 0.2606 & 0.0936 & 0.9953 & 1.00 & 94.64 & 100.00 & 0.00076 \\
 & HPGEE & 0.5 & 0.3303 & 0.0938 & 0.9941 & 1.00 & 94.60 & 100.00 & 0.00076 \\
 & PGEE & 0.1 & 0.3164 & 0.0937 & 0.9943 & 1.00 & 94.61 & 100.00 & 0.00076 \\
 & HPGEE & 0.1 & 0.3303 & 0.0938 & 0.9941 & 1.00 & 94.60 & 100.00 & 0.00076 \\
\addlinespace[2pt]
\multicolumn{10}{c}{\textit{Exchangeable}}\\
\addlinespace[1pt]
100 & GEE &  & 1.1326 & 0.2082 & 0.9956 & 1.00 & 94.84 & 100.00 & 0.00355 \\
 & RBR &  & 0.2011 & 0.2045 & 1.0093 & 1.04 & 95.05 & 100.00 & 0.00347 \\
 & PGEE & 0.5 & 0.5596 & 0.2056 & 1.0059 & 1.03 & 95.02 & 100.00 & 0.00350 \\
 & HPGEE & 0.5 & 1.1314 & 0.2081 & 0.9963 & 1.00 & 94.83 & 100.00 & 0.00355 \\
 & OPGEE & 0.5 & 0.5599 & 0.2055 & 1.0064 & 1.03 & 95.01 & 100.00 & 0.00350 \\
 & PGEE & 0.1 & 1.0171 & 0.2077 & 0.9977 & 1.01 & 94.88 & 100.00 & 0.00354 \\
 & HPGEE & 0.1 & 1.1320 & 0.2081 & 0.9962 & 1.00 & 94.83 & 100.00 & 0.00355 \\
 & OPGEE & 0.1 & 1.0170 & 0.2076 & 0.9982 & 1.01 & 94.86 & 100.00 & 0.00354 \\
\addlinespace[2pt]
500 & GEE &  & 0.3194 & 0.0937 & 0.9938 & 1.00 & 94.61 & 100.00 & 0.00071 \\
 & RBR &  & 0.1161 & 0.0933 & 0.9972 & 1.01 & 94.68 & 100.00 & 0.00071 \\
 & PGEE & 0.5 & 0.2036 & 0.0935 & 0.9959 & 1.01 & 94.64 & 100.00 & 0.00071 \\
 & HPGEE & 0.5 & 0.3223 & 0.0937 & 0.9939 & 1.00 & 94.60 & 100.00 & 0.00071 \\
 & OPGEE & 0.5 & 0.2065 & 0.0935 & 0.9960 & 1.01 & 94.64 & 100.00 & 0.00071 \\
 & PGEE & 0.1 & 0.2962 & 0.0937 & 0.9942 & 1.00 & 94.63 & 100.00 & 0.00071 \\
 & HPGEE & 0.1 & 0.3223 & 0.0937 & 0.9939 & 1.00 & 94.61 & 100.00 & 0.00071 \\
 & OPGEE & 0.1 & 0.2991 & 0.0937 & 0.9943 & 1.00 & 94.64 & 100.00 & 0.00071 \\
\addlinespace[2pt]
\multicolumn{10}{c}{\textit{Unstructured}}\\
\addlinespace[1pt]
100 & GEE &  & 0.6661 & 0.1986 & 0.9973 & 1.10 & 94.38 & 100.00 & 0.00318 \\
 & RBR &  & -0.1819 & 0.1952 & 1.0106 & 1.14 & 94.66 & 100.00 & 0.00311 \\
 & PGEE & 0.5 & 0.0556 & 0.1959 & 1.0080 & 1.13 & 94.65 & 100.00 & 0.00313 \\
 & HPGEE & 0.5 & 0.6958 & 0.1986 & 0.9975 & 1.10 & 94.40 & 100.00 & 0.00318 \\
 & OPGEE & 0.5 & 0.0839 & 0.1960 & 1.0081 & 1.13 & 94.66 & 100.00 & 0.00313 \\
 & PGEE & 0.1 & 0.5430 & 0.1980 & 0.9994 & 1.11 & 94.46 & 100.00 & 0.00317 \\
 & HPGEE & 0.1 & 0.6957 & 0.1986 & 0.9974 & 1.10 & 94.40 & 100.00 & 0.00318 \\
 & OPGEE & 0.1 & 0.5726 & 0.1981 & 0.9996 & 1.11 & 94.48 & 100.00 & 0.00317 \\
\addlinespace[2pt]
500 & GEE &  & 0.2153 & 0.0900 & 0.9923 & 1.08 & 94.62 & 100.00 & 0.00064 \\
 & RBR &  & 0.0295 & 0.0897 & 0.9955 & 1.09 & 94.70 & 100.00 & 0.00063 \\
 & PGEE & 0.5 & 0.0913 & 0.0898 & 0.9945 & 1.09 & 94.65 & 100.00 & 0.00064 \\
 & HPGEE & 0.5 & 0.2267 & 0.0900 & 0.9924 & 1.08 & 94.58 & 100.00 & 0.00064 \\
 & OPGEE & 0.5 & 0.1026 & 0.0898 & 0.9946 & 1.09 & 94.67 & 100.00 & 0.00064 \\
 & PGEE & 0.1 & 0.1905 & 0.0900 & 0.9927 & 1.09 & 94.63 & 100.00 & 0.00064 \\
 & HPGEE & 0.1 & 0.2267 & 0.0900 & 0.9924 & 1.08 & 94.58 & 100.00 & 0.00064 \\
 & OPGEE & 0.1 & 0.2019 & 0.0900 & 0.9928 & 1.09 & 94.60 & 100.00 & 0.00064 \\
\bottomrule
\end{tabular*}
\end{table}

\begin{table}[p]
\centering
\caption{Simulation results for estimation of $\beta_2=0.5$ in
$\Pr(Y_{ij}=1 \mid \mathbf{x}_{ij})=\Phi(\beta_{0}+\beta_{1}x_{1i}+\beta_{2}x_{2ij})$.
Columns report Bias ($\times 100$), empirical standard error (ESE), ratio of the mean bias-corrected standard error to the ESE (RE),
simulated relative efficiency (SRE), empirical coverage of the 95\% Wald confidence intervals based on the bias-corrected covariance matrix (EC),
convergence proportion (CP), and prediction mean squared error (PMSE), across sample sizes $N \in \{30, 50\}$ and working odds-ratio structures.}
\label{tab:app_regular_beta2_N30N50}
\footnotesize
\setlength{\tabcolsep}{2.8pt}
\renewcommand{\arraystretch}{1.0}
\begin{tabular*}{\textwidth}{@{\extracolsep{\fill}} c l c r r r r r r r @{}}
\toprule
\multicolumn{1}{c}{$N$} &
\multicolumn{1}{c}{Method} &
\multicolumn{1}{c}{$\delta$} &
\multicolumn{1}{c}{Bias ($\times 100$)} &
\multicolumn{1}{c}{ESE} &
\multicolumn{1}{c}{RE} &
\multicolumn{1}{c}{SRE} &
\multicolumn{1}{c}{EC} &
\multicolumn{1}{c}{CP} &
\multicolumn{1}{c}{PMSE} \\
\midrule
 
\multicolumn{10}{c}{\textit{Independence}}\\
\addlinespace[1pt]
30 & GEE &  & 3.9758 & 0.4400 & 0.9840 & 1.00 & 93.86 & 100.00 & 0.01265 \\
 & RBR &  & 0.8783 & 0.4170 & 1.0232 & 1.12 & 94.68 & 100.00 & 0.01173 \\
 & PGEE & 0.5 & 2.5522 & 0.4284 & 1.0044 & 1.06 & 94.38 & 100.00 & 0.01223 \\
 & HPGEE & 0.5 & 3.9558 & 0.4398 & 0.9845 & 1.00 & 93.88 & 100.00 & 0.01265 \\
 & PGEE & 0.1 & 3.6855 & 0.4376 & 0.9881 & 1.01 & 93.99 & 100.00 & 0.01257 \\
 & HPGEE & 0.1 & 3.9735 & 0.4400 & 0.9841 & 1.00 & 93.86 & 100.00 & 0.01265 \\
\addlinespace[2pt]
 
50 & GEE &  & 2.4048 & 0.3227 & 1.0039 & 1.00 & 94.71 & 100.00 & 0.00744 \\
 & RBR &  & 0.5071 & 0.3113 & 1.0318 & 1.08 & 95.23 & 100.00 & 0.00707 \\
 & PGEE & 0.5 & 1.6348 & 0.3177 & 1.0163 & 1.03 & 94.97 & 100.00 & 0.00729 \\
 & HPGEE & 0.5 & 2.4011 & 0.3226 & 1.0040 & 1.00 & 94.71 & 100.00 & 0.00744 \\
 & PGEE & 0.1 & 2.2490 & 0.3217 & 1.0064 & 1.01 & 94.77 & 100.00 & 0.00741 \\
 & HPGEE & 0.1 & 2.4045 & 0.3227 & 1.0039 & 1.00 & 94.71 & 100.00 & 0.00744 \\
\addlinespace[2pt]
 
\multicolumn{10}{c}{\textit{Exchangeable}}\\
\addlinespace[1pt]
30 & GEE &  & 1.9242 & 0.3582 & 1.0147 & 1.52 & 95.10 & 100.00 & 0.01151 \\
 & RBR &  & -0.5462 & 0.3422 & 1.0493 & 1.67 & 95.63 & 100.00 & 0.01076 \\
 & PGEE & 0.5 & 0.4130 & 0.3489 & 1.0339 & 1.60 & 95.49 & 100.00 & 0.01091 \\
 & HPGEE & 0.5 & 2.8523 & 0.3612 & 1.0088 & 1.49 & 94.87 & 100.00 & 0.01155 \\
 & OPGEE & 0.5 & 1.3505 & 0.3524 & 1.0263 & 1.57 & 95.27 & 100.00 & 0.01096 \\
 & PGEE & 0.1 & 1.6172 & 0.3563 & 1.0185 & 1.53 & 95.16 & 100.00 & 0.01139 \\
 & HPGEE & 0.1 & 2.8736 & 0.3614 & 1.0084 & 1.49 & 94.87 & 100.00 & 0.01156 \\
 & OPGEE & 0.1 & 2.5672 & 0.3596 & 1.0119 & 1.50 & 94.98 & 100.00 & 0.01143 \\
\addlinespace[2pt]
 
50 & GEE &  & 1.3810 & 0.2664 & 1.0087 & 1.47 & 95.32 & 100.00 & 0.00679 \\
 & RBR &  & -0.2102 & 0.2581 & 1.0339 & 1.57 & 95.66 & 100.00 & 0.00649 \\
 & PGEE & 0.5 & 0.5213 & 0.2622 & 1.0205 & 1.52 & 95.49 & 100.00 & 0.00657 \\
 & HPGEE & 0.5 & 1.9161 & 0.2676 & 1.0052 & 1.45 & 95.18 & 100.00 & 0.00681 \\
 & OPGEE & 0.5 & 1.0554 & 0.2637 & 1.0162 & 1.50 & 95.41 & 100.00 & 0.00659 \\
 & PGEE & 0.1 & 1.2074 & 0.2655 & 1.0110 & 1.48 & 95.34 & 100.00 & 0.00675 \\
 & HPGEE & 0.1 & 1.9226 & 0.2677 & 1.0050 & 1.45 & 95.18 & 100.00 & 0.00681 \\
 & OPGEE & 0.1 & 1.7485 & 0.2669 & 1.0072 & 1.46 & 95.21 & 100.00 & 0.00677 \\
\addlinespace[2pt]
 
\multicolumn{10}{c}{\textit{Unstructured}}\\
\addlinespace[1pt]
30 & GEE &  & -2.1734 & 0.3078 & 0.9732 & 2.05 & 92.98 & 100.00 & 0.01040 \\
 & RBR &  & -4.6161 & 0.2939 & 1.0057 & 2.21 & 93.09 & 99.99 & 0.00980 \\
 & PGEE & 0.5 & -4.4345 & 0.2978 & 0.9953 & 2.15 & 93.11 & 99.99 & 0.00985 \\
 & HPGEE & 0.5 & -0.2847 & 0.3136 & 0.9612 & 1.99 & 93.08 & 100.00 & 0.01046 \\
 & OPGEE & 0.5 & -2.4021 & 0.3038 & 0.9817 & 2.10 & 93.20 & 100.00 & 0.00991 \\
 & PGEE & 0.1 & -2.6278 & 0.3057 & 0.9776 & 2.07 & 93.00 & 99.99 & 0.01029 \\
 & HPGEE & 0.1 & -0.2411 & 0.3139 & 0.9604 & 1.98 & 93.08 & 100.00 & 0.01046 \\
 & OPGEE & 0.1 & -0.6704 & 0.3118 & 0.9645 & 2.01 & 93.05 & 100.00 & 0.01035 \\
\addlinespace[2pt]
 
50 & GEE &  & -1.2938 & 0.2302 & 0.9794 & 1.97 & 93.98 & 100.00 & 0.00611 \\
 & RBR &  & -2.8243 & 0.2230 & 1.0033 & 2.07 & 94.19 & 100.00 & 0.00587 \\
 & PGEE & 0.5 & -2.6481 & 0.2261 & 0.9915 & 2.02 & 94.06 & 100.00 & 0.00591 \\
 & HPGEE & 0.5 & -0.1361 & 0.2322 & 0.9744 & 1.94 & 93.86 & 100.00 & 0.00612 \\
 & OPGEE & 0.5 & -1.4365 & 0.2280 & 0.9863 & 2.01 & 94.04 & 100.00 & 0.00592 \\
 & PGEE & 0.1 & -1.5677 & 0.2294 & 0.9818 & 1.98 & 94.00 & 100.00 & 0.00607 \\
 & HPGEE & 0.1 & -0.1193 & 0.2323 & 0.9741 & 1.94 & 93.85 & 100.00 & 0.00612 \\
 & OPGEE & 0.1 & -0.3814 & 0.2314 & 0.9765 & 1.95 & 93.95 & 100.00 & 0.00608 \\
\bottomrule
\end{tabular*}
\end{table}

\begin{table}[p]
\centering
\caption{Simulation results for estimation of $\beta_2=0.5$ in
$\Pr(Y_{ij}=1 \mid \mathbf{x}_{ij})=\Phi(\beta_{0}+\beta_{1}x_{1i}+\beta_{2}x_{2ij})$.
Columns report Bias ($\times 100$), empirical standard error (ESE), ratio of the mean bias-corrected standard error to the ESE (RE),
simulated relative efficiency (SRE), empirical coverage of the 95\% Wald confidence intervals based on the bias-corrected covariance matrix (EC),
convergence proportion (CP), and prediction mean squared error (PMSE), across sample sizes $N \in \{100, 500\}$ and working odds-ratio structures.}
\label{tab:app_regular_beta2_N100N500}
\footnotesize
\setlength{\tabcolsep}{2.8pt}
\renewcommand{\arraystretch}{1.0}
\begin{tabular*}{\textwidth}{@{\extracolsep{\fill}} c l c r r r r r r r @{}}
\toprule
\multicolumn{1}{c}{$N$} &
\multicolumn{1}{c}{Method} &
\multicolumn{1}{c}{$\delta$} &
\multicolumn{1}{c}{Bias ($\times 100$)} &
\multicolumn{1}{c}{ESE} &
\multicolumn{1}{c}{RE} &
\multicolumn{1}{c}{SRE} &
\multicolumn{1}{c}{EC} &
\multicolumn{1}{c}{CP} &
\multicolumn{1}{c}{PMSE} \\
\midrule
 
\multicolumn{10}{c}{\textit{Independence}}\\
\addlinespace[1pt]
100 & GEE &  & 1.0573 & 0.1944 & 0.9977 & 1.00 & 94.95 & 100.00 & 0.00379 \\
 & RBR &  & 0.0230 & 0.1904 & 1.0139 & 1.05 & 95.10 & 100.00 & 0.00369 \\
 & PGEE & 0.5 & 0.6797 & 0.1928 & 1.0042 & 1.02 & 95.11 & 100.00 & 0.00375 \\
 & HPGEE & 0.5 & 1.0564 & 0.1944 & 0.9977 & 1.00 & 94.95 & 100.00 & 0.00379 \\
 & PGEE & 0.1 & 0.9814 & 0.1941 & 0.9990 & 1.00 & 94.96 & 100.00 & 0.00378 \\
 & HPGEE & 0.1 & 1.0573 & 0.1944 & 0.9977 & 1.00 & 94.95 & 100.00 & 0.00379 \\
\addlinespace[2pt]
500 & GEE &  & 0.2893 & 0.0854 & 1.0015 & 1.00 & 94.75 & 100.00 & 0.00076 \\
 & RBR &  & 0.0765 & 0.0851 & 1.0050 & 1.01 & 94.89 & 100.00 & 0.00076 \\
 & PGEE & 0.5 & 0.2156 & 0.0853 & 1.0028 & 1.00 & 94.81 & 100.00 & 0.00076 \\
 & HPGEE & 0.5 & 0.2893 & 0.0854 & 1.0015 & 1.00 & 94.75 & 100.00 & 0.00076 \\
 & PGEE & 0.1 & 0.2746 & 0.0854 & 1.0017 & 1.00 & 94.76 & 100.00 & 0.00076 \\
 & HPGEE & 0.1 & 0.2893 & 0.0854 & 1.0015 & 1.00 & 94.75 & 100.00 & 0.00076 \\
\addlinespace[2pt]
 
\multicolumn{10}{c}{\textit{Exchangeable}}\\
\addlinespace[1pt]
100 & GEE &  & 0.4340 & 0.1724 & 0.9970 & 1.28 & 94.66 & 100.00 & 0.00355 \\
 & RBR &  & -0.4639 & 0.1693 & 1.0109 & 1.32 & 94.90 & 100.00 & 0.00347 \\
 & PGEE & 0.5 & -0.0199 & 0.1708 & 1.0040 & 1.30 & 94.83 & 100.00 & 0.00350 \\
 & HPGEE & 0.5 & 0.6041 & 0.1726 & 0.9963 & 1.27 & 94.64 & 100.00 & 0.00355 \\
 & OPGEE & 0.5 & 0.1471 & 0.1711 & 1.0030 & 1.30 & 94.81 & 100.00 & 0.00350 \\
 & PGEE & 0.1 & 0.3427 & 0.1721 & 0.9984 & 1.28 & 94.70 & 100.00 & 0.00354 \\
 & HPGEE & 0.1 & 0.6056 & 0.1726 & 0.9963 & 1.27 & 94.64 & 100.00 & 0.00355 \\
 & OPGEE & 0.1 & 0.5136 & 0.1723 & 0.9976 & 1.28 & 94.66 & 100.00 & 0.00354 \\
\addlinespace[2pt]
500 & GEE &  & 0.1661 & 0.0753 & 1.0026 & 1.29 & 95.14 & 100.00 & 0.00071 \\
 & RBR &  & -0.0165 & 0.0750 & 1.0056 & 1.30 & 95.21 & 100.00 & 0.00071 \\
 & PGEE & 0.5 & 0.0760 & 0.0752 & 1.0039 & 1.29 & 95.19 & 100.00 & 0.00071 \\
 & HPGEE & 0.5 & 0.2042 & 0.0753 & 1.0023 & 1.29 & 95.11 & 100.00 & 0.00071 \\
 & OPGEE & 0.5 & 0.1140 & 0.0752 & 1.0036 & 1.29 & 95.17 & 100.00 & 0.00071 \\
 & PGEE & 0.1 & 0.1480 & 0.0753 & 1.0028 & 1.29 & 95.10 & 100.00 & 0.00071 \\
 & HPGEE & 0.1 & 0.2043 & 0.0753 & 1.0023 & 1.29 & 95.11 & 100.00 & 0.00071 \\
 & OPGEE & 0.1 & 0.1862 & 0.0753 & 1.0025 & 1.29 & 95.12 & 100.00 & 0.00071 \\
\addlinespace[2pt]
 
\multicolumn{10}{c}{\textit{Unstructured}}\\
\addlinespace[1pt]
100 & GEE &  & -0.7629 & 0.1548 & 0.9859 & 1.58 & 94.08 & 100.00 & 0.00318 \\
 & RBR &  & -1.5977 & 0.1521 & 0.9992 & 1.62 & 94.29 & 100.00 & 0.00311 \\
 & PGEE & 0.5 & -1.3564 & 0.1533 & 0.9927 & 1.60 & 94.12 & 100.00 & 0.00313 \\
 & HPGEE & 0.5 & -0.3578 & 0.1551 & 0.9857 & 1.58 & 94.16 & 100.00 & 0.00318 \\
 & OPGEE & 0.5 & -0.9515 & 0.1536 & 0.9923 & 1.60 & 94.14 & 100.00 & 0.00313 \\
 & PGEE & 0.1 & -0.8823 & 0.1545 & 0.9873 & 1.58 & 94.12 & 100.00 & 0.00317 \\
 & HPGEE & 0.1 & -0.3542 & 0.1551 & 0.9856 & 1.58 & 94.16 & 100.00 & 0.00318 \\
 & OPGEE & 0.1 & -0.4735 & 0.1548 & 0.9869 & 1.58 & 94.15 & 100.00 & 0.00317 \\
\addlinespace[2pt]
500 & GEE &  & -0.0744 & 0.0679 & 0.9963 & 1.58 & 94.99 & 100.00 & 0.00064 \\
 & RBR &  & -0.2422 & 0.0677 & 0.9992 & 1.59 & 94.97 & 100.00 & 0.00063 \\
 & PGEE & 0.5 & -0.1903 & 0.0678 & 0.9975 & 1.59 & 94.93 & 100.00 & 0.00064 \\
 & HPGEE & 0.5 & 0.0071 & 0.0679 & 0.9963 & 1.58 & 94.98 & 100.00 & 0.00064 \\
 & OPGEE & 0.5 & -0.1089 & 0.0678 & 0.9975 & 1.59 & 94.93 & 100.00 & 0.00064 \\
 & PGEE & 0.1 & -0.0976 & 0.0679 & 0.9965 & 1.59 & 95.00 & 100.00 & 0.00064 \\
 & HPGEE & 0.1 & 0.0072 & 0.0679 & 0.9962 & 1.58 & 94.98 & 100.00 & 0.00064 \\
 & OPGEE & 0.1 & -0.0160 & 0.0679 & 0.9965 & 1.58 & 94.98 & 100.00 & 0.00064 \\
\bottomrule
\end{tabular*}
\end{table}

\begin{table}[p]
\centering
\caption{Simulation results for estimation of $\beta_1=0.5$ in
$\Pr(Y_{ij}=1 \mid \mathbf{x}_{ij})=\Phi(\beta_{0}+\beta_{1}x_{1i}+\beta_{2}x_{2ij})$ under MCAR.
Columns report Bias ($\times 100$), empirical standard error (ESE), ratio of the mean bias-corrected standard error to the ESE (RE),
simulated relative efficiency (SRE), empirical coverage of the 95\% Wald confidence intervals based on the bias-corrected covariance matrix (EC),
convergence proportion (CP), and prediction mean squared error (PMSE), across sample sizes $N \in \{30,50\}$ and working odds-ratio structures.}
\label{tab:app_regular_beta1_N30N50_missing}
\footnotesize
\setlength{\tabcolsep}{2.8pt}
\renewcommand{\arraystretch}{1.0}
\begin{tabular*}{\textwidth}{@{\extracolsep{\fill}} c l c r r r r r r r @{}}
\toprule
\multicolumn{1}{c}{$N$} &
\multicolumn{1}{c}{Method} &
\multicolumn{1}{c}{$\delta$} &
\multicolumn{1}{c}{Bias ($\times 100$)} &
\multicolumn{1}{c}{ESE} &
\multicolumn{1}{c}{RE} &
\multicolumn{1}{c}{SRE} &
\multicolumn{1}{c}{EC} &
\multicolumn{1}{c}{CP} &
\multicolumn{1}{c}{PMSE} \\
\midrule
\multicolumn{10}{c}{\textit{Independence}}\\
\addlinespace[1pt]
30 & GEE &  & 4.6959 & 0.4741 & 0.9723 & 1.00 & 93.95 & 100.00 & 0.01581 \\
 & RBR &  & 1.2481 & 0.4459 & 1.0165 & 1.14 & 94.85 & 100.00 & 0.01445 \\
 & PGEE & 0.5 & 3.0024 & 0.4590 & 0.9963 & 1.07 & 94.55 & 100.00 & 0.01509 \\
 & HPGEE & 0.5 & 4.6686 & 0.4738 & 0.9730 & 1.00 & 93.99 & 100.00 & 0.01579 \\
 & PGEE & 0.1 & 4.3485 & 0.4710 & 0.9771 & 1.01 & 94.10 & 100.00 & 0.01566 \\
 & HPGEE & 0.1 & 4.6936 & 0.4741 & 0.9724 & 1.00 & 93.96 & 100.00 & 0.01581 \\
\addlinespace[2pt]
50 & GEE &  & 3.0118 & 0.3516 & 0.9850 & 1.00 & 94.45 & 100.00 & 0.00928 \\
 & RBR &  & 0.8268 & 0.3372 & 1.0168 & 1.09 & 95.14 & 100.00 & 0.00873 \\
 & PGEE & 0.5 & 2.0684 & 0.3448 & 1.0002 & 1.04 & 94.78 & 100.00 & 0.00902 \\
 & HPGEE & 0.5 & 3.0046 & 0.3515 & 0.9852 & 1.00 & 94.45 & 100.00 & 0.00928 \\
 & PGEE & 0.1 & 2.8203 & 0.3502 & 0.9880 & 1.01 & 94.52 & 100.00 & 0.00923 \\
 & HPGEE & 0.1 & 3.0111 & 0.3516 & 0.9850 & 1.00 & 94.45 & 100.00 & 0.00928 \\
\addlinespace[2pt]
\multicolumn{10}{c}{\textit{Exchangeable}}\\
\addlinespace[1pt]
30 & GEE &  & 3.8649 & 0.4581 & 0.9962 & 1.07 & 94.77 & 100.00 & 0.01389 \\
 & RBR &  & 0.7907 & 0.4331 & 1.0386 & 1.21 & 95.42 & 100.00 & 0.01282 \\
 & PGEE & 0.5 & 1.3831 & 0.4361 & 1.0360 & 1.19 & 95.36 & 100.00 & 0.01303 \\
 & HPGEE & 0.5 & 3.9457 & 0.4574 & 0.9984 & 1.08 & 94.86 & 100.00 & 0.01392 \\
 & OPGEE & 0.5 & 1.4544 & 0.4357 & 1.0376 & 1.19 & 95.40 & 100.00 & 0.01307 \\
 & PGEE & 0.1 & 3.3519 & 0.4536 & 1.0040 & 1.10 & 94.89 & 100.00 & 0.01371 \\
 & HPGEE & 0.1 & 3.9512 & 0.4575 & 0.9980 & 1.08 & 94.85 & 100.00 & 0.01393 \\
 & OPGEE & 0.1 & 3.4386 & 0.4530 & 1.0057 & 1.10 & 94.97 & 100.00 & 0.01375 \\
\addlinespace[2pt]
50 & GEE &  & 2.6373 & 0.3402 & 1.0032 & 1.07 & 95.03 & 100.00 & 0.00820 \\
 & RBR &  & 0.6876 & 0.3275 & 1.0331 & 1.16 & 95.49 & 100.00 & 0.00776 \\
 & PGEE & 0.5 & 1.1801 & 0.3299 & 1.0286 & 1.14 & 95.47 & 100.00 & 0.00787 \\
 & HPGEE & 0.5 & 2.6439 & 0.3399 & 1.0046 & 1.07 & 95.08 & 100.00 & 0.00821 \\
 & OPGEE & 0.5 & 1.1784 & 0.3297 & 1.0299 & 1.14 & 95.48 & 100.00 & 0.00789 \\
 & PGEE & 0.1 & 2.3400 & 0.3381 & 1.0082 & 1.08 & 95.08 & 100.00 & 0.00813 \\
 & HPGEE & 0.1 & 2.6425 & 0.3399 & 1.0045 & 1.07 & 95.06 & 100.00 & 0.00821 \\
 & OPGEE & 0.1 & 2.3448 & 0.3378 & 1.0095 & 1.09 & 95.12 & 100.00 & 0.00815 \\
\addlinespace[2pt]
\multicolumn{10}{c}{\textit{Unstructured}}\\
\addlinespace[1pt]
30 & GEE &  & 2.0077 & 0.4393 & 0.9942 & 1.17 & 94.83 & 99.92 & 0.01303 \\
 & RBR &  & -0.8647 & 0.4161 & 1.0364 & 1.31 & 95.59 & 99.91 & 0.01213 \\
 & PGEE & 0.5 & -0.7580 & 0.4162 & 1.0379 & 1.31 & 95.62 & 99.90 & 0.01216 \\
 & HPGEE & 0.5 & 2.3235 & 0.4400 & 0.9941 & 1.17 & 94.83 & 99.94 & 0.01316 \\
 & OPGEE & 0.5 & -0.4671 & 0.4171 & 1.0374 & 1.30 & 95.61 & 99.93 & 0.01230 \\
 & PGEE & 0.1 & 1.4368 & 0.4345 & 1.0028 & 1.20 & 95.07 & 99.92 & 0.01285 \\
 & HPGEE & 0.1 & 2.3238 & 0.4400 & 0.9940 & 1.17 & 94.80 & 99.94 & 0.01316 \\
 & OPGEE & 0.1 & 1.7514 & 0.4353 & 1.0023 & 1.20 & 95.02 & 99.93 & 0.01297 \\
\addlinespace[2pt]
50 & GEE &  & 1.6403 & 0.3277 & 1.0023 & 1.16 & 94.93 & 99.96 & 0.00769 \\
 & RBR &  & -0.1457 & 0.3162 & 1.0311 & 1.25 & 95.32 & 99.95 & 0.00733 \\
 & PGEE & 0.5 & 0.0413 & 0.3171 & 1.0297 & 1.24 & 95.30 & 99.95 & 0.00738 \\
 & HPGEE & 0.5 & 1.7364 & 0.3278 & 1.0026 & 1.16 & 94.87 & 99.96 & 0.00770 \\
 & OPGEE & 0.5 & 0.1187 & 0.3173 & 1.0300 & 1.24 & 95.30 & 99.96 & 0.00739 \\
 & PGEE & 0.1 & 1.3145 & 0.3255 & 1.0077 & 1.17 & 95.03 & 99.95 & 0.00762 \\
 & HPGEE & 0.1 & 1.7315 & 0.3278 & 1.0027 & 1.16 & 94.86 & 99.96 & 0.00770 \\
 & OPGEE & 0.1 & 1.4032 & 0.3257 & 1.0081 & 1.17 & 94.97 & 99.96 & 0.00763 \\
\bottomrule
\end{tabular*}
\end{table}

\begin{table}[p]
\centering
\caption{Simulation results for estimation of $\beta_1=0.5$ in
$\Pr(Y_{ij}=1 \mid \mathbf{x}_{ij})=\Phi(\beta_{0}+\beta_{1}x_{1i}+\beta_{2}x_{2ij})$ under MCAR.
Columns report Bias ($\times 100$), empirical standard error (ESE), ratio of the mean bias-corrected standard error to the ESE (RE),
simulated relative efficiency (SRE), empirical coverage of the 95\% Wald confidence intervals based on the bias-corrected covariance matrix (EC),
convergence proportion (CP), and prediction mean squared error (PMSE), across sample sizes $N \in \{100,500\}$ and working odds-ratio structures.}
\label{tab:app_regular_beta1_N100N500_missing}
\footnotesize
\setlength{\tabcolsep}{2.8pt}
\renewcommand{\arraystretch}{1.0}
\begin{tabular*}{\textwidth}{@{\extracolsep{\fill}} c l c r r r r r r r @{}}
\toprule
\multicolumn{1}{c}{$N$} &
\multicolumn{1}{c}{Method} &
\multicolumn{1}{c}{$\delta$} &
\multicolumn{1}{c}{Bias ($\times 100$)} &
\multicolumn{1}{c}{ESE} &
\multicolumn{1}{c}{RE} &
\multicolumn{1}{c}{SRE} &
\multicolumn{1}{c}{EC} &
\multicolumn{1}{c}{CP} &
\multicolumn{1}{c}{PMSE} \\
\midrule
\multicolumn{10}{c}{\textit{Independence}}\\
\addlinespace[1pt]
100 & GEE &  & 1.4242 & 0.2350 & 0.9834 & 1.00 & 94.60 & 100.00 & 0.00475 \\
 & RBR &  & 0.2311 & 0.2294 & 1.0015 & 1.05 & 94.94 & 100.00 & 0.00459 \\
 & PGEE & 0.5 & 0.9574 & 0.2326 & 0.9913 & 1.02 & 94.76 & 100.00 & 0.00468 \\
 & HPGEE & 0.5 & 1.4227 & 0.2350 & 0.9835 & 1.00 & 94.60 & 100.00 & 0.00475 \\
 & PGEE & 0.1 & 1.3301 & 0.2345 & 0.9850 & 1.00 & 94.64 & 100.00 & 0.00473 \\
 & HPGEE & 0.1 & 1.4240 & 0.2350 & 0.9834 & 1.00 & 94.60 & 100.00 & 0.00475 \\
\addlinespace[2pt]
500 & GEE &  & 0.4026 & 0.1041 & 0.9970 & 1.00 & 94.60 & 100.00 & 0.00094 \\
 & RBR &  & 0.1429 & 0.1035 & 1.0014 & 1.01 & 94.80 & 100.00 & 0.00093 \\
 & PGEE & 0.5 & 0.3076 & 0.1039 & 0.9987 & 1.00 & 94.64 & 100.00 & 0.00094 \\
 & HPGEE & 0.5 & 0.4025 & 0.1041 & 0.9970 & 1.00 & 94.60 & 100.00 & 0.00094 \\
 & PGEE & 0.1 & 0.3835 & 0.1041 & 0.9974 & 1.00 & 94.62 & 100.00 & 0.00094 \\
 & HPGEE & 0.1 & 0.4026 & 0.1041 & 0.9970 & 1.00 & 94.60 & 100.00 & 0.00094 \\
\addlinespace[2pt]
\multicolumn{10}{c}{\textit{Exchangeable}}\\
\addlinespace[1pt]
100 & GEE &  & 1.2443 & 0.2298 & 0.9855 & 1.05 & 94.56 & 100.00 & 0.00431 \\
 & RBR &  & 0.1385 & 0.2247 & 1.0020 & 1.10 & 94.93 & 100.00 & 0.00418 \\
 & PGEE & 0.5 & 0.5368 & 0.2262 & 0.9980 & 1.08 & 94.88 & 100.00 & 0.00422 \\
 & HPGEE & 0.5 & 1.2082 & 0.2295 & 0.9867 & 1.05 & 94.53 & 100.00 & 0.00431 \\
 & OPGEE & 0.5 & 0.4988 & 0.2260 & 0.9989 & 1.09 & 94.85 & 100.00 & 0.00422 \\
 & PGEE & 0.1 & 1.1014 & 0.2290 & 0.9880 & 1.05 & 94.58 & 100.00 & 0.00429 \\
 & HPGEE & 0.1 & 1.2079 & 0.2295 & 0.9866 & 1.05 & 94.53 & 100.00 & 0.00431 \\
 & OPGEE & 0.1 & 1.0649 & 0.2288 & 0.9890 & 1.06 & 94.60 & 100.00 & 0.00429 \\
\addlinespace[2pt]
500 & GEE &  & 0.3376 & 0.1019 & 0.9947 & 1.04 & 94.69 & 100.00 & 0.00085 \\
 & RBR &  & 0.0953 & 0.1013 & 0.9988 & 1.06 & 94.82 & 100.00 & 0.00084 \\
 & PGEE & 0.5 & 0.1932 & 0.1015 & 0.9973 & 1.05 & 94.80 & 100.00 & 0.00085 \\
 & HPGEE & 0.5 & 0.3330 & 0.1018 & 0.9949 & 1.05 & 94.69 & 100.00 & 0.00085 \\
 & OPGEE & 0.5 & 0.1884 & 0.1015 & 0.9975 & 1.05 & 94.82 & 100.00 & 0.00085 \\
 & PGEE & 0.1 & 0.3087 & 0.1018 & 0.9952 & 1.05 & 94.70 & 100.00 & 0.00085 \\
 & HPGEE & 0.1 & 0.3330 & 0.1019 & 0.9949 & 1.05 & 94.70 & 100.00 & 0.00085 \\
 & OPGEE & 0.1 & 0.3040 & 0.1018 & 0.9954 & 1.05 & 94.72 & 100.00 & 0.00085 \\
\addlinespace[2pt]
\multicolumn{10}{c}{\textit{Unstructured}}\\
\addlinespace[1pt]
100 & GEE &  & 0.6899 & 0.2216 & 0.9869 & 1.13 & 94.46 & 100.00 & 0.00400 \\
 & RBR &  & -0.3329 & 0.2170 & 1.0031 & 1.18 & 94.80 & 100.00 & 0.00389 \\
 & PGEE & 0.5 & -0.0725 & 0.2180 & 1.0002 & 1.17 & 94.81 & 100.00 & 0.00392 \\
 & HPGEE & 0.5 & 0.6958 & 0.2216 & 0.9876 & 1.13 & 94.47 & 100.00 & 0.00399 \\
 & OPGEE & 0.5 & -0.0726 & 0.2179 & 1.0008 & 1.17 & 94.75 & 100.00 & 0.00391 \\
 & PGEE & 0.1 & 0.5359 & 0.2209 & 0.9895 & 1.14 & 94.53 & 100.00 & 0.00398 \\
 & HPGEE & 0.1 & 0.6945 & 0.2216 & 0.9876 & 1.13 & 94.47 & 100.00 & 0.00399 \\
 & OPGEE & 0.1 & 0.5396 & 0.2208 & 0.9902 & 1.14 & 94.48 & 100.00 & 0.00397 \\
\addlinespace[2pt]
500 & GEE &  & 0.2188 & 0.0990 & 0.9931 & 1.11 & 94.75 & 100.00 & 0.00078 \\
 & RBR &  & -0.0072 & 0.0985 & 0.9970 & 1.12 & 94.87 & 100.00 & 0.00078 \\
 & PGEE & 0.5 & 0.0631 & 0.0987 & 0.9959 & 1.11 & 94.83 & 100.00 & 0.00078 \\
 & HPGEE & 0.5 & 0.2232 & 0.0990 & 0.9933 & 1.11 & 94.79 & 100.00 & 0.00078 \\
 & OPGEE & 0.5 & 0.0672 & 0.0987 & 0.9961 & 1.11 & 94.84 & 100.00 & 0.00078 \\
 & PGEE & 0.1 & 0.1876 & 0.0990 & 0.9937 & 1.11 & 94.78 & 100.00 & 0.00078 \\
 & HPGEE & 0.1 & 0.2232 & 0.0990 & 0.9933 & 1.11 & 94.79 & 100.00 & 0.00078 \\
 & OPGEE & 0.1 & 0.1919 & 0.0989 & 0.9939 & 1.11 & 94.78 & 100.00 & 0.00078 \\
\bottomrule
\end{tabular*}
\end{table}

\begin{table}[p]
\centering
\caption{Simulation results for estimation of $\beta_2=0.5$ in
$\Pr(Y_{ij}=1 \mid \mathbf{x}_{ij})=\Phi(\beta_{0}+\beta_{1}x_{1i}+\beta_{2}x_{2ij})$ under MCAR.
Columns report Bias ($\times 100$), empirical standard error (ESE), ratio of the mean bias-corrected standard error to the ESE (RE),
simulated relative efficiency (SRE), empirical coverage of the 95\% Wald confidence intervals based on the bias-corrected covariance matrix (EC),
convergence proportion (CP), and prediction mean squared error (PMSE), across sample sizes $N \in \{30,50\}$ and working odds-ratio structures.}
\label{tab:app_regular_beta2_N30N50_missing}
\footnotesize
\setlength{\tabcolsep}{2.8pt}
\renewcommand{\arraystretch}{1.0}
\begin{tabular*}{\textwidth}{@{\extracolsep{\fill}} c l c r r r r r r r @{}}
\toprule
\multicolumn{1}{c}{$N$} &
\multicolumn{1}{c}{Method} &
\multicolumn{1}{c}{$\delta$} &
\multicolumn{1}{c}{Bias ($\times 100$)} &
\multicolumn{1}{c}{ESE} &
\multicolumn{1}{c}{RE} &
\multicolumn{1}{c}{SRE} &
\multicolumn{1}{c}{EC} &
\multicolumn{1}{c}{CP} &
\multicolumn{1}{c}{PMSE} \\
\midrule
\multicolumn{10}{c}{\textit{Independence}}\\
\addlinespace[1pt]
30 & GEE &  & 5.3443 & 0.5148 & 0.9586 & 1.00 & 93.74 & 100.00 & 0.01581 \\
 & RBR &  & 1.5666 & 0.4821 & 1.0047 & 1.15 & 94.48 & 100.00 & 0.01445 \\
 & PGEE & 0.5 & 3.3257 & 0.4954 & 0.9867 & 1.09 & 94.13 & 100.00 & 0.01509 \\
 & HPGEE & 0.5 & 5.3013 & 0.5142 & 0.9595 & 1.00 & 93.77 & 100.00 & 0.01579 \\
 & PGEE & 0.1 & 4.9294 & 0.5108 & 0.9642 & 1.02 & 93.82 & 100.00 & 0.01566 \\
 & HPGEE & 0.1 & 5.3397 & 0.5147 & 0.9587 & 1.00 & 93.74 & 100.00 & 0.01581 \\
\addlinespace[2pt]
50 & GEE &  & 3.1252 & 0.3660 & 0.9967 & 1.00 & 94.68 & 100.00 & 0.00928 \\
 & RBR &  & 0.7703 & 0.3502 & 1.0305 & 1.10 & 95.42 & 100.00 & 0.00873 \\
 & PGEE & 0.5 & 2.0462 & 0.3581 & 1.0139 & 1.05 & 95.15 & 100.00 & 0.00902 \\
 & HPGEE & 0.5 & 3.1172 & 0.3659 & 0.9969 & 1.00 & 94.71 & 100.00 & 0.00928 \\
 & PGEE & 0.1 & 2.9059 & 0.3644 & 1.0001 & 1.01 & 94.78 & 100.00 & 0.00923 \\
 & HPGEE & 0.1 & 3.1245 & 0.3660 & 0.9967 & 1.00 & 94.69 & 100.00 & 0.00928 \\
\addlinespace[2pt]
\multicolumn{10}{c}{\textit{Exchangeable}}\\
\addlinespace[1pt]
30 & GEE &  & 2.6428 & 0.4218 & 0.9918 & 1.50 & 94.47 & 100.00 & 0.01389 \\
 & RBR &  & -0.4374 & 0.3987 & 1.0327 & 1.68 & 95.08 & 100.00 & 0.01282 \\
 & PGEE & 0.5 & 0.5671 & 0.4060 & 1.0198 & 1.62 & 94.93 & 100.00 & 0.01303 \\
 & HPGEE & 0.5 & 3.6754 & 0.4256 & 0.9858 & 1.47 & 94.30 & 100.00 & 0.01392 \\
 & OPGEE & 0.5 & 1.6056 & 0.4102 & 1.0122 & 1.59 & 94.82 & 100.00 & 0.01307 \\
 & PGEE & 0.1 & 2.2174 & 0.4186 & 0.9973 & 1.52 & 94.53 & 100.00 & 0.01371 \\
 & HPGEE & 0.1 & 3.7106 & 0.4260 & 0.9849 & 1.46 & 94.26 & 100.00 & 0.01393 \\
 & OPGEE & 0.1 & 3.2837 & 0.4228 & 0.9903 & 1.49 & 94.35 & 100.00 & 0.01375 \\
\addlinespace[2pt]
50 & GEE &  & 1.7491 & 0.3064 & 0.9977 & 1.43 & 95.00 & 100.00 & 0.00820 \\
 & RBR &  & -0.2524 & 0.2946 & 1.0279 & 1.55 & 95.60 & 100.00 & 0.00776 \\
 & PGEE & 0.5 & 0.5644 & 0.2995 & 1.0147 & 1.50 & 95.37 & 100.00 & 0.00787 \\
 & HPGEE & 0.5 & 2.3123 & 0.3080 & 0.9940 & 1.41 & 94.92 & 100.00 & 0.00821 \\
 & OPGEE & 0.5 & 1.1221 & 0.3012 & 1.0103 & 1.48 & 95.20 & 100.00 & 0.00789 \\
 & PGEE & 0.1 & 1.5088 & 0.3050 & 1.0011 & 1.45 & 95.07 & 100.00 & 0.00813 \\
 & HPGEE & 0.1 & 2.3210 & 0.3080 & 0.9938 & 1.41 & 94.92 & 100.00 & 0.00821 \\
 & OPGEE & 0.1 & 2.0790 & 0.3067 & 0.9970 & 1.43 & 94.95 & 100.00 & 0.00815 \\
\addlinespace[2pt]
\multicolumn{10}{c}{\textit{Unstructured}}\\
\addlinespace[1pt]
30 & GEE &  & -2.4000 & 0.3736 & 0.9391 & 1.91 & 92.30 & 99.92 & 0.01303 \\
 & RBR &  & -5.4949 & 0.3527 & 0.9770 & 2.10 & 92.54 & 99.91 & 0.01213 \\
 & PGEE & 0.5 & -5.3590 & 0.3563 & 0.9702 & 2.06 & 92.43 & 99.90 & 0.01216 \\
 & HPGEE & 0.5 & -0.2215 & 0.3821 & 0.9261 & 1.84 & 92.24 & 99.94 & 0.01316 \\
 & OPGEE & 0.5 & -3.0061 & 0.3654 & 0.9540 & 1.99 & 92.36 & 99.93 & 0.01230 \\
 & PGEE & 0.1 & -3.0080 & 0.3701 & 0.9451 & 1.94 & 92.34 & 99.92 & 0.01285 \\
 & HPGEE & 0.1 & -0.1506 & 0.3826 & 0.9249 & 1.83 & 92.21 & 99.94 & 0.01316 \\
 & OPGEE & 0.1 & -0.7117 & 0.3790 & 0.9305 & 1.86 & 92.20 & 99.93 & 0.01297 \\
\addlinespace[2pt]
50 & GEE &  & -1.7305 & 0.2764 & 0.9502 & 1.76 & 92.88 & 99.96 & 0.00769 \\
 & RBR &  & -3.6561 & 0.2657 & 0.9782 & 1.88 & 93.19 & 99.95 & 0.00733 \\
 & PGEE & 0.5 & -3.5083 & 0.2695 & 0.9668 & 1.83 & 92.98 & 99.95 & 0.00738 \\
 & HPGEE & 0.5 & -0.2523 & 0.2780 & 0.9489 & 1.75 & 93.09 & 99.96 & 0.00770 \\
 & OPGEE & 0.5 & -1.9614 & 0.2713 & 0.9646 & 1.82 & 93.27 & 99.96 & 0.00739 \\
 & PGEE & 0.1 & -2.0892 & 0.2750 & 0.9534 & 1.77 & 92.93 & 99.95 & 0.00762 \\
 & HPGEE & 0.1 & -0.2235 & 0.2782 & 0.9485 & 1.74 & 93.08 & 99.96 & 0.00770 \\
 & OPGEE & 0.1 & -0.5691 & 0.2768 & 0.9516 & 1.76 & 93.14 & 99.96 & 0.00763 \\
\bottomrule
\end{tabular*}
\end{table}

\begin{table}[p]
\centering
\caption{Simulation results for estimation of $\beta_2=0.5$ in
$\Pr(Y_{ij}=1 \mid \mathbf{x}_{ij})=\Phi(\beta_{0}+\beta_{1}x_{1i}+\beta_{2}x_{2ij})$ under MCAR.
Columns report Bias ($\times 100$), empirical standard error (ESE), ratio of the mean bias-corrected standard error to the ESE (RE),
simulated relative efficiency (SRE), empirical coverage of the 95\% Wald confidence intervals based on the bias-corrected covariance matrix (EC),
convergence proportion (CP), and prediction mean squared error (PMSE), across sample sizes $N \in \{100,500\}$ and working odds-ratio structures.}
\label{tab:app_regular_beta2_N100N500_missing}
\footnotesize
\setlength{\tabcolsep}{2.8pt}
\renewcommand{\arraystretch}{1.0}
\begin{tabular*}{\textwidth}{@{\extracolsep{\fill}} c l c r r r r r r r @{}}
\toprule
\multicolumn{1}{c}{$N$} &
\multicolumn{1}{c}{Method} &
\multicolumn{1}{c}{$\delta$} &
\multicolumn{1}{c}{Bias ($\times 100$)} &
\multicolumn{1}{c}{ESE} &
\multicolumn{1}{c}{RE} &
\multicolumn{1}{c}{SRE} &
\multicolumn{1}{c}{EC} &
\multicolumn{1}{c}{CP} &
\multicolumn{1}{c}{PMSE} \\
\midrule
\multicolumn{10}{c}{\textit{Independence}}\\
\addlinespace[1pt]
100 & GEE &  & 1.2218 & 0.2158 & 0.9980 & 1.00 & 94.62 & 100.00 & 0.00475 \\
 & RBR &  & -0.0429 & 0.2103 & 1.0177 & 1.06 & 94.91 & 100.00 & 0.00459 \\
 & PGEE & 0.5 & 0.7084 & 0.2134 & 1.0069 & 1.02 & 94.80 & 100.00 & 0.00468 \\
 & HPGEE & 0.5 & 1.2203 & 0.2158 & 0.9981 & 1.00 & 94.63 & 100.00 & 0.00475 \\
 & PGEE & 0.1 & 1.1184 & 0.2153 & 0.9998 & 1.01 & 94.65 & 100.00 & 0.00473 \\
 & HPGEE & 0.1 & 1.2217 & 0.2158 & 0.9980 & 1.00 & 94.63 & 100.00 & 0.00475 \\
\addlinespace[2pt]
500 & GEE &  & 0.3267 & 0.0958 & 0.9999 & 1.00 & 95.12 & 100.00 & 0.00094 \\
 & RBR &  & 0.0614 & 0.0953 & 1.0042 & 1.01 & 95.24 & 100.00 & 0.00093 \\
 & PGEE & 0.5 & 0.2264 & 0.0956 & 1.0017 & 1.01 & 95.15 & 100.00 & 0.00094 \\
 & HPGEE & 0.5 & 0.3267 & 0.0958 & 0.9999 & 1.00 & 95.12 & 100.00 & 0.00094 \\
 & PGEE & 0.1 & 0.3066 & 0.0958 & 1.0003 & 1.00 & 95.13 & 100.00 & 0.00094 \\
 & HPGEE & 0.1 & 0.3267 & 0.0958 & 0.9999 & 1.00 & 95.12 & 100.00 & 0.00094 \\
\addlinespace[2pt]
\multicolumn{10}{c}{\textit{Exchangeable}}\\
\addlinespace[1pt]
100 & GEE &  & 0.5479 & 0.1918 & 0.9998 & 1.27 & 95.00 & 100.00 & 0.00431 \\
 & RBR &  & -0.5520 & 0.1876 & 1.0168 & 1.33 & 95.19 & 100.00 & 0.00418 \\
 & PGEE & 0.5 & -0.0591 & 0.1894 & 1.0095 & 1.30 & 95.12 & 100.00 & 0.00422 \\
 & HPGEE & 0.5 & 0.6955 & 0.1920 & 0.9992 & 1.27 & 94.96 & 100.00 & 0.00431 \\
 & OPGEE & 0.5 & 0.0834 & 0.1896 & 1.0086 & 1.30 & 95.15 & 100.00 & 0.00422 \\
 & PGEE & 0.1 & 0.4255 & 0.1913 & 1.0017 & 1.28 & 95.03 & 100.00 & 0.00429 \\
 & HPGEE & 0.1 & 0.6968 & 0.1920 & 0.9991 & 1.27 & 94.96 & 100.00 & 0.00431 \\
 & OPGEE & 0.1 & 0.5733 & 0.1915 & 1.0010 & 1.27 & 94.99 & 100.00 & 0.00429 \\
\addlinespace[2pt]
500 & GEE &  & 0.1830 & 0.0844 & 1.0022 & 1.29 & 95.02 & 100.00 & 0.00085 \\
 & RBR &  & -0.0438 & 0.0840 & 1.0059 & 1.30 & 95.10 & 100.00 & 0.00084 \\
 & PGEE & 0.5 & 0.0633 & 0.0841 & 1.0041 & 1.30 & 95.03 & 100.00 & 0.00085 \\
 & HPGEE & 0.5 & 0.2171 & 0.0844 & 1.0020 & 1.29 & 95.01 & 100.00 & 0.00085 \\
 & OPGEE & 0.5 & 0.0972 & 0.0842 & 1.0039 & 1.30 & 95.03 & 100.00 & 0.00085 \\
 & PGEE & 0.1 & 0.1590 & 0.0843 & 1.0026 & 1.29 & 95.00 & 100.00 & 0.00085 \\
 & HPGEE & 0.1 & 0.2171 & 0.0844 & 1.0020 & 1.29 & 95.02 & 100.00 & 0.00085 \\
 & OPGEE & 0.1 & 0.1931 & 0.0843 & 1.0024 & 1.29 & 95.02 & 100.00 & 0.00085 \\
\addlinespace[2pt]
\multicolumn{10}{c}{\textit{Unstructured}}\\
\addlinespace[1pt]
100 & GEE &  & -0.9956 & 0.1777 & 0.9823 & 1.47 & 94.40 & 100.00 & 0.00400 \\
 & RBR &  & -2.0210 & 0.1738 & 0.9987 & 1.52 & 94.48 & 100.00 & 0.00389 \\
 & PGEE & 0.5 & -1.7583 & 0.1754 & 0.9914 & 1.50 & 94.42 & 100.00 & 0.00392 \\
 & HPGEE & 0.5 & -0.5522 & 0.1777 & 0.9835 & 1.48 & 94.51 & 100.00 & 0.00399 \\
 & OPGEE & 0.5 & -1.3167 & 0.1755 & 0.9923 & 1.51 & 94.47 & 100.00 & 0.00391 \\
 & PGEE & 0.1 & -1.1494 & 0.1772 & 0.9841 & 1.48 & 94.41 & 100.00 & 0.00398 \\
 & HPGEE & 0.1 & -0.5470 & 0.1777 & 0.9834 & 1.48 & 94.51 & 100.00 & 0.00399 \\
 & OPGEE & 0.1 & -0.7009 & 0.1773 & 0.9852 & 1.48 & 94.50 & 100.00 & 0.00397 \\
\addlinespace[2pt]
500 & GEE &  & -0.1227 & 0.0779 & 0.9981 & 1.52 & 95.05 & 100.00 & 0.00078 \\
 & RBR &  & -0.3332 & 0.0775 & 1.0017 & 1.53 & 95.14 & 100.00 & 0.00078 \\
 & PGEE & 0.5 & -0.2701 & 0.0777 & 0.9998 & 1.52 & 95.10 & 100.00 & 0.00078 \\
 & HPGEE & 0.5 & -0.0362 & 0.0779 & 0.9981 & 1.52 & 95.04 & 100.00 & 0.00078 \\
 & OPGEE & 0.5 & -0.1838 & 0.0777 & 0.9999 & 1.52 & 95.11 & 100.00 & 0.00078 \\
 & PGEE & 0.1 & -0.1523 & 0.0778 & 0.9984 & 1.52 & 95.08 & 100.00 & 0.00078 \\
 & HPGEE & 0.1 & -0.0360 & 0.0779 & 0.9981 & 1.52 & 95.04 & 100.00 & 0.00078 \\
 & OPGEE & 0.1 & -0.0655 & 0.0778 & 0.9985 & 1.52 & 95.05 & 100.00 & 0.00078 \\
\bottomrule
\end{tabular*}
\end{table}

\hypertarget{sec:sandwich}{%
\subsection{Model-Based, Sandwich and Bias-Corrected Standard Errors}\label{sec:sandwich}}

This section supports Section 3.3 of the main paper. Every fit of the quasi-complete separation and regular regimes was stored with three standard errors: model-based, from \(\widehat{\boldsymbol \Sigma}_{0}^{-1}\); sandwich, from \(\widehat{\boldsymbol \Sigma}_{R}\) in (3); and bias-corrected, from \(\widehat{\boldsymbol \Sigma}_{M}\) in (4). The datasets, failure rule, and successful fits are those of Tables S1--S12. \ref{supp:sandwich} reports, for \(N\in\{30,100,500\}\), the empirical coverage (EC, \%) of nominal 95\% Wald intervals and the ratio of the mean estimated standard error to the empirical standard error (RE; values below 1 indicate that the standard errors are too small on average), for PGEE with \(\delta=0.5\) under each working structure and, in the regular regime, for ordinary GEE. Results for \(N=50\), for OPGEE, HPGEE, and RBR, and for the MCAR scenario are produced by the scripts below.

The accompanying zip folder also contains the scripts that generate all simulation results, using version 0.1.1 of the \textsf{R} package \texttt{geer}. The scripts \texttt{05\_separation\_pgee.R} and \texttt{07\_regular\_pgee.R} simulate the separation and regular regimes and store every fit, and \texttt{06\_separation\_pgee\_summary.R} and \texttt{08\_regular\_pgee\_summary.R} produce Tables S1--S12 and \ref{supp:sandwich} and the corresponding tables of the main paper.

\hypertarget{sec:example}{%
\section{Additional Results for the Respiratory Trial}\label{sec:example}}

This Appendix provides additional numerical results for the analysis in Section 4 of the main paper. \ref{sec:example-penalty} reports full coefficient tables for PGEE, OPGEE, and HPGEE under odds-ratio working structures for \(\delta\in\{0.5,0.1\}\). \ref{sec:example-logit} presents a sensitivity analysis under a logit link. \ref{sec:corrcoef} documents the inadmissibility of correlation-coefficient parameterizations during variable selection. \ref{sec:example-cv} reports cross-validation and QIC across values of the penalty strength \(\delta\).

\begin{figure}[p]
\centering
\includegraphics[width=\textwidth]{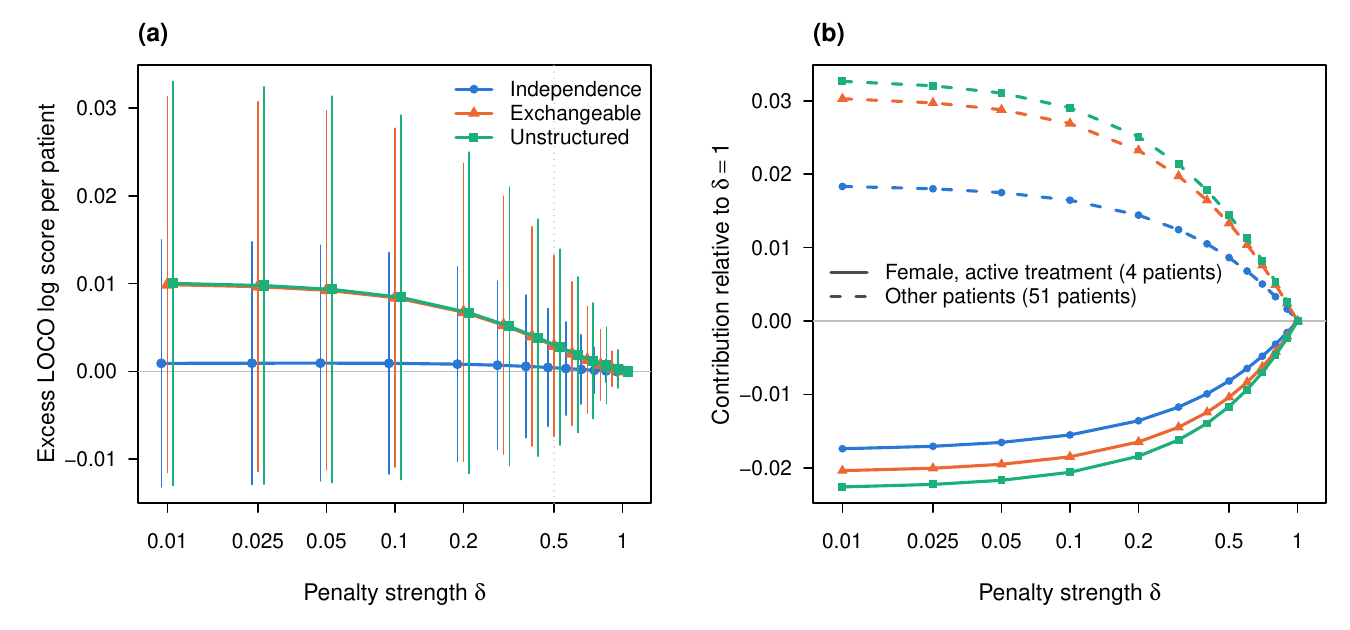}
\caption{Cross-validation of the penalty strength $\delta$ for PGEE in the respiratory illness trial (selected model, probit link). (a) Mean LOCO log score per patient minus its minimum over the grid, with $\pm 1$ paired standard error bars (points slightly offset horizontally); the dotted line marks the default $\delta=1/2$. (b) Contributions to the mean LOCO log score of the four female patients on active treatment (solid) and of the remaining 51 patients (dashed), relative to their values at $\delta=1$. Colors and symbols identify the working odds-ratio structure.}
\label{fig:cv_delta}
\end{figure}

The accompanying zip folder contains the following self-contained \textsf{R} scripts:

\begin{enumerate}
    \item \texttt{01\_logit\_sensitivity.R}: reproduces the logit link sensitivity analysis of \ref{sec:example-logit},
    \item \texttt{02\_corrcoef\_inadmissibility.R}: documents the inadmissibility of correlation-coefficient parameterizations during variable selection,
    \item \texttt{03\_cv\_delta.R}: reproduces the cross-validation and QIC results of \ref{sec:example-cv}, including \ref{supp:cv_delta} and Figure \ref{fig:cv_delta}.
\end{enumerate}

Scripts 1 and 2 require the \texttt{geer} package, and script 3 requires \texttt{geer}, \texttt{foreach}, \texttt{doParallel}, and \texttt{doRNG}, and runs in about three minutes on a single core.

\hypertarget{sec:example-penalty}{%
\subsection{Sensitivity to Penalty Parameter and One-Step Approximations}\label{sec:example-penalty}}

\ref{supp:results_dataset} extends Table 3 of the main paper by adding OPGEE and HPGEE results and sensitivity to \(\delta=0.1\). Ordinary GEE did not converge for any working association structure. Regression coefficient estimates were broadly consistent across working structures and penalty values.

\begin{table}[p]
\caption{Results for PGEE, OPGEE, and HPGEE using odds ratios to characterize the working association structure for the respiratory illness clinical trial.\label{supp:results_dataset}}
\centering
\footnotesize
\setlength{\tabcolsep}{2.5pt}
\renewcommand{\arraystretch}{0.92}
\begin{tabular*}{\textwidth}{@{\extracolsep{\fill}} l r r r r r r r r r @{}}
  \toprule
 & \multicolumn{3}{c}{Independence} & \multicolumn{3}{c}{Exchangeable} 
 & \multicolumn{3}{c}{Unstructured}\\
  \cmidrule(lr){2-4}\cmidrule(lr){5-7}\cmidrule(lr){8-10}
Parameter & Estimate & SE & $p$-value & Estimate & SE & $p$-value 
          & Estimate & SE & $p$-value\\
  \midrule
\addlinespace[0.3ex]
& \multicolumn{9}{c}{PGEE with $\delta = 0.5$}\\
\addlinespace[0.3ex]
$\beta_0$ & $-$1.1848 & 0.7797 & 0.1286 & $-$1.3384 & 0.7559 & 0.0766 & $-$1.2292 & 0.7164 & 0.0862 \\
$\beta_1$ &    2.3727 & 0.5196 & $<.0001$ &    2.3341 & 0.5096 & $<.0001$ &    2.3229 & 0.5129 & $<.0001$ \\
$\beta_2$ &    0.2927 & 0.4345 & 0.5005 &    0.2795 & 0.4458 & 0.5307 &    0.3130 & 0.4383 & 0.4751 \\
$\beta_3$ &    0.3542 & 0.2204 & 0.1080 &    0.3994 & 0.2044 & 0.0507 &    0.3509 & 0.1957 & 0.0730 \\
$\beta_4$ &    0.0203 & 0.0162 & 0.2103 &    0.0250 & 0.0155 & 0.1082 &    0.0195 & 0.0146 & 0.1803 \\
$\beta_5$ &    0.6650 & 0.3175 & 0.0362 &    0.6604 & 0.3194 & 0.0387 &    0.6976 & 0.3136 & 0.0261 \\
$\beta_6$ & $-$1.7077 & 0.6293 & 0.0067 & $-$1.6485 & 0.6228 & 0.0081 & $-$1.6341 & 0.6221 & 0.0086 \\
$\beta_7$ & $-$0.0106 & 0.0053 & 0.0449 & $-$0.0119 & 0.0049 & 0.0145 & $-$0.0100 & 0.0046 & 0.0295 \\
\addlinespace[0.3ex]
& \multicolumn{9}{c}{OPGEE with $\delta = 0.5$}\\
\addlinespace[0.3ex]
$\beta_0$ &  &  &  & $-$1.3182 & 0.7546 & 0.0807 & $-$1.2286 & 0.7162 & 0.0863 \\
$\beta_1$ &  &  &  &    2.3327 & 0.5091 & $<.0001$ &    2.3214 & 0.5129 & $<.0001$ \\
$\beta_2$ &  &  &  &    0.2759 & 0.4456 & 0.5358 &    0.3150 & 0.4380 & 0.4721 \\
$\beta_3$ &  &  &  &    0.3956 & 0.2041 & 0.0526 &    0.3502 & 0.1955 & 0.0732 \\
$\beta_4$ &  &  &  &    0.0243 & 0.0155 & 0.1162 &    0.0196 & 0.0146 & 0.1789 \\
$\beta_5$ &  &  &  &    0.6572 & 0.3193 & 0.0395 &    0.6954 & 0.3133 & 0.0264 \\
$\beta_6$ &  &  &  & $-$1.6461 & 0.6222 & 0.0082 & $-$1.6368 & 0.6220 & 0.0085 \\
$\beta_7$ &  &  &  & $-$0.0118 & 0.0049 & 0.0156 & $-$0.0100 & 0.0046 & 0.0295 \\
\addlinespace[0.3ex]
& \multicolumn{9}{c}{HPGEE with $\delta = 0.5$}\\
\addlinespace[0.3ex]
$\beta_0$ & $-$1.2320 & 0.7881 & 0.1180 & $-$1.3735 & 0.7681 & 0.0737 & $-$1.2872 & 0.7296 & 0.0777 \\ 
$\beta_1$ &    2.8188 & 0.6490 & $<.0001$ &    2.8138 & 0.5972 & $<.0001$ &    2.8208 & 0.6044 & $<.0001$\\ 
$\beta_2$ &    0.3006 & 0.4367 & 0.4912 &    0.2926 & 0.4487 & 0.5142 &    0.3316 & 0.4412 & 0.4524 \\ 
$\beta_3$ &    0.3680 & 0.2238 & 0.1002 &    0.4077 & 0.2085 & 0.0505 &    0.3658 & 0.2002 & 0.0677 \\ 
$\beta_4$ &    0.0213 & 0.0164 & 0.1936 &    0.0251 & 0.0158 & 0.1113 &    0.0204 & 0.0148 & 0.1684 \\ 
$\beta_5$ &    0.6825 & 0.3230 & 0.0346 &    0.6915 & 0.3277 & 0.0349 &    0.7287 & 0.3215 & 0.0234 \\ 
$\beta_6$ & $-$2.1385 & 0.7411 & 0.0039 & $-$2.1006 & 0.6992 & 0.0027 & $-$2.1099 & 0.7019 & 0.0026 \\ 
$\beta_7$ & $-$0.0110 & 0.0054 & 0.0410 & $-$0.0121 & 0.0050 & 0.0150 & $-$0.0104 & 0.0047 & 0.0271 \\ 
\addlinespace[0.3ex]
& \multicolumn{9}{c}{PGEE with $\delta = 0.1$}\\
\addlinespace[0.3ex]
$\beta_0$ & $-$1.2265 & 0.7878 & 0.1195 & $-$1.3996 & 0.7700 & 0.0691 & $-$1.2958 & 0.7305 & 0.0761 \\
$\beta_1$ &    2.9839 & 0.7360 & 0.0001 &    2.9848 & 0.6607 & $<.0001$ &    2.9959 & 0.6677 & $<.0001$ \\
$\beta_2$ &    0.2997 & 0.4364 & 0.4922 &    0.2929 & 0.4486 & 0.5137 &    0.3283 & 0.4412 & 0.4568 \\
$\beta_3$ &    0.3658 & 0.2237 & 0.1020 &    0.4121 & 0.2083 & 0.0479 &    0.3662 & 0.2000 & 0.0670 \\
$\beta_4$ &    0.0212 & 0.0164 & 0.1956 &    0.0260 & 0.0158 & 0.1003 &    0.0206 & 0.0148 & 0.1657 \\
$\beta_5$ &    0.6807 & 0.3234 & 0.0353 &    0.6914 & 0.3290 & 0.0356 &    0.7303 & 0.3229 & 0.0237 \\
$\beta_6$ & $-$2.3064 & 0.8181 & 0.0048 & $-$2.2739 & 0.7542 & 0.0026 & $-$2.2822 & 0.7570 & 0.0026 \\
$\beta_7$ & $-$0.0109 & 0.0054 & 0.0421 & $-$0.0123 & 0.0050 & 0.0138 & $-$0.0104 & 0.0047 & 0.0271 \\
\addlinespace[0.3ex]
& \multicolumn{9}{c}{OPGEE with $\delta = 0.1$}\\
\addlinespace[0.3ex]
$\beta_0$ &  &  &  & $-$1.3753 & 0.7684 & 0.0735 & $-$1.2881 & 0.7298 & 0.0776 \\
$\beta_1$ &  &  &  &    2.9825 & 0.6593 & $<.0001$ &    2.9874 & 0.6656 & $<.0001$ \\
$\beta_2$ &  &  &  &    0.2900 & 0.4483 & 0.5177 &    0.3290 & 0.4409 & 0.4555 \\
$\beta_3$ &  &  &  &    0.4078 & 0.2081 & 0.0500 &    0.3645 & 0.1996 & 0.0679 \\
$\beta_4$ &  &  &  &    0.0253 & 0.0158 & 0.1088 &    0.0205 & 0.0148 & 0.1667 \\
$\beta_5$ &  &  &  &    0.6878 & 0.3287 & 0.0364 &    0.7264 & 0.3224 & 0.0242 \\
$\beta_6$ &  &  &  & $-$2.2728 & 0.7528 & 0.0025 & $-$2.2798 & 0.7550 & 0.0025 \\
$\beta_7$ &  &  &  & $-$0.0121 & 0.0050 & 0.0148 & $-$0.0104 & 0.0047 & 0.0274 \\
\addlinespace[0.3ex]
& \multicolumn{9}{c}{HPGEE with $\delta = 0.1$}\\
\addlinespace[0.3ex]
$\beta_0$ & $-$1.2360 & 0.7896 & 0.1175 & $-$1.3867 & 0.7720 & 0.0725 & $-$1.3002 & 0.7333 & 0.0762 \\ 
$\beta_1$ &    3.3413 & 1.0455 & 0.0014 &    3.3407 & 0.8841 & 0.0002 &    3.3482 & 0.8885 & 0.0002 \\ 
$\beta_2$ &    0.3013 & 0.4369 & 0.4903 &    0.2935 & 0.4490 & 0.5133 &    0.3324 & 0.4416 & 0.4516 \\ 
$\beta_3$ &    0.3686 & 0.2245 & 0.1007 &    0.4103 & 0.2091 & 0.0497 &    0.3677 & 0.2007 & 0.0669 \\ 
$\beta_4$ &    0.0214 & 0.0164 & 0.1924 &    0.0255 & 0.0158 & 0.1079 &    0.0207 & 0.0149 & 0.1647 \\ 
$\beta_5$ &    0.6842 & 0.3249 & 0.0352 &    0.6948 & 0.3316 & 0.0362 &    0.7333 & 0.3250 & 0.0241 \\ 
$\beta_6$ & $-$2.6606 & 1.1050 & 0.0161 & $-$2.6254 & 0.9565 & 0.0061 & $-$2.6351 & 0.9578 & 0.0059 \\ 
$\beta_7$ & $-$0.0110 & 0.0054 & 0.0416 & $-$0.0122 & 0.0050 & 0.0148 & $-$0.0105 & 0.0047 & 0.0270 \\ 
\bottomrule
\end{tabular*}
\end{table}

\hypertarget{sec:example-logit}{%
\subsection{Sensitivity to the Link Function}\label{sec:example-logit}}

\ref{supp:results_logit} reports PGEE, OPGEE, and HPGEE estimates under a logit link for the selected Clinic 2 model, for \(\delta\in\{0.5, 0.1\}\) and all three working structures. Backward elimination was repeated under the logit link and yielded the same selected model as under the probit link. Substantive conclusions were qualitatively unchanged relative to the probit results. Baseline respiratory status, the treatment-by-gender interaction, and the time-by-age interaction remained statistically significant at the \(5\%\) level, and the direction and relative magnitude of all estimates were consistent across link functions. The one exception is HPGEE under the independence working structure with \(\delta=0.1\), for which \(\beta_6\) does not reach the \(5\%\) significance threshold.

\begin{table}[p]
\caption{Results for PGEE, OPGEE, and HPGEE using a logit link and 
odds ratios to characterize the working association structure for the 
respiratory illness clinical trial. \label{supp:results_logit}}
\centering
\footnotesize
\setlength{\tabcolsep}{2.5pt}
\renewcommand{\arraystretch}{0.92}
\begin{tabular*}{\textwidth}{@{\extracolsep{\fill}} l r r r r r r r r r @{}}
  \toprule
 & \multicolumn{3}{c}{Independence} & \multicolumn{3}{c}{Exchangeable} 
 & \multicolumn{3}{c}{Unstructured}\\
  \cmidrule(lr){2-4}\cmidrule(lr){5-7}\cmidrule(lr){8-10}
Parameter & Estimate & SE & $p$-value & Estimate & SE & $p$-value 
          & Estimate & SE & $p$-value\\
  \midrule
\addlinespace[0.3ex]
& \multicolumn{9}{c}{PGEE with $\delta = 0.5$}\\
\addlinespace[0.3ex]
$\beta_0$ & $-$1.9221 & 1.3081 & 0.1417 & $-$2.2082 & 1.2681 & 0.0816 & $-$2.0181 & 1.1979 & 0.0920 \\
$\beta_1$ &    4.0218 & 1.0487 & 0.0001 &    3.7609 & 1.0031 & 0.0002 &    3.7346 & 1.0065 & 0.0002 \\
$\beta_2$ &    0.4511 & 0.7081 & 0.5241 &    0.4039 & 0.7236 & 0.5767 &    0.4638 & 0.7128 & 0.5153 \\
$\beta_3$ &    0.5791 & 0.3728 & 0.1203 &    0.6636 & 0.3422 & 0.0525 &    0.5804 & 0.3254 & 0.0745 \\
$\beta_4$ &    0.0330 & 0.0273 & 0.2267 &    0.0423 & 0.0263 & 0.1073 &    0.0329 & 0.0245 & 0.1801 \\
$\beta_5$ &    1.0700 & 0.5323 & 0.0444 &    1.0314 & 0.5334 & 0.0531 &    1.1019 & 0.5252 & 0.0359 \\
$\beta_6$ & $-$2.9284 & 1.2071 & 0.0153 & $-$2.6086 & 1.1691 & 0.0257 & $-$2.5928 & 1.1690 & 0.0266 \\
$\beta_7$ & $-$0.0171 & 0.0089 & 0.0547 & $-$0.0197 & 0.0082 & 0.0160 & $-$0.0164 & 0.0076 & 0.0319 \\
\addlinespace[0.3ex]
& \multicolumn{9}{c}{OPGEE with $\delta = 0.5$}\\
\addlinespace[0.3ex]
$\beta_0$ &  &  &  & $-$2.1744 & 1.2641 & 0.0854 & $-$2.0151 & 1.1971 & 0.0923 \\
$\beta_1$ &  &  &  &    3.7749 & 1.0056 & 0.0002 &    3.7475 & 1.0088 & 0.0002 \\
$\beta_2$ &  &  &  &    0.3916 & 0.7227 & 0.5879 &    0.4638 & 0.7123 & 0.5150 \\
$\beta_3$ &  &  &  &    0.6560 & 0.3411 & 0.0545 &    0.5789 & 0.3249 & 0.0748 \\
$\beta_4$ &  &  &  &    0.0413 & 0.0261 & 0.1146 &    0.0329 & 0.0245 & 0.1794 \\
$\beta_5$ &  &  &  &    1.0261 & 0.5330 & 0.0542 &    1.0967 & 0.5247 & 0.0366 \\
$\beta_6$ &  &  &  & $-$2.6216 & 1.1705 & 0.0251 & $-$2.6095 & 1.1706 & 0.0258 \\
$\beta_7$ &  &  &  & $-$0.0194 & 0.0081 & 0.0173 & $-$0.0163 & 0.0076 & 0.0320 \\
\addlinespace[0.3ex]
& \multicolumn{9}{c}{HPGEE with $\delta = 0.5$}\\
\addlinespace[0.3ex]
$\beta_0$ & $-$2.0293 & 1.3292 & 0.1268 & $-$2.2888 & 1.2985 & 0.0780 & $-$2.1404 & 1.2309 & 0.0821 \\
$\beta_1$ &    5.0865 & 1.4407 & 0.0004 &    5.0711 & 1.3262 & 0.0001 &    5.0918 & 1.3421 & 0.0001 \\
$\beta_2$ &    0.4710 & 0.7141 & 0.5096 &    0.4342 & 0.7316 & 0.5528 &    0.5056 & 0.7217 & 0.4836 \\
$\beta_3$ &    0.6125 & 0.3813 & 0.1082 &    0.6844 & 0.3535 & 0.0529 &    0.6154 & 0.3379 & 0.0686 \\
$\beta_4$ &    0.0352 & 0.0278 & 0.2050 &    0.0426 & 0.0269 & 0.1134 &    0.0345 & 0.0252 & 0.1716 \\
$\beta_5$ &    1.1105 & 0.5434 & 0.0410 &    1.1076 & 0.5512 & 0.0445 &    1.1765 & 0.5428 & 0.0302 \\
$\beta_6$ & $-$3.9539 & 1.5632 & 0.0114 & $-$3.8429 & 1.4629 & 0.0086 & $-$3.8801 & 1.4744 & 0.0085 \\
$\beta_7$ & $-$0.0181 & 0.0092 & 0.0477 & $-$0.0202 & 0.0084 & 0.0169 & $-$0.0173 & 0.0079 & 0.0290 \\
\addlinespace[0.3ex]
& \multicolumn{9}{c}{PGEE with $\delta = 0.1$}\\
\addlinespace[0.3ex]
$\beta_0$ & $-$2.0102 & 1.3266 & 0.1297 & $-$2.3325 & 1.3012 & 0.0730 & $-$2.1540 & 1.2307 & 0.0801 \\
$\beta_1$ &    5.6580 & 1.8012 & 0.0017 &    5.6142 & 1.5763 & 0.0004 &    5.6321 & 1.5892 & 0.0004 \\
$\beta_2$ &    0.4671 & 0.7130 & 0.5123 &    0.4315 & 0.7306 & 0.5548 &    0.4975 & 0.7207 & 0.4900 \\
$\beta_3$ &    0.6063 & 0.3802 & 0.1108 &    0.6936 & 0.3530 & 0.0494 &    0.6158 & 0.3366 & 0.0673 \\
$\beta_4$ &    0.0348 & 0.0277 & 0.2089 &    0.0443 & 0.0270 & 0.1010 &    0.0348 & 0.0252 & 0.1671 \\
$\beta_5$ &    1.1033 & 0.5431 & 0.0422 &    1.0966 & 0.5519 & 0.0469 &    1.1716 & 0.5436 & 0.0311 \\
$\beta_6$ & $-$4.5327 & 1.9000 & 0.0170 & $-$4.3886 & 1.6927 & 0.0095 & $-$4.4198 & 1.7017 & 0.0094 \\
$\beta_7$ & $-$0.0180 & 0.0091 & 0.0494 & $-$0.0205 & 0.0085 & 0.0152 & $-$0.0173 & 0.0079 & 0.0288 \\
\addlinespace[0.3ex]
& \multicolumn{9}{c}{OPGEE with $\delta = 0.1$}\\
\addlinespace[0.3ex]
$\beta_0$ &  &  &  & $-$2.2867 & 1.2963 & 0.0777 & $-$2.1351 & 1.2285 & 0.0822 \\
$\beta_1$ &  &  &  &    5.6135 & 1.5761 & 0.0004 &    5.6258 & 1.5875 & 0.0004 \\
$\beta_2$ &  &  &  &    0.4241 & 0.7299 & 0.5612 &    0.4961 & 0.7200 & 0.4908 \\
$\beta_3$ &  &  &  &    0.6839 & 0.3519 & 0.0520 &    0.6123 & 0.3359 & 0.0683 \\
$\beta_4$ &  &  &  &    0.0429 & 0.0269 & 0.1100 &    0.0346 & 0.0252 & 0.1694 \\
$\beta_5$ &  &  &  &    1.0927 & 0.5511 & 0.0474 &    1.1642 & 0.5425 & 0.0319 \\
$\beta_6$ &  &  &  & $-$4.3936 & 1.6919 & 0.0094 & $-$4.4228 & 1.6997 & 0.0093 \\
$\beta_7$ &  &  &  & $-$0.0202 & 0.0084 & 0.0166 & $-$0.0172 & 0.0079 & 0.0292 \\
\addlinespace[0.3ex]
& \multicolumn{9}{c}{HPGEE with $\delta = 0.1$}\\
\addlinespace[0.3ex]
$\beta_0$ & $-$2.0328 & 1.3311 & 0.1267 & $-$2.3107 & 1.3040 & 0.0764 & $-$2.1614 & 1.2359 & 0.0803 \\
$\beta_1$ &    6.6714 & 2.8166 & 0.0179 &    6.6558 & 2.3466 & 0.0046 &    6.6794 & 2.3599 & 0.0046 \\
$\beta_2$ &    0.4713 & 0.7142 & 0.5094 &    0.4329 & 0.7319 & 0.5542 &    0.5048 & 0.7221 & 0.4845 \\
$\beta_3$ &    0.6134 & 0.3821 & 0.1084 &    0.6900 & 0.3545 & 0.0516 &    0.6201 & 0.3387 & 0.0671 \\
$\beta_4$ &    0.0353 & 0.0278 & 0.2045 &    0.0432 & 0.0270 & 0.1098 &    0.0349 & 0.0253 & 0.1678 \\
$\beta_5$ &    1.1117 & 0.5455 & 0.0416 &    1.1095 & 0.5554 & 0.0458 &    1.1806 & 0.5467 & 0.0308 \\
$\beta_6$ & $-$5.5379 & 2.8812 & 0.0546 & $-$5.4203 & 2.4270 & 0.0255 & $-$5.4611 & 2.4376 & 0.0251 \\
$\beta_7$ & $-$0.0182 & 0.0092 & 0.0481 & $-$0.0203 & 0.0085 & 0.0164 & $-$0.0174 & 0.0080 & 0.0285 \\
\bottomrule
\end{tabular*}
\end{table}

\hypertarget{sec:corrcoef}{%
\subsection{Inadmissibility of Correlation-Coefficient Parameterizations During Variable Selection}\label{sec:corrcoef}}

To document the inadmissibility of working correlation estimates produced by correlation-coefficient-based PGEE during backward elimination, we repeated the variable selection procedure described in Section 4 of the main paper using a correlation-coefficient parameterization of the working association structure in place of the odds-ratio parameterization. The procedure was run under the exchangeable and unstructured working structures, with the dispersion parameter \(\phi\) both estimated from the data and fixed at \(\phi=1\), as these are the nonindependence working structures for which inadmissibility was observed. For completeness, the procedure was also run under the independence working structure with \(\phi\) estimated, which differs from the odds-ratio independence working structure because \(\phi\) enters the working variance and must be estimated separately, whereas under the odds-ratio parameterization \(\phi\) is fixed at its canonical value of one by default.

Under all nonindependence working structures and dispersion specifications, the initial model and/or candidate models evaluated during backward elimination produced working correlation estimates that were inadmissible for binary responses. Specifically, some estimated correlations lay outside the interval \((-1,1)\) for the working unstructured correlation matrix, while others were numerically within \((-1,1)\) but violated the Fréchet bounds implied by the fitted marginal probabilities \citep{Prentice_Correlated_1988}, rendering the working correlation matrix invalid.

The \textsf{R} code in script \texttt{02\_corrcoef\_inadmissibility.R} of the accompanying zip folder reproduces these findings. Note that correlation-coefficient-based working structures are fitted using the \texttt{geewa} function, which provides the generic interface of the \texttt{geer} package, rather than the \texttt{geewa\_binary} function, which is designed specifically for correlated binary responses under an odds-ratio parameterization.

\hypertarget{sec:example-cv}{%
\subsection{Cross-Validation of the Penalty Strength}\label{sec:example-cv}}

To assess the choice of \(\delta\) for the respiratory illness trial, we computed cross-validated predictive scores and QIC \citep{Pan2001} for the selected model of Section 4 of the main paper, over the grid \(\delta\in\{0.01, 0.025, 0.05, 0.1, 0.2, \ldots, 1\}\) and under each working odds-ratio structure. In leave-one-cluster-out (LOCO) cross-validation, PGEE was fitted to 54 patients, and the four responses of the held-out patient were scored by the log score and by the Brier score \(\sum_{j}(y_{ij}-\widehat{\pi}_{ij})^2\). Both scores were summed over visits and averaged over the 55 patients, so smaller values are better. Repeated stratified five-fold cross-validation used 50 random partitions of the patients, stratified by treatment and gender (seed 20261105). QIC and QICu were computed on the full data using the robust covariance estimator, as implemented in the function \texttt{geecriteria()} of \texttt{geer}. All 2,145 LOCO fits and 9,750 five-fold fits converged with all estimates below 100 in absolute value.

\ref{supp:cv_delta} and Figure \ref{fig:cv_delta}(a) show that all cross-validated criteria are nearly constant in \(\delta\). Across the grid, the LOCO log score per patient varies by at most \(0.010\), and every value of \(\delta\) lies within half a paired standard error of the minimum. The five-fold curves have the same shape as the LOCO curves, but the value of \(\delta\) minimizing the five-fold log score varies between partitions: \(\delta=1\) is selected in 36 to 38 of the 50 partitions, depending on the working structure, and \(\delta=0.01\) in 10 partitions under independence. QIC and QICu vary by less than 2 units across \(\delta\). QICu, and QIC under independence, increase with \(\delta\), favoring the weakest penalty. QIC under the exchangeable and unstructured structures is smallest at \(\delta=0.1\). Values of \(\delta\) above 1, which lie outside the range covered by the existence results, changed the LOCO log score by at most \(0.0004\) for \(\delta\in\{1.25,1.5\}\), and increased it at \(\delta=2\).

Figure \ref{fig:cv_delta}(b) explains this flatness. It decomposes the mean LOCO log score into the contributions of the four female patients on active treatment, whose responses were all good, and of the remaining 51 patients. The first contribution increases with \(\delta\), because weaker penalization predicts the all-good responses of a held-out patient from this group better. The second decreases with \(\delta\). Relative to \(\delta=1\), the two contributions at \(\delta=0.01\) are \(-0.017\) and \(+0.018\) under independence, \(-0.020\) and \(+0.030\) under the exchangeable structure, and \(-0.023\) and \(+0.033\) under the unstructured structure. They nearly cancel, so the observed data carry little information about \(\delta\). This supports retaining the default \(\delta=1/2\), as in the main paper.

\begin{table}[p]
\caption{Cross-validated predictive scores and QIC across the penalty strength $\delta$ for PGEE in the respiratory illness trial (selected model, probit link). LOCO: leave-one-cluster-out; 5-fold: mean over 50 repeated stratified five-fold partitions. Scores are per patient (summed over the four visits); smaller values are better for all columns.\label{supp:cv_delta}}
\centering
\small
\setlength{\tabcolsep}{4pt}
\begin{tabular*}{\textwidth}{@{\extracolsep{\fill}} c r r r r r @{}}
\toprule
$\delta$ & LOCO log score & LOCO Brier score & 5-fold log score & QIC & QICu \\
\midrule
\multicolumn{6}{c}{\textit{Independence}}\\
\addlinespace[1pt]
0.01 & 2.3741 & 0.8344 & 2.3923 & 255.78 & 246.84 \\
0.05 & 2.3741 & 0.8341 & 2.3919 & 255.82 & 246.90 \\
0.1  & 2.3741 & 0.8337 & 2.3914 & 255.87 & 246.98 \\
0.5  & 2.3736 & 0.8312 & 2.3876 & 256.34 & 247.59 \\
1    & 2.3731 & 0.8289 & 2.3841 & 257.01 & 248.36 \\
\addlinespace[2pt]
\multicolumn{6}{c}{\textit{Exchangeable}}\\
\addlinespace[1pt]
0.01 & 2.3769 & 0.8341 & 2.3964 & 255.88 & 246.97 \\
0.05 & 2.3763 & 0.8336 & 2.3951 & 255.86 & 247.02 \\
0.1  & 2.3755 & 0.8330 & 2.3935 & 255.85 & 247.09 \\
0.5  & 2.3700 & 0.8285 & 2.3834 & 256.17 & 247.73 \\
1    & 2.3670 & 0.8248 & 2.3767 & 257.03 & 248.74 \\
\addlinespace[2pt]
\multicolumn{6}{c}{\textit{Unstructured}}\\
\addlinespace[1pt]
0.01 & 2.3706 & 0.8323 & 2.3863 & 254.77 & 247.11 \\
0.05 & 2.3699 & 0.8318 & 2.3849 & 254.75 & 247.15 \\
0.1  & 2.3690 & 0.8311 & 2.3832 & 254.73 & 247.21 \\
0.5  & 2.3634 & 0.8263 & 2.3733 & 254.99 & 247.78 \\
1    & 2.3606 & 0.8224 & 2.3670 & 255.85 & 248.78 \\
\bottomrule
\end{tabular*}
\end{table}

\begin{table}[p]
\caption{Estimates of $\beta_1$ (bias-corrected standard errors in parentheses) in the bivariate design of \ref{sec:twobytwo} with $N=20$, for ordinary GEE and PGEE with $\delta\in\{0.1,0.5,1\}$. Ind.: independence; Exch.: exchangeable odds-ratio structure, with the estimated marginalized odds ratio in parentheses. ``fails'': nonconvergence or an estimate exceeding 100 in absolute value.\label{supp:twobytwo}}
\centering
\small
\setlength{\tabcolsep}{4pt}
\begin{tabular*}{\textwidth}{@{\extracolsep{\fill}} l l r r r r @{}}
\toprule
Link & Structure & GEE & PGEE, $\delta=0.1$ & PGEE, $\delta=0.5$ & PGEE, $\delta=1$ \\
\midrule
\multicolumn{6}{l}{\textit{A. no zero cell}}\\
logit & Ind. & 0.405 (0.554) & 0.401 (0.553) & 0.386 (0.553) & 0.368 (0.552) \\
logit & Exch. (4.91) & 0.405 (0.538) & 0.402 (0.538) & 0.388 (0.537) & 0.373 (0.536) \\
probit & Ind. & 0.253 (0.345) & 0.252 (0.345) & 0.246 (0.345) & 0.239 (0.345) \\
probit & Exch. (4.91) & 0.253 (0.336) & 0.252 (0.336) & 0.246 (0.335) & 0.239 (0.335) \\
\addlinespace[2pt]
\multicolumn{6}{l}{\textit{B. one zero cell (quasi-complete)}}\\
logit & Ind. & fails & 4.465 (1.210) & 2.911 (0.752) & 2.282 (0.667) \\
logit & Exch. (2.23) & fails & 4.467 (1.204) & 2.920 (0.739) & 2.298 (0.650) \\
probit & Ind. & fails & 2.134 (0.518) & 1.594 (0.401) & 1.346 (0.378) \\
probit & Exch. (2.23) & fails & 2.136 (0.515) & 1.600 (0.396) & 1.356 (0.371) \\
\addlinespace[2pt]
\multicolumn{6}{l}{\textit{C. two zero cells (complete)}}\\
logit & Ind. & fails & 10.607 (1.573) & 7.427 (0.840) & 6.089 (0.702) \\
logit & Exch. (0.02) & fails & 10.267 (1.561) & 6.546 (0.829) & 5.041 (0.709) \\
probit & Ind. & fails & 5.311 (0.585) & 4.204 (0.370) & 3.677 (0.330) \\
probit & Exch. (0.02) & fails & 5.217 (0.591) & 3.882 (0.383) & 3.231 (0.352) \\
\addlinespace[2pt]
\multicolumn{6}{l}{\textit{D. no successes}}\\
logit & Ind. & fails & 0.000 (1.502) & 0.000 (0.691) & 0.000 (0.506) \\
logit & Exch. (41.00) & fails & 0.000 (1.263) & 0.000 (0.420) & 0.000 (0.269) \\
probit & Ind. & fails & 0.000 (0.564) & 0.000 (0.318) & 0.000 (0.256) \\
probit & Exch. (41.00) & fails & 0.000 (0.496) & 0.000 (0.217) & 0.000 (0.154) \\
\bottomrule
\end{tabular*}
\end{table}

\begin{table}[p]
\caption{Empirical coverage (EC, \%) of nominal 95\% Wald intervals and ratio of the mean estimated standard error to the empirical standard error (RE) for the model-based, sandwich, and bias-corrected covariance estimators (\ref{sec:sandwich}), computed from the simulations of Section 3 of the main paper (10,000 datasets per configuration, probit link). Ind., Exch., Unstr.: independence, exchangeable, and unstructured odds-ratio working structures. PGEE uses $\delta=0.5$.\label{supp:sandwich}}
\centering
\scriptsize
\setlength{\tabcolsep}{3pt}
\renewcommand{\arraystretch}{0.92}
\begin{tabular*}{\textwidth}{@{\extracolsep{\fill}} l l r rrr rrr @{}}
\toprule
 & & & \multicolumn{3}{c}{EC} & \multicolumn{3}{c}{RE}\\
\cmidrule(lr){4-6}\cmidrule(lr){7-9}
Method & Structure & $N$ & Model-based & Sandwich & Bias-corrected & Model-based & Sandwich & Bias-corrected\\
\midrule
\multicolumn{9}{l}{\textit{Quasi-complete A, $\beta_2$}}\\
PGEE & Ind. & 30 & 85.23 & 78.37 & 82.51 & 0.46 & 0.42 & 0.47 \\
PGEE & Ind. & 100 & 84.17 & 92.09 & 92.53 & 0.67 & 0.87 & 0.88 \\
PGEE & Ind. & 500 & 83.39 & 94.21 & 94.33 & 0.70 & 0.97 & 0.98 \\
PGEE & Exch. & 30 & 96.93 & 72.85 & 83.39 & 1.26 & 0.61 & 0.76 \\
PGEE & Exch. & 100 & 98.49 & 89.28 & 90.81 & 1.27 & 0.83 & 0.87 \\
PGEE & Exch. & 500 & 98.74 & 93.95 & 94.16 & 1.30 & 0.97 & 0.97 \\
PGEE & Unstr. & 30 & 96.20 & 71.17 & 80.96 & 1.25 & 0.62 & 0.77 \\
PGEE & Unstr. & 100 & 98.22 & 89.07 & 90.53 & 1.27 & 0.84 & 0.88 \\
PGEE & Unstr. & 500 & 98.66 & 93.81 & 93.99 & 1.30 & 0.97 & 0.97 \\
\addlinespace[2pt]
\multicolumn{9}{l}{\textit{Quasi-complete B, $\beta_2$}}\\
PGEE & Ind. & 30 & 92.91 & 87.87 & 90.88 & 0.74 & 0.68 & 0.74 \\
PGEE & Ind. & 100 & 91.31 & 93.34 & 93.93 & 0.85 & 0.92 & 0.94 \\
PGEE & Ind. & 500 & 91.29 & 94.54 & 94.63 & 0.86 & 0.99 & 0.99 \\
PGEE & Exch. & 30 & 93.34 & 87.55 & 90.55 & 0.76 & 0.68 & 0.74 \\
PGEE & Exch. & 100 & 91.91 & 93.35 & 93.95 & 0.86 & 0.92 & 0.94 \\
PGEE & Exch. & 500 & 91.98 & 94.59 & 94.64 & 0.88 & 0.99 & 0.99 \\
PGEE & Unstr. & 30 & 93.02 & 86.92 & 90.08 & 0.75 & 0.67 & 0.74 \\
PGEE & Unstr. & 100 & 91.95 & 93.33 & 93.96 & 0.86 & 0.92 & 0.94 \\
PGEE & Unstr. & 500 & 92.00 & 94.57 & 94.64 & 0.88 & 0.99 & 0.99 \\
\addlinespace[2pt]
\multicolumn{9}{l}{\textit{Regular, $\beta_1$}}\\
GEE & Ind. & 30 & 79.68 & 92.72 & 94.69 & 0.63 & 0.92 & 0.99 \\
GEE & Ind. & 100 & 79.98 & 94.31 & 94.77 & 0.65 & 0.97 & 0.99 \\
GEE & Ind. & 500 & 80.25 & 94.52 & 94.60 & 0.65 & 0.99 & 0.99 \\
GEE & Exch. & 30 & 94.81 & 92.74 & 94.80 & 0.96 & 0.93 & 1.00 \\
GEE & Exch. & 100 & 95.29 & 94.22 & 94.84 & 1.00 & 0.97 & 1.00 \\
GEE & Exch. & 500 & 95.05 & 94.43 & 94.61 & 1.01 & 0.99 & 0.99 \\
GEE & Unstr. & 30 & 94.78 & 92.74 & 94.82 & 0.96 & 0.93 & 1.00 \\
GEE & Unstr. & 100 & 94.88 & 93.80 & 94.38 & 1.00 & 0.97 & 1.00 \\
GEE & Unstr. & 500 & 95.00 & 94.49 & 94.62 & 1.00 & 0.99 & 0.99 \\
PGEE & Ind. & 30 & 80.53 & 93.25 & 94.94 & 0.65 & 0.94 & 1.01 \\
PGEE & Ind. & 100 & 80.28 & 94.43 & 94.93 & 0.65 & 0.98 & 1.00 \\
PGEE & Ind. & 500 & 80.28 & 94.55 & 94.64 & 0.65 & 0.99 & 1.00 \\
PGEE & Exch. & 30 & 95.54 & 93.53 & 95.22 & 1.00 & 0.96 & 1.04 \\
PGEE & Exch. & 100 & 95.52 & 94.42 & 95.02 & 1.01 & 0.98 & 1.01 \\
PGEE & Exch. & 500 & 95.13 & 94.57 & 94.64 & 1.01 & 0.99 & 1.00 \\
PGEE & Unstr. & 30 & 95.42 & 93.53 & 95.51 & 1.00 & 0.96 & 1.04 \\
PGEE & Unstr. & 100 & 94.98 & 94.01 & 94.65 & 1.01 & 0.98 & 1.01 \\
PGEE & Unstr. & 500 & 95.12 & 94.59 & 94.65 & 1.01 & 0.99 & 0.99 \\
\addlinespace[2pt]
\multicolumn{9}{l}{\textit{Regular, $\beta_2$}}\\
GEE & Ind. & 30 & 83.84 & 91.25 & 93.86 & 0.71 & 0.90 & 0.98 \\
GEE & Ind. & 100 & 82.25 & 94.28 & 94.95 & 0.69 & 0.97 & 1.00 \\
GEE & Ind. & 500 & 83.03 & 94.70 & 94.75 & 0.70 & 1.00 & 1.00 \\
GEE & Exch. & 30 & 94.47 & 92.90 & 95.10 & 0.95 & 0.94 & 1.01 \\
GEE & Exch. & 100 & 94.74 & 94.18 & 94.66 & 0.98 & 0.97 & 1.00 \\
GEE & Exch. & 500 & 94.99 & 95.03 & 95.14 & 0.99 & 1.00 & 1.00 \\
GEE & Unstr. & 30 & 94.39 & 90.47 & 92.98 & 0.98 & 0.89 & 0.97 \\
GEE & Unstr. & 100 & 94.83 & 93.41 & 94.08 & 1.00 & 0.96 & 0.99 \\
GEE & Unstr. & 500 & 95.36 & 94.82 & 94.99 & 1.01 & 0.99 & 1.00 \\
PGEE & Ind. & 30 & 84.87 & 91.80 & 94.38 & 0.72 & 0.92 & 1.00 \\
PGEE & Ind. & 100 & 82.53 & 94.44 & 95.11 & 0.69 & 0.98 & 1.00 \\
PGEE & Ind. & 500 & 83.11 & 94.69 & 94.81 & 0.70 & 1.00 & 1.00 \\
PGEE & Exch. & 30 & 95.00 & 93.29 & 95.49 & 0.97 & 0.96 & 1.03 \\
PGEE & Exch. & 100 & 94.83 & 94.39 & 94.83 & 0.99 & 0.98 & 1.00 \\
PGEE & Exch. & 500 & 94.99 & 95.04 & 95.19 & 0.99 & 1.00 & 1.00 \\
PGEE & Unstr. & 30 & 94.44 & 90.48 & 93.11 & 1.00 & 0.91 & 1.00 \\
PGEE & Unstr. & 100 & 94.88 & 93.55 & 94.12 & 1.01 & 0.97 & 0.99 \\
PGEE & Unstr. & 500 & 95.39 & 94.85 & 94.93 & 1.01 & 0.99 & 1.00 \\
\bottomrule
\end{tabular*}
\end{table}

\begin{table}[p]
\caption{PGEE with $\delta=0.5$ in the bivariate design of \ref{sec:twobytwo} with $N=20$, under the exchangeable odds-ratio structure and under an exchangeable correlation-coefficient structure with the dispersion $\phi$ estimated or fixed at 1. $\widehat{\alpha}$: estimated marginalized odds ratio; $\widehat{\rho}$: estimated correlation, with $\widehat{\phi}$ in parentheses; range: correlations compatible with the fitted marginal probabilities of the fit with $\phi=1$. ``fails'': the moment estimate of the correlation fell outside $(-1,1)$.\label{supp:twobytwo_cc}}
\centering
\small
\setlength{\tabcolsep}{4pt}
\begin{tabular*}{\textwidth}{@{\extracolsep{\fill}} l l r r r r @{}}
\toprule
 & & Odds ratio & \multicolumn{3}{c}{Correlation coefficient}\\
\cmidrule(lr){3-3}\cmidrule(lr){4-6}
Configuration & Link & $\widehat{\alpha}$ & $\widehat{\rho}$ ($\widehat{\phi}$), $\phi$ estimated & $\widehat{\rho}$, $\phi=1$ & Range, $\phi=1$\\
\midrule
A. no zero cell & logit & 4.91 & 0.431 (1.051) & 0.453 & $[-0.828, 0.821]$\\
 & probit & 4.91 & 0.431 (1.051) & 0.453 & $[-0.823, 0.819]$\\
B. one zero cell & logit & 2.23 & 0.003 (0.528) & 0.004 & $[-0.105, 0.233]$\\
 & probit & 2.23 & 0.001 (0.529) & 0.002 & $[-0.089, 0.203]$\\
C. two zero cells & logit & 0.02 & fails & $-0.028$ & $[-1.000, 0.025]$\\
 & probit & 0.02 & fails & $-0.021$ & $[-1.000, 0.018]$\\
D. no successes & logit & 41.00 & fails & 0.028 & $[-0.025, 1.000]$\\
 & probit & 41.00 & fails & 0.021 & $[-0.018, 1.000]$\\
\bottomrule
\end{tabular*}
\end{table}

\hypertarget{sec:rcode}{%
\section{R Code and Computing Environment}\label{sec:rcode}}

This Appendix provides \textsf{R} code to reproduce the respiratory clinical trial analysis in Section 4 of the main paper. The code below (i) subsets the trial data, (ii) checks separation under a probit GLM, (iii) fits GEE and PGEE models, (iv) performs backward elimination, (v) reports summaries of the fitted models across working structures and penalty values, and (vi) provides one-step approximations.

\hypertarget{sec:packages}{%
\subsection{Packages and Data}\label{sec:packages}}

\begin{Shaded}
\begin{Highlighting}[]
\FunctionTok{library}\NormalTok{(}\StringTok{"geer"}\NormalTok{)}
\FunctionTok{library}\NormalTok{(}\StringTok{"detectseparation"}\NormalTok{)}
\FunctionTok{data}\NormalTok{(}\StringTok{"respiratory"}\NormalTok{, }\AttributeTok{package =} \StringTok{"geer"}\NormalTok{)}
\NormalTok{respiratory2 }\OtherTok{\textless{}{-}} \FunctionTok{subset}\NormalTok{(respiratory, center }\SpecialCharTok{==} \StringTok{"C2"}\NormalTok{)}
\end{Highlighting}
\end{Shaded}

\hypertarget{sec:baseline}{%
\subsection{Initial GEE Fit under Working Independence}\label{sec:baseline}}

\begin{Shaded}
\begin{Highlighting}[]
\NormalTok{fitted\_model\_gee }\OtherTok{\textless{}{-}}
  \FunctionTok{geewa\_binary}\NormalTok{(}
    \AttributeTok{formula =}\NormalTok{ status }\SpecialCharTok{\textasciitilde{}}\NormalTok{ (}\FunctionTok{I}\NormalTok{(treatment }\SpecialCharTok{==} \StringTok{"active"}\NormalTok{) }\SpecialCharTok{+}\NormalTok{ gender }\SpecialCharTok{+}\NormalTok{ visit }\SpecialCharTok{+}\NormalTok{ age }\SpecialCharTok{+}\NormalTok{ baseline)}\SpecialCharTok{\^{}}\DecValTok{2}\NormalTok{,}
    \AttributeTok{id =}\NormalTok{ id,}
    \AttributeTok{link =} \StringTok{"probit"}\NormalTok{,}
    \AttributeTok{method =} \StringTok{"gee"}\NormalTok{,}
    \AttributeTok{data =}\NormalTok{ respiratory2,}
    \AttributeTok{orstr =} \StringTok{"independence"}\NormalTok{)}
\FunctionTok{coef}\NormalTok{(fitted\_model\_gee) }\SpecialCharTok{|\textgreater{}} \FunctionTok{abs}\NormalTok{() }\SpecialCharTok{|\textgreater{}} \FunctionTok{range}\NormalTok{()}
\end{Highlighting}
\end{Shaded}

\hypertarget{separation-diagnostics}{%
\subsection{Separation Diagnostics}\label{separation-diagnostics}}

\begin{Shaded}
\begin{Highlighting}[]
\NormalTok{fitted\_model\_glm }\OtherTok{\textless{}{-}}
  \FunctionTok{glm}\NormalTok{(}
    \AttributeTok{formula =}\NormalTok{ status }\SpecialCharTok{\textasciitilde{}}\NormalTok{ (}\FunctionTok{I}\NormalTok{(treatment }\SpecialCharTok{==} \StringTok{"active"}\NormalTok{) }\SpecialCharTok{+}\NormalTok{ gender }\SpecialCharTok{+}\NormalTok{ visit }\SpecialCharTok{+}\NormalTok{ age }\SpecialCharTok{+}\NormalTok{ baseline)}\SpecialCharTok{\^{}}\DecValTok{2}\NormalTok{,}
    \AttributeTok{data =}\NormalTok{ respiratory2,}
    \AttributeTok{family =} \FunctionTok{binomial}\NormalTok{(}\AttributeTok{link =} \StringTok{"probit"}\NormalTok{))}
\FunctionTok{update}\NormalTok{(fitted\_model\_glm,}
       \AttributeTok{method =} \StringTok{"detect\_separation"}\NormalTok{)}
\end{Highlighting}
\end{Shaded}

\hypertarget{pgee-fits-and-variable-selection}{%
\subsection{PGEE Fits and Variable Selection}\label{pgee-fits-and-variable-selection}}

\begin{Shaded}
\begin{Highlighting}[]
\DocumentationTok{\#\# backward elimination with delta = 0.5}
\NormalTok{fitted\_model\_pgee }\OtherTok{\textless{}{-}}
  \FunctionTok{update}\NormalTok{(fitted\_model\_gee,}
         \AttributeTok{method =} \StringTok{"pgee{-}jeffreys"}\NormalTok{,}
         \AttributeTok{control =} \FunctionTok{list}\NormalTok{(}\AttributeTok{jeffreys\_power =} \FloatTok{0.5}\NormalTok{))}
\NormalTok{fitted\_model\_pgee\_inde05 }\OtherTok{\textless{}{-}}
\NormalTok{  fitted\_model\_pgee }\SpecialCharTok{|\textgreater{}}
  \FunctionTok{step\_p}\NormalTok{(}
    \AttributeTok{scope =} \FunctionTok{list}\NormalTok{(}\AttributeTok{lower =}\NormalTok{  status }\SpecialCharTok{\textasciitilde{}} \DecValTok{1}\NormalTok{, }\AttributeTok{upper =}\NormalTok{ status }\SpecialCharTok{\textasciitilde{}}\NormalTok{ .}\SpecialCharTok{\^{}}\DecValTok{2}\NormalTok{),}
    \AttributeTok{test =} \StringTok{"wald"}\NormalTok{,}
    \AttributeTok{cov\_type =} \StringTok{"bias{-}corrected"}\NormalTok{,}
    \AttributeTok{p\_remove =} \FloatTok{0.1}\NormalTok{,}
    \AttributeTok{direction =} \StringTok{"backward"}\NormalTok{)}
\NormalTok{fitted\_model\_pgee\_inde05}
\DocumentationTok{\#\# backward elimination with delta = 0.1}
\NormalTok{fitted\_model\_pgee\_inde01 }\OtherTok{\textless{}{-}}
\NormalTok{  fitted\_model\_pgee }\SpecialCharTok{|\textgreater{}}
  \FunctionTok{update}\NormalTok{(}\AttributeTok{control =} \FunctionTok{list}\NormalTok{(}\AttributeTok{jeffreys\_power =} \FloatTok{0.1}\NormalTok{)) }\SpecialCharTok{|\textgreater{}}
  \FunctionTok{step\_p}\NormalTok{(}
    \AttributeTok{scope =} \FunctionTok{list}\NormalTok{(}\AttributeTok{lower =}\NormalTok{  status }\SpecialCharTok{\textasciitilde{}} \DecValTok{1}\NormalTok{, }\AttributeTok{upper =}\NormalTok{ status }\SpecialCharTok{\textasciitilde{}}\NormalTok{ .}\SpecialCharTok{\^{}}\DecValTok{2}\NormalTok{),}
    \AttributeTok{test =} \StringTok{"wald"}\NormalTok{,}
    \AttributeTok{cov\_type =} \StringTok{"bias{-}corrected"}\NormalTok{,}
    \AttributeTok{p\_remove =} \FloatTok{0.1}\NormalTok{,}
    \AttributeTok{direction =} \StringTok{"backward"}\NormalTok{)}
\end{Highlighting}
\end{Shaded}

\hypertarget{summaries-across-working-associations}{%
\subsection{Summaries across Working Associations}\label{summaries-across-working-associations}}

\begin{Shaded}
\begin{Highlighting}[]
\DocumentationTok{\#\# delta = 0.5}
\NormalTok{fitted\_model\_pgee\_inde05 }\SpecialCharTok{|\textgreater{}} \FunctionTok{summary}\NormalTok{(}\AttributeTok{cov\_type =} \StringTok{"bias{-}corrected"}\NormalTok{) }
\NormalTok{fitted\_model\_pgee\_inde05 }\SpecialCharTok{|\textgreater{}} \FunctionTok{update}\NormalTok{(}\AttributeTok{orstr =} \StringTok{"exchangeable"}\NormalTok{) }\SpecialCharTok{|\textgreater{}} \FunctionTok{summary}\NormalTok{(}\AttributeTok{cov\_type =} \StringTok{"bias{-}corrected"}\NormalTok{) }
\NormalTok{fitted\_model\_pgee\_inde05 }\SpecialCharTok{|\textgreater{}} \FunctionTok{update}\NormalTok{(}\AttributeTok{orstr =} \StringTok{"unstructured"}\NormalTok{) }\SpecialCharTok{|\textgreater{}} \FunctionTok{summary}\NormalTok{(}\AttributeTok{cov\_type =} \StringTok{"bias{-}corrected"}\NormalTok{) }
\DocumentationTok{\#\# delta = 0.1}
\NormalTok{fitted\_model\_pgee\_inde01 }\SpecialCharTok{|\textgreater{}} \FunctionTok{summary}\NormalTok{(}\AttributeTok{cov\_type =} \StringTok{"bias{-}corrected"}\NormalTok{) }
\NormalTok{fitted\_model\_pgee\_inde01 }\SpecialCharTok{|\textgreater{}} \FunctionTok{update}\NormalTok{(}\AttributeTok{orstr =} \StringTok{"exchangeable"}\NormalTok{) }\SpecialCharTok{|\textgreater{}} \FunctionTok{summary}\NormalTok{(}\AttributeTok{cov\_type =} \StringTok{"bias{-}corrected"}\NormalTok{) }
\NormalTok{fitted\_model\_pgee\_inde01 }\SpecialCharTok{|\textgreater{}} \FunctionTok{update}\NormalTok{(}\AttributeTok{orstr =} \StringTok{"unstructured"}\NormalTok{) }\SpecialCharTok{|\textgreater{}} \FunctionTok{summary}\NormalTok{(}\AttributeTok{cov\_type =} \StringTok{"bias{-}corrected"}\NormalTok{) }
\end{Highlighting}
\end{Shaded}

\hypertarget{sec:opgee}{%
\subsection{One-Step Approximations}\label{sec:opgee}}

\begin{Shaded}
\begin{Highlighting}[]
\DocumentationTok{\#\# OPGEE: delta = 0.5}
\NormalTok{fitted\_model\_pgee\_inde05 }\SpecialCharTok{|\textgreater{}}
  \FunctionTok{update}\NormalTok{(}\AttributeTok{orstr =} \StringTok{"exchangeable"}\NormalTok{, }\AttributeTok{method =} \StringTok{"opgee{-}jeffreys"}\NormalTok{) }\SpecialCharTok{|\textgreater{}} 
  \FunctionTok{summary}\NormalTok{(}\AttributeTok{cov\_type =} \StringTok{"bias{-}corrected"}\NormalTok{) }
\NormalTok{fitted\_model\_pgee\_inde05 }\SpecialCharTok{|\textgreater{}}   
  \FunctionTok{update}\NormalTok{(}\AttributeTok{orstr =} \StringTok{"unstructured"}\NormalTok{, }\AttributeTok{method =} \StringTok{"opgee{-}jeffreys"}\NormalTok{) }\SpecialCharTok{|\textgreater{}}  
  \FunctionTok{summary}\NormalTok{(}\AttributeTok{cov\_type =} \StringTok{"bias{-}corrected"}\NormalTok{) }
\DocumentationTok{\#\# HPGEE: delta = 0.5}
\NormalTok{fitted\_model\_pgee\_inde05 }\SpecialCharTok{|\textgreater{}} 
  \FunctionTok{update}\NormalTok{(}\AttributeTok{method =} \StringTok{"hpgee{-}jeffreys"}\NormalTok{) }\SpecialCharTok{|\textgreater{}} 
  \FunctionTok{summary}\NormalTok{(}\AttributeTok{cov\_type =} \StringTok{"bias{-}corrected"}\NormalTok{) }
\NormalTok{fitted\_model\_pgee\_inde05 }\SpecialCharTok{|\textgreater{}}
  \FunctionTok{update}\NormalTok{(}\AttributeTok{orstr =} \StringTok{"exchangeable"}\NormalTok{, }\AttributeTok{method =} \StringTok{"hpgee{-}jeffreys"}\NormalTok{) }\SpecialCharTok{|\textgreater{}} 
  \FunctionTok{summary}\NormalTok{(}\AttributeTok{cov\_type =} \StringTok{"bias{-}corrected"}\NormalTok{) }
\NormalTok{fitted\_model\_pgee\_inde05 }\SpecialCharTok{|\textgreater{}}   
  \FunctionTok{update}\NormalTok{(}\AttributeTok{orstr =} \StringTok{"unstructured"}\NormalTok{, }\AttributeTok{method =} \StringTok{"hpgee{-}jeffreys"}\NormalTok{) }\SpecialCharTok{|\textgreater{}} 
  \FunctionTok{summary}\NormalTok{(}\AttributeTok{cov\_type =} \StringTok{"bias{-}corrected"}\NormalTok{) }
\DocumentationTok{\#\# OPGEE: delta = 0.1}
\NormalTok{fitted\_model\_pgee\_inde01 }\SpecialCharTok{|\textgreater{}}
  \FunctionTok{update}\NormalTok{(}\AttributeTok{orstr =} \StringTok{"exchangeable"}\NormalTok{, }\AttributeTok{method =} \StringTok{"opgee{-}jeffreys"}\NormalTok{) }\SpecialCharTok{|\textgreater{}} 
  \FunctionTok{summary}\NormalTok{(}\AttributeTok{cov\_type =} \StringTok{"bias{-}corrected"}\NormalTok{) }
\NormalTok{fitted\_model\_pgee\_inde01 }\SpecialCharTok{|\textgreater{}}   
  \FunctionTok{update}\NormalTok{(}\AttributeTok{orstr =} \StringTok{"unstructured"}\NormalTok{, }\AttributeTok{method =} \StringTok{"opgee{-}jeffreys"}\NormalTok{) }\SpecialCharTok{|\textgreater{}} 
  \FunctionTok{summary}\NormalTok{(}\AttributeTok{cov\_type =} \StringTok{"bias{-}corrected"}\NormalTok{) }
\DocumentationTok{\#\# HPGEE: delta = 0.1}
\NormalTok{fitted\_model\_pgee\_inde01 }\SpecialCharTok{|\textgreater{}} 
  \FunctionTok{update}\NormalTok{(}\AttributeTok{method =} \StringTok{"hpgee{-}jeffreys"}\NormalTok{) }\SpecialCharTok{|\textgreater{}} 
  \FunctionTok{summary}\NormalTok{(}\AttributeTok{cov\_type =} \StringTok{"bias{-}corrected"}\NormalTok{) }
\NormalTok{fitted\_model\_pgee\_inde01 }\SpecialCharTok{|\textgreater{}}
  \FunctionTok{update}\NormalTok{(}\AttributeTok{orstr =} \StringTok{"exchangeable"}\NormalTok{, }\AttributeTok{method =} \StringTok{"hpgee{-}jeffreys"}\NormalTok{) }\SpecialCharTok{|\textgreater{}} 
  \FunctionTok{summary}\NormalTok{(}\AttributeTok{cov\_type =} \StringTok{"bias{-}corrected"}\NormalTok{) }
\NormalTok{fitted\_model\_pgee\_inde01 }\SpecialCharTok{|\textgreater{}}   
  \FunctionTok{update}\NormalTok{(}\AttributeTok{orstr =} \StringTok{"unstructured"}\NormalTok{, }\AttributeTok{method =} \StringTok{"hpgee{-}jeffreys"}\NormalTok{) }\SpecialCharTok{|\textgreater{}} 
  \FunctionTok{summary}\NormalTok{(}\AttributeTok{cov\_type =} \StringTok{"bias{-}corrected"}\NormalTok{) }
\end{Highlighting}
\end{Shaded}

\hypertarget{sec:cvcode}{%
\subsection{Cross-Validation of the Penalty Strength}\label{sec:cvcode}}

The following code computes the LOCO log score of \ref{sec:example-cv} for one working structure and one value of \(\delta\). The full analysis, including the Brier score, repeated five-fold cross-validation, QIC, and Figure \ref{fig:cv_delta}, is in script \texttt{03\_cv\_delta.R} of the accompanying zip folder.

\begin{Shaded}
\begin{Highlighting}[]
\NormalTok{respiratory2}\SpecialCharTok{$}\NormalTok{active }\OtherTok{\textless{}{-}} \FunctionTok{as.numeric}\NormalTok{(respiratory2}\SpecialCharTok{$}\NormalTok{treatment }\SpecialCharTok{==} \StringTok{"active"}\NormalTok{)}
\NormalTok{formula\_selected }\OtherTok{\textless{}{-}}\NormalTok{ status }\SpecialCharTok{\textasciitilde{}}\NormalTok{ active }\SpecialCharTok{+}\NormalTok{ gender }\SpecialCharTok{+}\NormalTok{ visit }\SpecialCharTok{+}\NormalTok{ age }\SpecialCharTok{+}\NormalTok{ baseline }\SpecialCharTok{+}
\NormalTok{  active}\SpecialCharTok{:}\NormalTok{gender }\SpecialCharTok{+}\NormalTok{ visit}\SpecialCharTok{:}\NormalTok{age}
\NormalTok{loco\_log\_score }\OtherTok{\textless{}{-}} \ControlFlowTok{function}\NormalTok{(delta, orstr) \{}
\NormalTok{  scores }\OtherTok{\textless{}{-}} \FunctionTok{sapply}\NormalTok{(}\FunctionTok{unique}\NormalTok{(respiratory2}\SpecialCharTok{$}\NormalTok{id), }\ControlFlowTok{function}\NormalTok{(i) \{}
\NormalTok{    fit }\OtherTok{\textless{}{-}} \FunctionTok{geewa\_binary}\NormalTok{(formula\_selected, }\AttributeTok{id =}\NormalTok{ id, }\AttributeTok{repeated =}\NormalTok{ visit,}
                        \AttributeTok{link =} \StringTok{"probit"}\NormalTok{, }\AttributeTok{data =} \FunctionTok{subset}\NormalTok{(respiratory2, id }\SpecialCharTok{!=}\NormalTok{ i),}
                        \AttributeTok{orstr =}\NormalTok{ orstr, }\AttributeTok{method =} \StringTok{"pgee{-}jeffreys"}\NormalTok{,}
                        \AttributeTok{jeffreys\_power =}\NormalTok{ delta)}
\NormalTok{    held\_out }\OtherTok{\textless{}{-}} \FunctionTok{subset}\NormalTok{(respiratory2, id }\SpecialCharTok{==}\NormalTok{ i)}
\NormalTok{    p }\OtherTok{\textless{}{-}} \FunctionTok{predict}\NormalTok{(fit, }\AttributeTok{newdata =}\NormalTok{ held\_out, }\AttributeTok{type =} \StringTok{"response"}\NormalTok{)}
    \SpecialCharTok{{-}}\FunctionTok{sum}\NormalTok{(held\_out}\SpecialCharTok{$}\NormalTok{status }\SpecialCharTok{*} \FunctionTok{log}\NormalTok{(p) }\SpecialCharTok{+}\NormalTok{ (}\DecValTok{1} \SpecialCharTok{{-}}\NormalTok{ held\_out}\SpecialCharTok{$}\NormalTok{status) }\SpecialCharTok{*} \FunctionTok{log}\NormalTok{(}\DecValTok{1} \SpecialCharTok{{-}}\NormalTok{ p))}
\NormalTok{  \})}
  \FunctionTok{mean}\NormalTok{(scores)}
\NormalTok{\}}
\FunctionTok{loco\_log\_score}\NormalTok{(}\AttributeTok{delta =} \FloatTok{0.5}\NormalTok{, }\AttributeTok{orstr =} \StringTok{"independence"}\NormalTok{)}
\end{Highlighting}
\end{Shaded}

\hypertarget{additional-reproduction-code}{%
\subsection{Additional Reproduction Code}\label{additional-reproduction-code}}

Additional \textsf{R} scripts reproducing the logit link sensitivity analysis, the correlation-coefficient instability documentation, and the cross-validation of the penalty strength are provided in the accompanying zip folder, as described in \ref{sec:example}.

\hypertarget{sec:session}{%
\subsection{Computing Environment}\label{sec:session}}

The computing environment and package versions are reported below to support reproducibility.

\begin{Shaded}
\begin{Highlighting}[]
\FunctionTok{sessionInfo}\NormalTok{()}
\CommentTok{\#\textgreater{} R version 4.3.3 (2024{-}02{-}29)}
\CommentTok{\#\textgreater{} Platform: x86\_64{-}pc{-}linux{-}gnu (64{-}bit)}
\CommentTok{\#\textgreater{} Running under: Ubuntu 24.04.5 LTS}
\CommentTok{\#\textgreater{} }
\CommentTok{\#\textgreater{} Matrix products: default}
\CommentTok{\#\textgreater{} BLAS:   /usr/lib/x86\_64{-}linux{-}gnu/blas/libblas.so.3.12.0 }
\CommentTok{\#\textgreater{} LAPACK: /usr/lib/x86\_64{-}linux{-}gnu/lapack/liblapack.so.3.12.0}
\CommentTok{\#\textgreater{} }
\CommentTok{\#\textgreater{} locale:}
\CommentTok{\#\textgreater{} [1] C}
\CommentTok{\#\textgreater{} }
\CommentTok{\#\textgreater{} time zone: Europe/London}
\CommentTok{\#\textgreater{} tzcode source: system (glibc)}
\CommentTok{\#\textgreater{} }
\CommentTok{\#\textgreater{} attached base packages:}
\CommentTok{\#\textgreater{} [1] stats     graphics  grDevices utils     datasets  methods   base     }
\CommentTok{\#\textgreater{} }
\CommentTok{\#\textgreater{} other attached packages:}
\CommentTok{\#\textgreater{} [1] detectseparation\_0.4.0 geer\_0.1.1            }
\CommentTok{\#\textgreater{} }
\CommentTok{\#\textgreater{} loaded via a namespace (and not attached):}
\CommentTok{\#\textgreater{}  [1] Matrix\_1.6{-}5             compiler\_4.3.3           Rcpp\_1.0.12             }
\CommentTok{\#\textgreater{}  [4] tinytex\_0.49             slam\_0.1{-}56              nleqslv\_3.3.5           }
\CommentTok{\#\textgreater{}  [7] ROI.plugin.lpsolve\_1.0{-}2 splines\_4.3.3            yaml\_2.3.8              }
\CommentTok{\#\textgreater{} [10] fastmap\_1.1.1            statmod\_1.5.0            lattice\_0.22{-}5          }
\CommentTok{\#\textgreater{} [13] coda\_0.19{-}4.1            TH.data\_1.1{-}2            generics\_0.1.3          }
\CommentTok{\#\textgreater{} [16] knitr\_1.45               MASS\_7.3{-}60.0.1          backports\_1.4.1         }
\CommentTok{\#\textgreater{} [19] checkmate\_2.3.4          ROI\_1.0{-}2                bookdown\_0.37           }
\CommentTok{\#\textgreater{} [22] nnet\_7.3{-}19              rlang\_1.1.3              enrichwith\_0.6          }
\CommentTok{\#\textgreater{} [25] multcomp\_1.4{-}25          xfun\_0.41                registry\_0.5{-}1          }
\CommentTok{\#\textgreater{} [28] estimability\_1.4.1       cli\_3.6.2                emmeans\_1.10.0          }
\CommentTok{\#\textgreater{} [31] digest\_0.6.34            grid\_4.3.3               lpSolveAPI\_5.5.2.0{-}17.15}
\CommentTok{\#\textgreater{} [34] mvtnorm\_1.2{-}4            xtable\_1.8{-}4             rticles\_0.27            }
\CommentTok{\#\textgreater{} [37] sandwich\_3.1{-}0           evaluate\_0.23            numDeriv\_2016.8{-}1.1     }
\CommentTok{\#\textgreater{} [40] codetools\_0.2{-}19         zoo\_1.8{-}12               survival\_3.5{-}8          }
\CommentTok{\#\textgreater{} [43] brglm2\_1.1.1             rmarkdown\_2.25           tools\_4.3.3             }
\CommentTok{\#\textgreater{} [46] htmltools\_0.5.7}
\end{Highlighting}
\end{Shaded}

\renewcommand\refname{References}
  \bibliography{bibliography-supp.bib}